\documentclass{aa}  

\usepackage{natbib}
\usepackage{graphicx}
\usepackage{txfonts}
\usepackage{hyperref}
\usepackage{color}
\usepackage{upquote} 
\usepackage{xspace}
\usepackage{ulem}  

\def\todo#1{}

\def\Gaia{\textit{Gaia}\xspace}
\def\gmag{\ensuremath{G}\xspace}
\def\gbp{\ensuremath{G_{\rm BP}}\xspace}
\def\grp{\ensuremath{G_{\rm RP}}\xspace}
\newcommand{\bpminrp}{\ensuremath{G_\mathrm{BP}-G_\mathrm{RP}}\xspace}

\begin{document} 

  \title{\Gaia Data Release 2: The first \Gaia catalogue of long-period variable candidates}
  \subtitle{}
  \titlerunning{\Gaia DR2. Long-period variable candidates}
  
  \author{N.~Mowlavi\inst{\ref{inst1},\ref{inst2}}\fnmsep\thanks{Corresponding author: N. Mowlavi
(\href{mailto:Nami.Mowlavi@unige.ch}{\tt Nami.Mowlavi@unige.ch})},
          I.~Lecoeur-Ta\"{i}bi\inst{\ref{inst2}},
          T.~Lebzelter\inst{\ref{inst3}},
          L.~Rimoldini\inst{\ref{inst2}},
          D.~Lorenz\inst{\ref{inst3}},
          M.~Audard\inst{\ref{inst1},\ref{inst2}},
          J.~De~Ridder\inst{\ref{inst8}},
          L.~Eyer\inst{\ref{inst1},\ref{inst2}},
          L.P.~Guy\inst{\ref{inst2},\ref{inst5}},
          B.~Holl\inst{\ref{inst1},\ref{inst2}},
          G.~Jevardat~de~Fombelle\inst{\ref{inst6}},
          O.~Marchal\inst{\ref{inst2},\ref{inst7}},
          K.~Nienartowicz\inst{\ref{inst6}},
          S.~Regibo\inst{\ref{inst8}},
          M.~Roelens\inst{\ref{inst1},\ref{inst2}},
          L.~M.~Sarro\inst{\ref{inst9}}
         }
         
  \authorrunning{Mowlavi et al.}
  
  \institute{Department of Astronomy, University of Geneva, Ch. des Maillettes 51, CH-1290 Versoix, Switzerland\label{inst1}
             \and
             Department of Astronomy, University of Geneva, Ch. d'Ecogia 16, CH-1290 Versoix, Switzerland\label{inst2}
             \and
             University of Vienna, Department of Astrophysics, Tuerkenschanz\-strasse 17, A1180 Vienna, Austria\label{inst3}
             \and
             Large Synoptic Survey Telescope, 950 N. Cherry Avenue, Tucson, AZ 85719, USA\label{inst5}
             \and
             SixSq, Rue du Bois-du-Lan 8, CH-1217 Geneva, Switzerland\label{inst6}
             \and
             GEPI, Observatoire de Paris, Universit{\'e} PSL, CNRS,  5 Place Jules Janssen, 92190 Meudon, France \label{inst7}
             \and
             Institute of Astronomy, KU Leuven, Celestijnenlaan 200D, 3001 Leuven, Belgium \label{inst8}
             \and
             Dpto. Inteligencia Artificial, UNED, c/ Juan del Rosal 16, 28040 Madrid, Spain \label{inst9}
            }

  \date{April 2018}

  \abstract
   {\Gaia Data Release 2 (DR2) provides a unique all-sky catalogue of 550\,737 variable stars, of which 151\,761 are long-period variable (LPV) candidates with \gmag variability amplitudes larger than 0.2~mag (5-95\% quantile range).
   About one-fifth of the LPV candidates are Mira candidates, the majority of the rest are semi-regular variable candidates.
   For each source, \gmag, \gbp , and \grp photometric time-series are published, together with some LPV-specific attributes for the subset of 89\,617 candidates with periods in \gmag longer than 60 days.
   }
   {We describe this first \Gaia catalogue of LPV candidates, give an overview of its content, and present various validation checks.
   }
   {Various samples of LPVs were used to validate the catalogue:
   a sample of  well-studied very bright LPVs with light curves from the American Association of Variable Star Observers that are partly contemporaneous with \Gaia light curves,
   a sample of \Gaia LPV candidates with good parallaxes,
   the All-Sky Automated Survey for Supernovae catalogue of LPVs,
   and the Optical Gravitational Lensing Experiment catalogues of LPVs towards the Magellanic Clouds and the Galactic bulge.
   }
   {The analyses of these samples show a good agreement between \Gaia DR2 and literature periods.
   The same is globally true for bolometric corrections of M-type stars.
   The main contaminant of our DR2 catalogue comes from young stellar objects (YSOs) in the solar vicinity (within $\sim$1~kpc), although their number in the whole catalogue is only at the percent level.
   A cautionary note is provided about parallax-dependent LPV attributes published in the catalogue.
   }
   {This first \Gaia catalogue of LPVs approximately doubles the number of known LPVs with amplitudes larger than 0.2~mag, despite the conservative candidate selection criteria that prioritise low contamination over high completeness, and despite the limited DR2 time coverage compared to the long periods characteristic of LPVs.
   It also contains a small set of YSO candidates, which offers the serendipitous opportunity to study these objects at an early stage of the \Gaia data releases.
   }


\maketitle

\section{Introduction}

The late evolutionary stages of low- and intermediate-mass stars are accompanied by large-amplitude and long-period variations in radius, brightness, and temperature of the star.
Accordingly, this group of variables has been named {\it long-period variables} (LPVs).
Because of their large variability amplitudes, in particular in the visual range, they have been known and studied for a long time.
Classically, two types of LPVs have been distinguished: Miras, with large amplitude and long periods (names after Mira, their prototype star), and semi-regular variables (SRVs) that show smaller amplitudes (a few tenths of a magnitude) and a wider range of periods. 
The light variations of SRVs do not follow the strict periodicity seen in Miras.
In the 1990s, it became apparent that fainter red giants are variable as well down to K giants \citep{Grenon93,EyerGrenonFalin_etal94,EdmondsGilliland96,JorissenMowlaviSterken_etal97}.
The advent of surveys originally dedicated to the search of dark compact objects in the Galactic halo provided evidence for a continuous sequence of pulsating stars from Miras down to SRVs and red giant branch stars \citep[e.g.][]{1999IAUS..191..151W,TaburBeddingKiss_etal10}.

\begin{figure}
\centering
\includegraphics[width=\hsize]{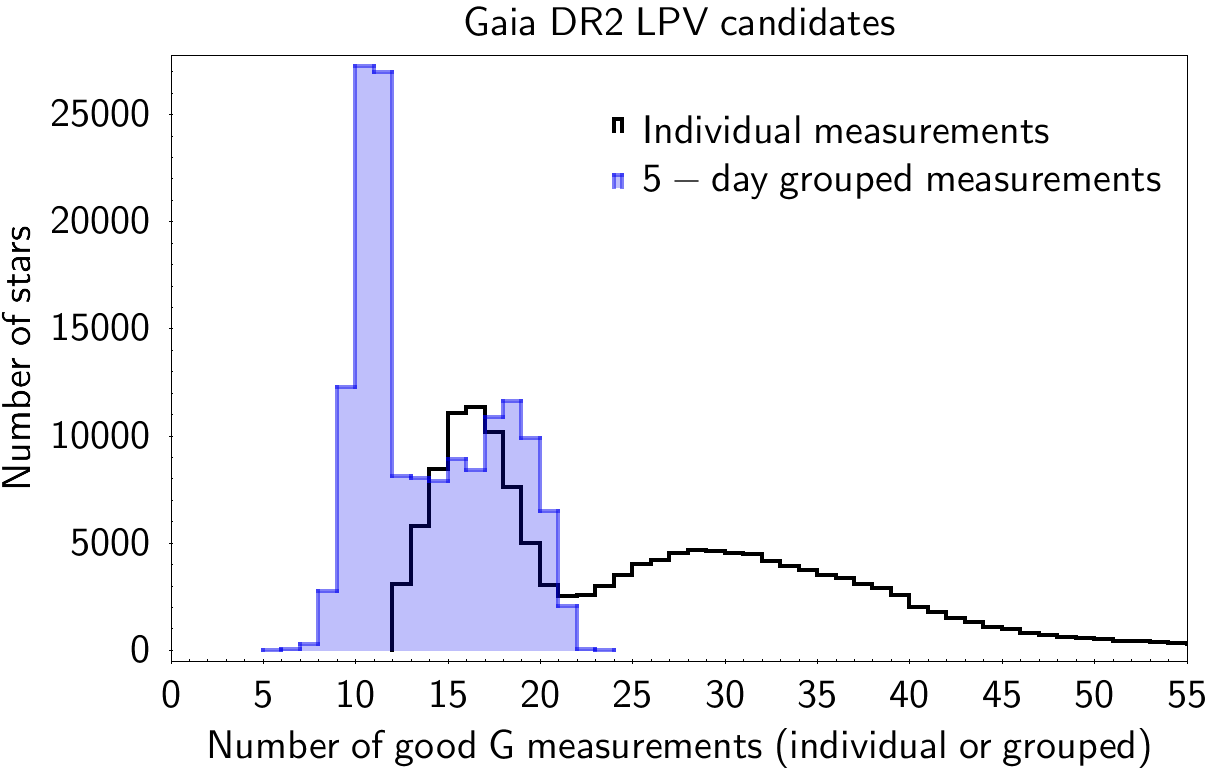}
\caption{Histogram of the total number of measurements (black histogram) and of the number of five-day clumped measurements (filled blue histogram) in the cleaned \gmag-band time-series of all \Gaia DR2 LPV candidates (see text).
        }
\label{Fig:histoNumMeasurements}
\end{figure}

LPVs are primarily stars on the upper giant branch. 
During this phase, nucleosynthesis, mixing, and mass-loss events take place that not only conclude the evolution of these individual stars before they end their lives as white dwarfs, but also enrich the interstellar medium with heavy elements.
Stellar variability plays a key role in this context, since periodic levitation of the atmosphere enhances mass loss 
\citep{2018A&ARv..26....1H}.

Owing to their high intrinsic brightness, LPVs can be detected in an extensive volume of space. 
As a consequence, large catalogues of LPVs have been produced as by-products of various sky surveys. 
The first major catalogue in this context was produced by the MACHO search for massive compact halo objects \citep{1993ASPC...43..291A} in the direction of the Magellanic Clouds. 
This catalogue of LPVs in the Clouds significantly improved our understanding of LPV pulsation through the detection of several parallel pulsation sequences on the upper giant branch.
We refer to the ground-breaking papers by \citet{1999IAUS..191..151W} and \citet{2017ApJ...847..139T} in this context for a further description of these sequences and the related pulsation modes and properties.
This success triggered further studies using the catalogues produced by similar surveys covering the Magellanic Clouds, the Galactic bulge, and further stellar systems.
One of the most influential catalogues in this context was produced by the Optical Gravitational Lensing Experiment (OGLE) team \citep{2009AcA....59..239S}. 
These catalogues were extended by the correlation with infrared colours \citep[for example][]{2004MNRAS.353..705I,2010ApJ...723.1195R}, forming extended databases that allowed studying key aspects such as the connection between pulsation properties and stellar chemistry or mass loss. 

The \Gaia all-sky survey is expected to add a major contribution to the study of the populations of LPVs during its five-year nominal mission plus extensions (mission extension has already been approved, to date, until the end of 2020), in particular, by covering Galactic field and halo stars. 
These groups of stars can only be studied in a way comparable to the Magellanic Cloud surveys if the distance is obtained for the stars. 
Furthermore, the high level of completeness of LPVs expected from the \Gaia survey will offer the opportunity to study, among other subjects, the frequencies of various groups of LPVs in the extended solar neighbourhood and other parts of the Galaxy.
Thus, the \Gaia database of LPVs will be unique and will provide a great step forward in understanding these variables. 

\Gaia Data Release 2 (DR2) offers the first opportunity to provide a \Gaia catalogue of LPV candidates to the scientific community, based on \Gaia data collected over a time span of 22 months.
However, knowing that LPVs have periods that can exceed 1000 days, and given the sparsity of \Gaia measurements due to the spacecraft scanning law, this first \Gaia catalogue of LPV candidates can only be limited.
Therefore, we have placed priority on ensuring the lowest possible level of contamination, without targeting completeness.
We have thus limited the catalogue to Mira and SRVs with variability amplitudes larger than 0.2~mag in the \Gaia \gmag band.
Small-amplitude red giant variables, detected as a large group in the OGLE database \citep[][]{2009AcA....59..239S}, were excluded at this stage.

\begin{figure}
\centering
\includegraphics[width=\hsize,height=0.5\hsize]{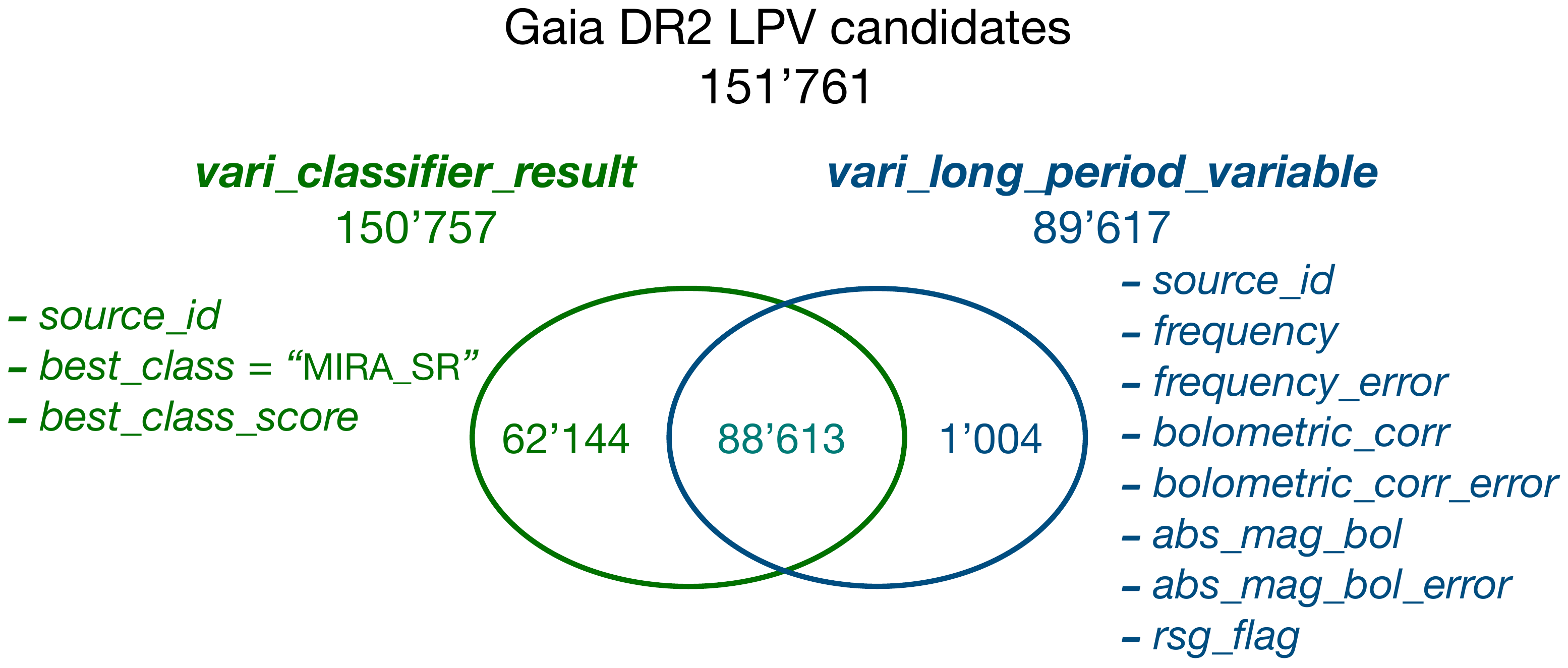}
\caption{Venn diagram of the repartition of the \Gaia DR2 LPV candidates in the \texttt{vari\_classifier\_result} (left in the figure) and \texttt{vari\_long\_period\_variable} (right in the figure) tables published in \Gaia DR2.
         The data fields available in each table are listed in the figure below the corresponding table name.
        }
\label{Fig:numLPVs}
\end{figure}

\begin{table*}
\caption{Data fields available in the \Gaia DR2 tables related to LPVs.}
\centering
\begin{tabular}{l l c}
\hline\hline
Data field name & Description & Affected by \\
                &             & parallax? \\
\hline\hline
\multicolumn{3}{l}{Table \texttt{vari\_classifier\_result}} \\
\hline
\texttt{source\_id}
  & Unique source identifier of the LPV candidate
  & No \\
\texttt{best\_class}
  & Equals \texttt{'MIRA\_SR'} for all LPV candidates
  & No \\
\texttt{best\_class\_score}
  & Confidence (between 0 and 1) of the classifier in the identification of an LPV
  & No \\
\hline\hline
\multicolumn{3}{l}{Table \texttt{vari\_classifier\_result}} \\
\hline
\texttt{source\_id}
  & Unique source identifier of the LPV candidate
  & No \\
\texttt{frequency}
  & Frequency of the LPV [day$^{-1}$]
  & No \\
\texttt{frequency\_error}
  & Uncertainty on the frequency [day$^{-1}$]
  & No \\
\texttt{bolometric\_corr}
  & Bolometric correction [mag]
  & ~~~Yes\tablefootmark{(a)} \\
\texttt{bolometric\_corr\_error}
  & Uncertainty on the bolometric correction [mag]
  & ~~~Yes\tablefootmark{(a)} \\
\texttt{abs\_mag\_bol}
  & Absolute bolometric magnitude [mag]
  & Yes \\
\texttt{abs\_mag\_bol\_error}
  & Uncertainty on the absolute bolometric magnitude  [mag]
  & Yes \\
\texttt{rsg\_flag}
  & Red supergiant flag [boolean]
  & Yes \\
\hline
\end{tabular}
\tablefoot{\tablefoottext{a}{Bolometric corrections are affected by parallax for stars identified as red super giants, see Appendix~\ref{Sect:SOS_BCs}.}
}
\label{Tab:dataFields}
\end{table*}

A general description of \Gaia DR2 is provided in several separate papers produced by the \Gaia consortium \citep[see e.g.][and references therein]{DR2-DPACP-14,DR2-DPACP-36,DR2-DPACP-40}, with an overview of the DR2 processing of variable stars given in \citet{Holl_etal18}.
In this paper, we focus on the DR2 catalogue of LPV candidates.

The catalogue data are first described in Sect.~\ref{Sect:catalog}, and an overview of the catalgoue content is presented in Sect.~\ref{Sect:overview}.
Analyses of different subsets of the catalogue are then performed in Sects.~\ref{Sect:individualLPVs} to \ref{Sect:OGLE} for validation and completeness/contamination estimation purposes.
We start in Sect.~\ref{Sect:individualLPVs} with the analysis of known individual LPVs that are well studied in the literature, including sources for which light curves from the American Association of Variable Star Observers (AAVSO) have a time interval in common with \Gaia light curves.
This sample of bright sources will allow us in particular to check the availability of bright LPVs in the catalogue and to validate the periods and bolometric corrections (BCs) published in DR2.
Section~\ref{Sect:goodParallaxes} analyses a sample of LPV candidates that have good \Gaia parallaxes.
Their positioning in the observational Hertzsprung-Russell (HR) diagram will enable identifying non-red giant contaminants in the sample.
Sections~\ref{Sect:ASAS} and \ref{Sect:OGLE} then compare the DR2 catalogue with specific LPV catalogues from ground-based large-scale surveys.
Section~\ref{Sect:ASAS} compares the DR2 catalogue with the recently published all-sky LPV catalogue from the All-Sky Automated Survey for Supernovae (ASAS\_SN) survey \citep{Jayasinghe_etal18}.
The \Gaia-derived periods published in DR2 are, in particular, compared to the ASAS\_SN-derived periods.
Section~\ref{Sect:OGLE} then checks the DR2 LPV candidates towards the Galactic bulge and the Magellanic Clouds using the OGLE-III LPV catalogues \citep{2009AcA....59..239S,2011AcA....61..217S,2013AcA....63...21S}.
In addition to validating the periods, the comparison will allow us to estimate the level of completeness of our \Gaia DR2 catalogue of LPV candidates.
The main results of the different validation sections are summarised in Sect.~\ref{Sect:summary}.
Finally, Sect.~\ref{Sect:conclusions} lists the main limitations of the current DR2 catalogue, and concludes with the improvements expected for LPVs in future \Gaia data releases.

Three appendices complete the main body of the paper.
Appendix~\ref{Sect:SOS} details the procedure implemented within the \Gaia variability pipeline of the DPAC for selecting LPV candidates and for computing the LPV-specific attributes published in DR2.
Appendix~\ref{Sect:catalogRetrieval} provides examples of queries, written in the Astronomical Data Query Langage (ADQL), to be submitted to the web interface of the Gaia DR2 archive in order to retrieve LPV-related DR2 data.
Finally, Appendix~\ref{Sect:parallaxes} explains why the parallaxes used in the preparation of the DR2 catalogue were not the final parallaxes, which in turn necessitated recomputing some of the parallax-dependent LPV attributes published in DR2 (the procedures for recomputing these attributes are provided in Appendix~\ref{Sect:SOS}).

\section{Gaia DR2 catalogue of LPV candidates}
\label{Sect:catalog}

\Gaia DR 2 is based on data collected between July 25, 2014 (10:30 UTC), and May 23, 2016 (11:35 UTC), spanning an observation duration of 668 days.
In this time span, the mean number of \gmag-band measurements per source is 26 for the 151\,761 LPV candidates published in DR2.
The observations are, however, not evenly distributed over the observation duration.
\Gaia has two fields of view, leading to 1h\,46min and 4h\,14min pairs of observations that may be repeated several times because of the 6-hour rotation period of the spacecraft \citep[see Sect.~A.4.1 of][for more details on the \Gaia scanning law]{DR1-DPACP-15}.
For LPVs, which have periods longer than tens of days and up to several thousand days, these multiple successive observations contribute to a unique point in phase, thereby reducing their relevance as multiple observations for the characterisation of their light curves.
The number of times \Gaia returns to a given region of the sky is more informative because of its 63-day precession period combined to its 6-hour rotation period.
A relevant information for LPVs is then the number of groups of clumped observations present in a light curve.
Taking a clumping time interval of 5 days, there are, in the mean, 13.4 groups of observations per \gmag-band light curve in the DR2 set of LPV candidates.
The distribution of the number of measurements is shown in Fig.~\ref{Fig:histoNumMeasurements}.

The observation duration of DR2 was shorter than two years. Combined with the relatively low number of clumped observations per light curve, this means that it is impossible to aim at a high level of LPV completeness in \Gaia DR2.
We instead focused in this release on limiting the contamination by non-LPV stars.
The resulting catalogue nevertheless contains more than 150\,000 LPV candidates, which is about twice the number of LPVs currently known with amplitudes larger than 0.2~mag, as shown in this paper.
This shows the great potential of \Gaia for the detection and characterisation of a still-larger set of LPVs in future data releases.

The data processing steps of the variability pipeline leading to the production of this catalogue are described in Appendix~\ref{Sect:SOS}.
In particular, the filtering criteria applied for LPV selection to minimise contamination are listed in Sect.~\ref{Sect:SOS_filters}.

In this section, we detail the content of the catalogue (Sect.~\ref{Sect:catalogContent}) and provide our recommendations for its usage (Sect.~\ref{Sect:catalogUsage}).
Procedures for retrieving the data are provided in Appendix~\ref{Sect:catalogRetrieval}.

\subsection{Catalogue content}
\label{Sect:catalogContent}

Long-period variable candidates are published in \Gaia DR2 in two tables, \texttt{vari\_classifier\_result} and \texttt{vari\_long\_period\_variable}.
As described in \citet{Holl_etal18}, the first table contains all candidates identified by the \texttt{'nTransits:2+'} classifier \citep[the 'geq2' path of processing described in][]{Holl_etal18}, and the second table contains all candidates for which LPV-specific attributes have been computed.

The number of LPV candidates in the two tables is summarised in Fig.~\ref{Fig:numLPVs}.
Table \texttt{vari\_long\_period\_variable} represents a subset of the table \texttt{vari\_classifier\_result}, except for about 1\% of sources that were identified as LPV candidates through a parallel classification procedure \citep[the 'geq20' path of processing described in][]{Holl_etal18}.
All LPV candidates in \texttt{vari\_long\_period\_variable} have periods longer than 60 days, which was the defining selection criterion to populate the table.

The data fields available in \texttt{vari\_classifier\_result} and \texttt{vari\_long\_period\_variable} are summarised in Fig.~\ref{Fig:numLPVs} and described in Table~\ref{Tab:dataFields}.
The last column of the table indicates whether the computation of a given LPV attribute required knowing the parallax.
Since the variability pipeline was run at the same time as the last version of the parallaxes was computed within the DPAC, the variability results published in DR2 relied on a preliminary version of the parallaxes and are not compliant with the published parallaxes.
Therefore, the variability attributes that depend on the parallax needed to be recomputed (see Appendix~\ref{Sect:parallaxes}).
They are flagged in the last column of Table~\ref{Tab:dataFields}, and the prescription for recomputing them is given in Sect.~\ref{Sect:catalogUsage}.


\subsection{Catalogue usage}
\label{Sect:catalogUsage}

The \Gaia DR2 catalogue provides a show case to the astronomical community of the unique potential of \Gaia for all-sky LPV studies.
Being provided at a relatively early stage of the \Gaia data releases (the second out of the expected four data releases of the nominal mission, and covering less than two years of \Gaia data), it necessarily contains limitations inherent to such early catalogues.

We describe here the points to be kept in mind
by any user of the DR2 catalogue of LPV candidates, listed by data fields published in the catalogue.
These remarks come in addition to those applicable to \Gaia DR2 products in general \citep[see][and references therein]{DR2-DPACP-36}.

\paragraph{Best class.} The all-sky classification of high-amplitude pulsating stars, made in the variability pipeline of the \Gaia DPAC, is described in \citet{Rimoldini_etal18}.
The classification of LPVs has mainly been driven by their red colours and their slow light-variability rates \citep[mainly the attributes \texttt{\small{BP\_MINUS\_RP\_COLOUR}}, \texttt{\small{G\_MINUS\_RP\_COLOUR}} and  \texttt{\small{MEDIAN\_RANGE\_HALFDAY\_TO\_ALL}} described in the \Gaia DR2 Documentation, see][]{Rimoldini18}.
From this classification, the subset of LPV candidates with \gmag-band amplitudes larger than 0.2~mag has been extracted for publication in DR2, after several filters were applied to reduce the level of contamination by non-LPV variables (see Appendix~\ref{Sect:SOS}).
Contamination is, however, still present, although at a low level.
An example is given in Sect.~\ref{Sect:goodParallaxes} with nearby young stellar objects (YSOs), which can also display large variability amplitudes on long timescales in addition to their short-timescale variability that is not always caught with the \Gaia scanning law.

\paragraph{Best class scores.}
The attribute \texttt{best\_class\_score} is a quantity between 0 and 1
provided by the classification pipeline to estimate the confidence of the classifier in the identification of various variability types.
For LPVs, which are mainly identified from their colour and absence of prominent variability on short timescales (see above), the value of \texttt{best\_class\_score} is less relevant.
We therefore do not use it
in the analyses presented in this paper.

\paragraph{Frequencies.} All LPV candidates with periods greater than 60~d have LPV-specific attributes published in DR2.
Comparison with published periods of individual, ASAS\_SN, and OGLE-III LPVs shows good agreement over the whole period range (see Table~\ref{Tab:sampleBrightLPVs} in Sect.~\ref{Sect:individualLPVs_wellStudiedCases}, Fig.~\ref{Fig:periodsGaiaAsas} in Sect.~\ref{Sect:ASAS}, and Figs.~\ref{Fig:periodsGaiaVsOgle_Clouds}-\ref{Fig:periodsGaiaVsOgle_Bulge} in Sect.~\ref{Sect:OGLE}, respectively).
Periods below 100~d, however, may suffer from more severe aliasing issues in the frequencygram than at longer periods, and caution must be taken when using them.
This also motivated the setting of a lower cutoff at 60~d (see Appendix~\ref{Sect:SOS_period}) for LPV candidates to populate the \texttt{vari\_long\_period\_variable} table.
It must also be stressed that multi-periodicity was not tackled in DR2, even though many LPVs are known to be multi-periodic, and that long secondary periods were beyond the reach of this release given the DR2 observation time span.

\paragraph{Bolometric corrections.} The \gmag-band BCs are computed using the formula given in Sect.~\ref{Sect:SOS_BCs} of Appendix~\ref{Sect:SOS}, using mean values of \gbp and \grp magnitudes.
In brief, the relations are based on the paper by \citet{2010A&A...524A..87K} and were adapted to the \Gaia photometric system.
For M-type red giants with amplitudes below 3 mag, the apparent bolometric magnitudes $m_{\rm bol}$ derived with the BCs published in DR2 show a satisfactory agreement, within a few tenths of a magnitude, with the bolometric magnitudes derived from photometric multi-band fits (see Sect.~\ref{Sect:individualLPVs_BCs}).
The following points have nevertheless to be kept in mind:
\begin{itemize}
    \item The relations used in DR2 are only valid for M-type stars, since no distinction between M-, S- and C-type stars could be made based on the \Gaia data available for DR2.
    
    \vskip 2mm
    \item Specific BC values have been adopted for large-amplitude red giants and for red supergiant stars (see Sect~\ref{Sect:SOS_BCs} in Appendix~\ref{Sect:SOS}).
    Variability over the light cycle of large-amplitude red giants and effects due to their atmospheric structure and to possible circumstellar material have to be taken into account when a larger database of LPV candidates is available.
    
    \vskip 2mm
    \item The BCs do not take into account extinction, which was not available at the time of our variability processing.
    
    \vskip 2mm
    \item If the star suffers from a misclassification between red supergiant (RSG) and asymptotic giant branch (AGB) due to the parallax issue described below (see next paragraph on absolute bolometric magnitudes), its BC must be updated using the published parallaxes, with the procedure described in Sect.~\ref{Sect:SOS_BCs} of Appendix~\ref{Sect:SOS}.
\end{itemize}

Finally, it must be noted that a small fraction of the LPV candidates published in DR2 are YSO contaminants (see Sect.~\ref{Sect:individualLPVs}).
The BCs published in the catalogue are not adapted for these objects.

\begin{figure*}
\centering
\includegraphics[width=\hsize]{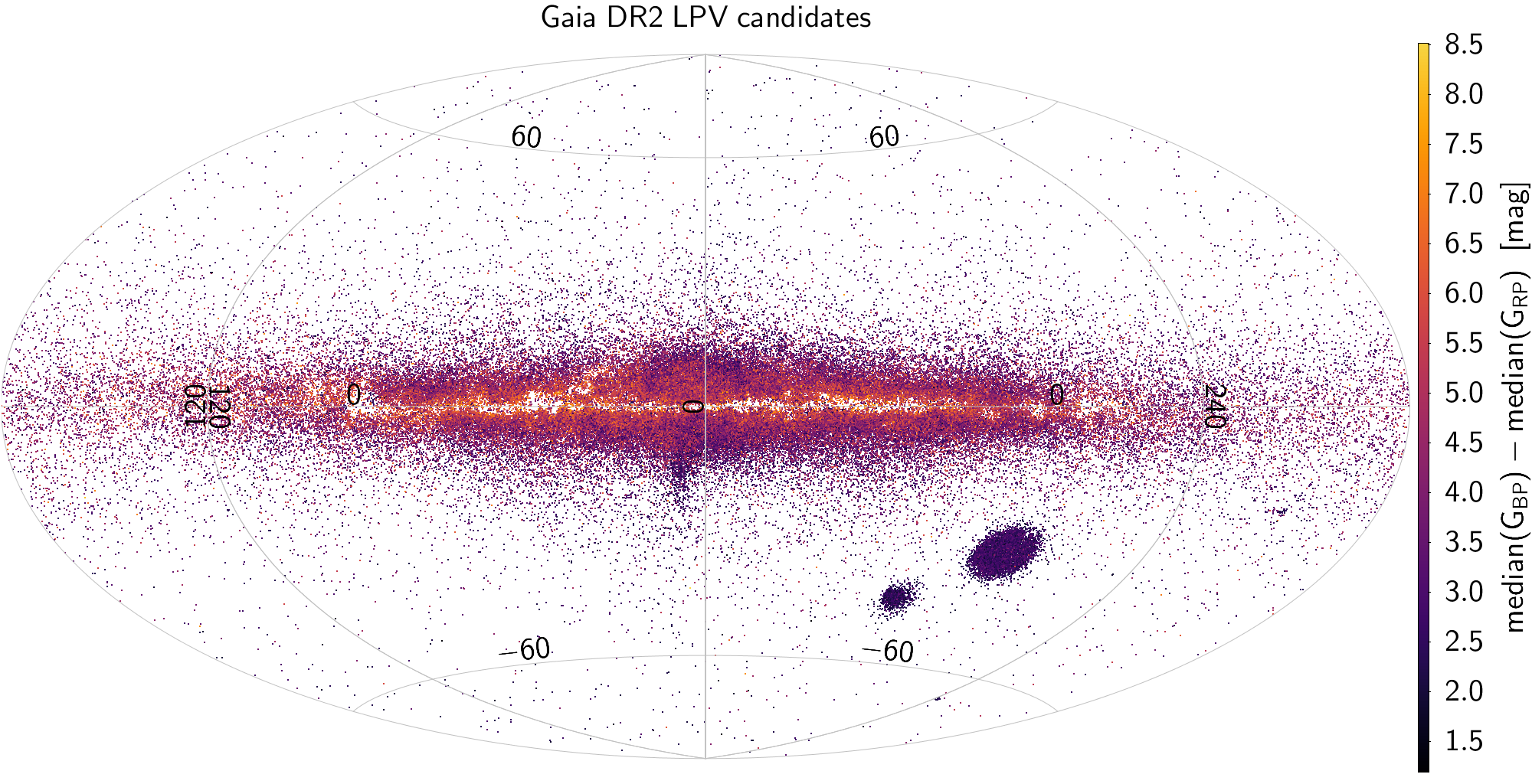}
\caption{Sky distribution in Galactic coordinates of all LPV candidates published in \Gaia DR2.
         Each star is drawn with a purple-red-yellow colour-grading scale according to the $\mathrm{median}(\gbp)-\mathrm{median}(\grp)$ colour scale shown on the right of the figure.
        }
\label{Fig:sky_all}
\end{figure*}

\begin{figure}
\centering
\includegraphics[width=\hsize]{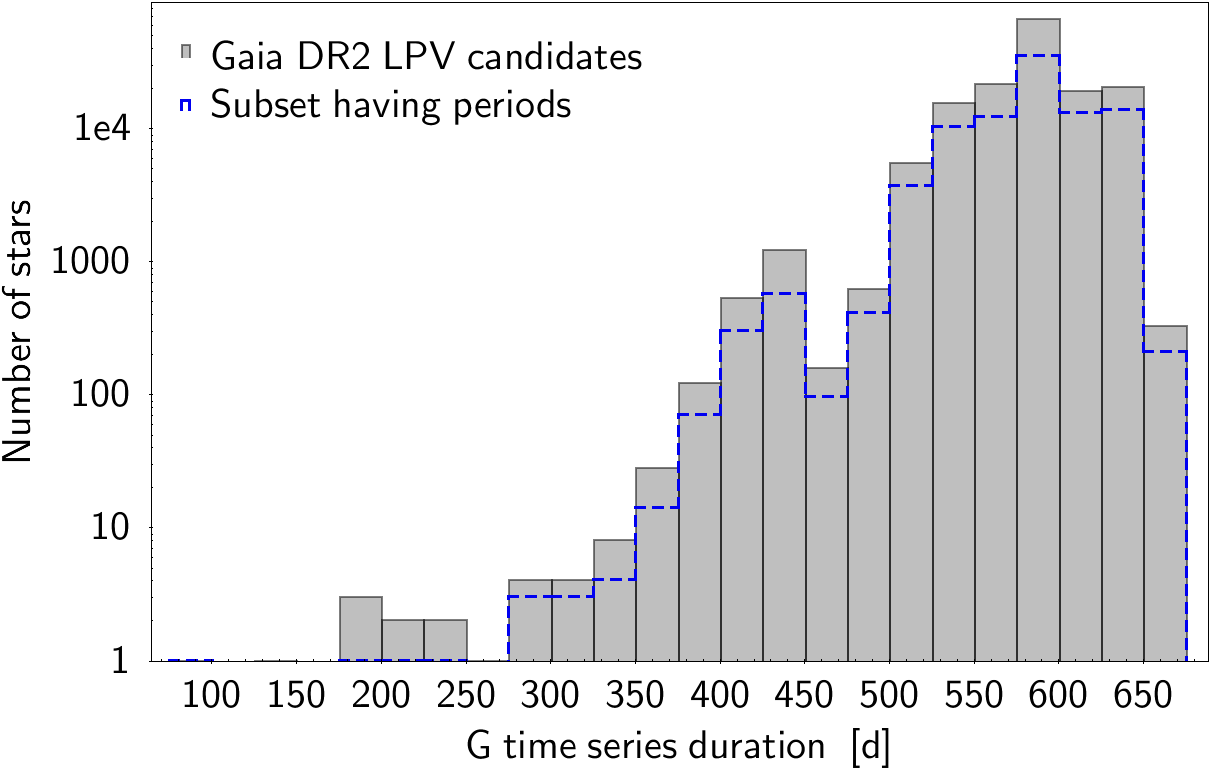}
\caption{Histogram of $\gmag$ time-series durations, in days, of all \Gaia DR2 LPV candidates.
         The dashed blue histogram represents the subset of the catalogue published with periods for the sources.
        }
\label{Fig:histoDuration}
\end{figure}

\begin{figure}
\centering
\includegraphics[width=\hsize]{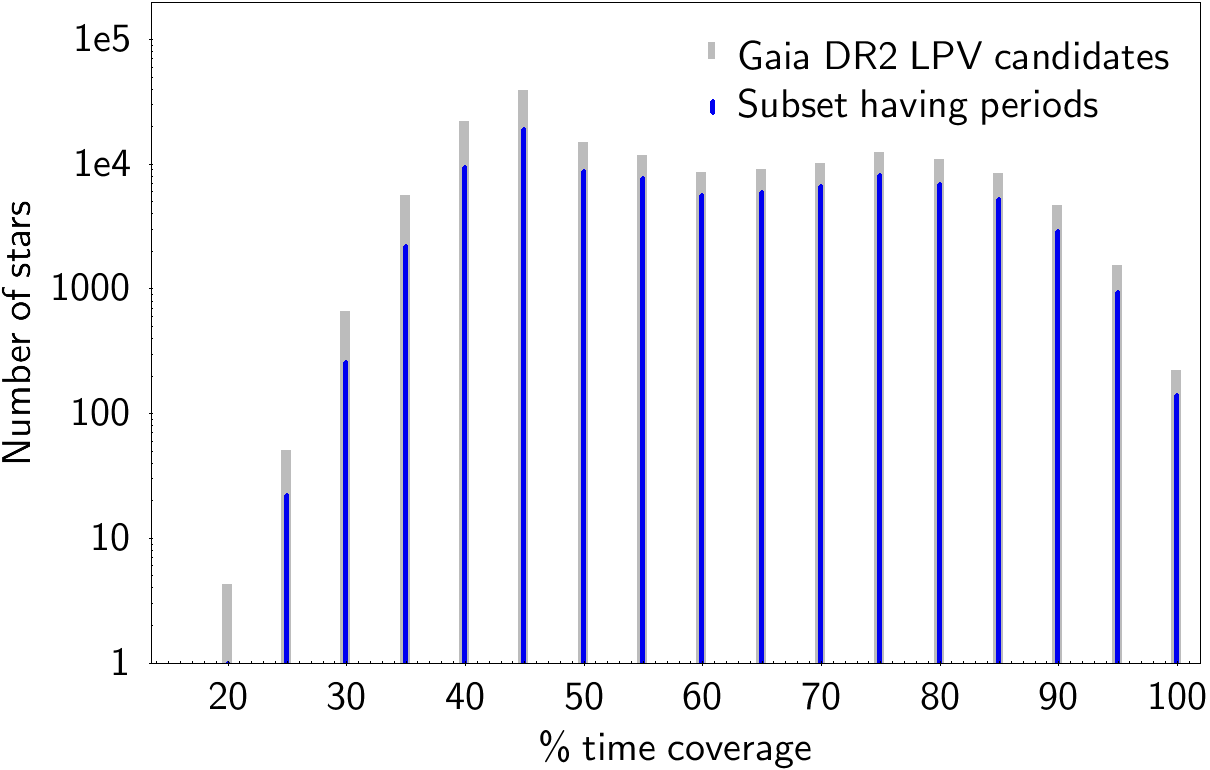}
\caption{Histogram of the time coverage (see Sect.~\ref{Sect:overview_lcs}), in percentage, of measurements in the cleaned \gmag time-series of all \Gaia DR2 LPV candidates (grey bars) and of the subset with published periods (blue bars).
        }
\label{Fig:histoTimeCoverage}
\end{figure}

\paragraph{Absolute bolometric magnitudes (and the parallax issue).}
Reliable absolute bolometric magnitudes can be computed only for the subset of LPV candidates with good parallaxes.
For parallax uncertainties satisfying $\varpi/\epsilon(\varpi)>10$, this concerns 3093 sources based on the parallaxes published in DR2, that is, 2\% of all DR2 LPV candidates, or 3.5\% of the DR2 candidates with published LPV attributes (see Sect.~\ref{Sect:goodParallaxes}).
The bolometric magnitudes published in DR2 inherit the limitations affecting BCs.
More important, however, is the fact that they are based on preliminary parallaxes that were available within the \Gaia DPAC at the time of variability processing.
These preliminary parallaxes have been much improved within the consortium by the time of DR2 data finalisation (see Appendix~\ref{Sect:parallaxes}), but were not made available to the variability processing pipeline.
As a result, the bolometric magnitudes published in the DR2 catalogue of LPV candidates do not comply with the published DR2 parallaxes, and  they must be recomputed with the prescription given in Sect.~\ref{Sect:SOS_Mbol} of Appendix~\ref{Sect:SOS}.
All results shown in this paper are based on bolometric magnitudes recomputed in this way.

\paragraph{Is RSG flag.} Red supergiant stars are identified from their location in the $P-M_\mathrm{bol}$ diagram, as described in Sect.~\ref{Sect:SOS_Mbol} of Appendix~\ref{Sect:SOS}.
This therefore inherits the limitations affecting the published values of $M_\mathrm{bol}$.
The RSG flag must therefore be re-evaluated based on the published DR2 parallaxes, according to the prescription given in Sect.~\ref{Sect:SOS_Mbol}.
Sources flagged as RSG stars in DR2 may turn out not to be RSG stars, and vice versa, sources flagged as standard red giant candidates may turn out to be RSG stars.

\paragraph{Light curves} All light curves haven been cleaned at the start of the variability processing pipeline from outliers, from measurements with too large uncertainties, and from duplicate measurements.
The cleaning procedure is described in \citet[][Sect.~4.1]{Holl_etal18}.
All time-series quantities described in this paper and/or published in DR2, such as the mean and median values, are derived from the cleaned time-series,
but all data points available in the original time-series are published in DR2.
Consequently, all measurements are shown in the figures displaying \Gaia light curves in this paper, and the disregarded measurements are identified by magenta crosses.

\section{Overview of the \Gaia all-sky LPV survey}
\label{Sect:overview}

We present in this section an overview of the all-sky \Gaia DR2 catalogue of LPV candidates, together with light-curve examples.
Validations on catalogue subsets according to distance to the Sun or stellar populations are presented in the next sections.

\subsection{Sky distribution overview}
\label{Sect:overview_sky}

The sky distribution of all DR2 LPV candidates is shown in Fig.~\ref{Fig:sky_all}, colour-coded with the $\gbp-\grp$ colours\footnote{
All \gmag, \gbp , and \grp values shown in this paper are taken equal to the median values of the respective time-series, unless otherwise indicated.
}
of the stars according to the colour-scale on the right of the figure.
The Galactic disc, bulge, and halo are clearly visible, as well as the extra-galactic populations of the Magellanic Clouds (the two concentration areas below the Galactic plane at Galactic longitudes of about 285$^o$ for the Large Magellanic Cloud (LMC) and 303$^o$ for the Small Magellanic Cloud (SMC)) and of the Sagittarius dwarf galaxy (elongated tail below the Galactic bulge).
The reddest LPV candidates (yellowish colour in the figure) are seen to be distributed in the Galactic plane, where dust leads to high extinctions.

\begin{figure}
\centering
\includegraphics[width=\hsize]{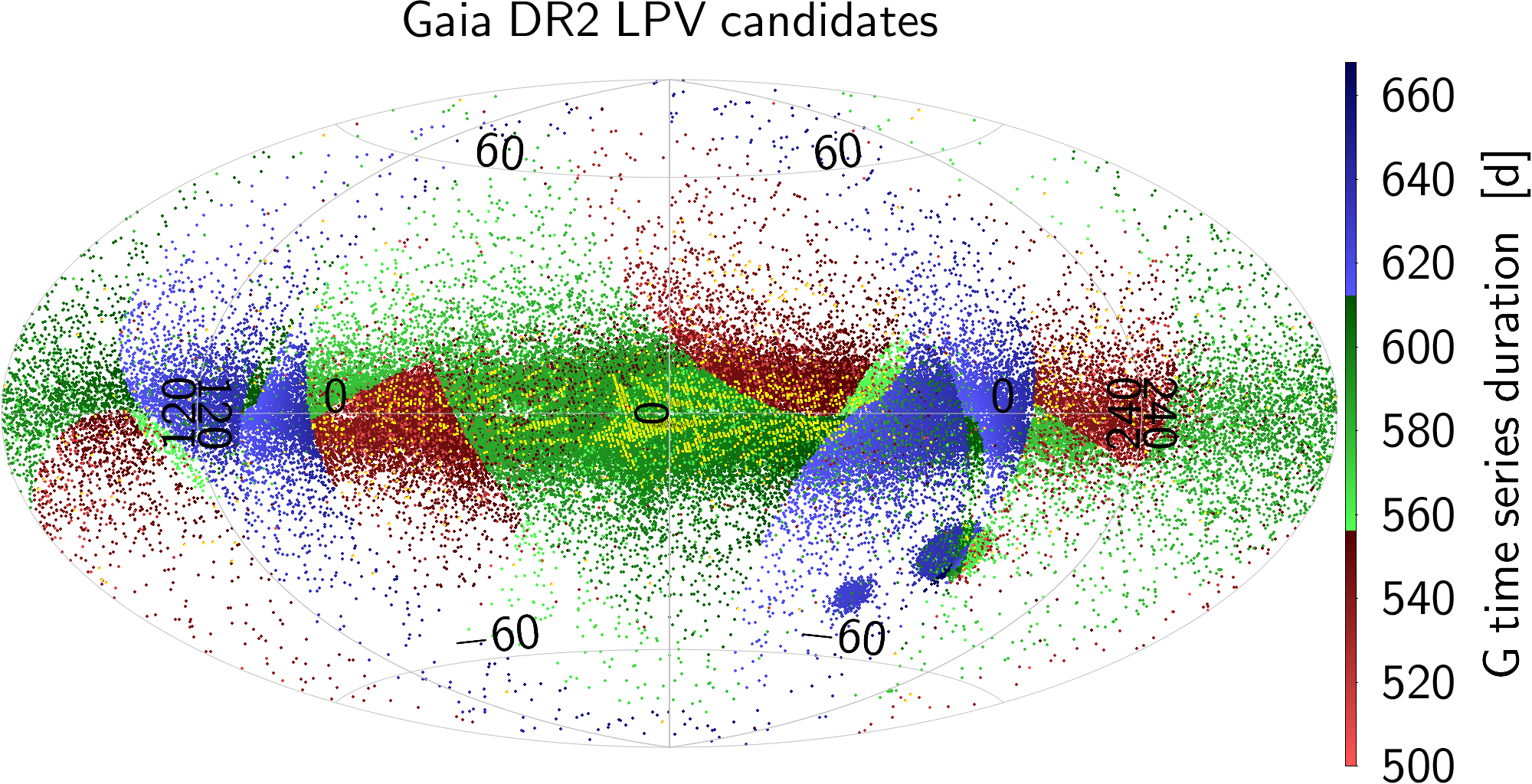}
\caption{Sky distribution of the \gmag-band time-series durations of each DR2 LPV candidate, according to the colour scale shown on the right of the figure.
         time-series with durations shorter than 500~d are plotted in yellow (they are mostly located in the bulge region).
        }
\label{Fig:skyDuration}
\end{figure}

\begin{figure}
\centering
\includegraphics[width=\hsize]{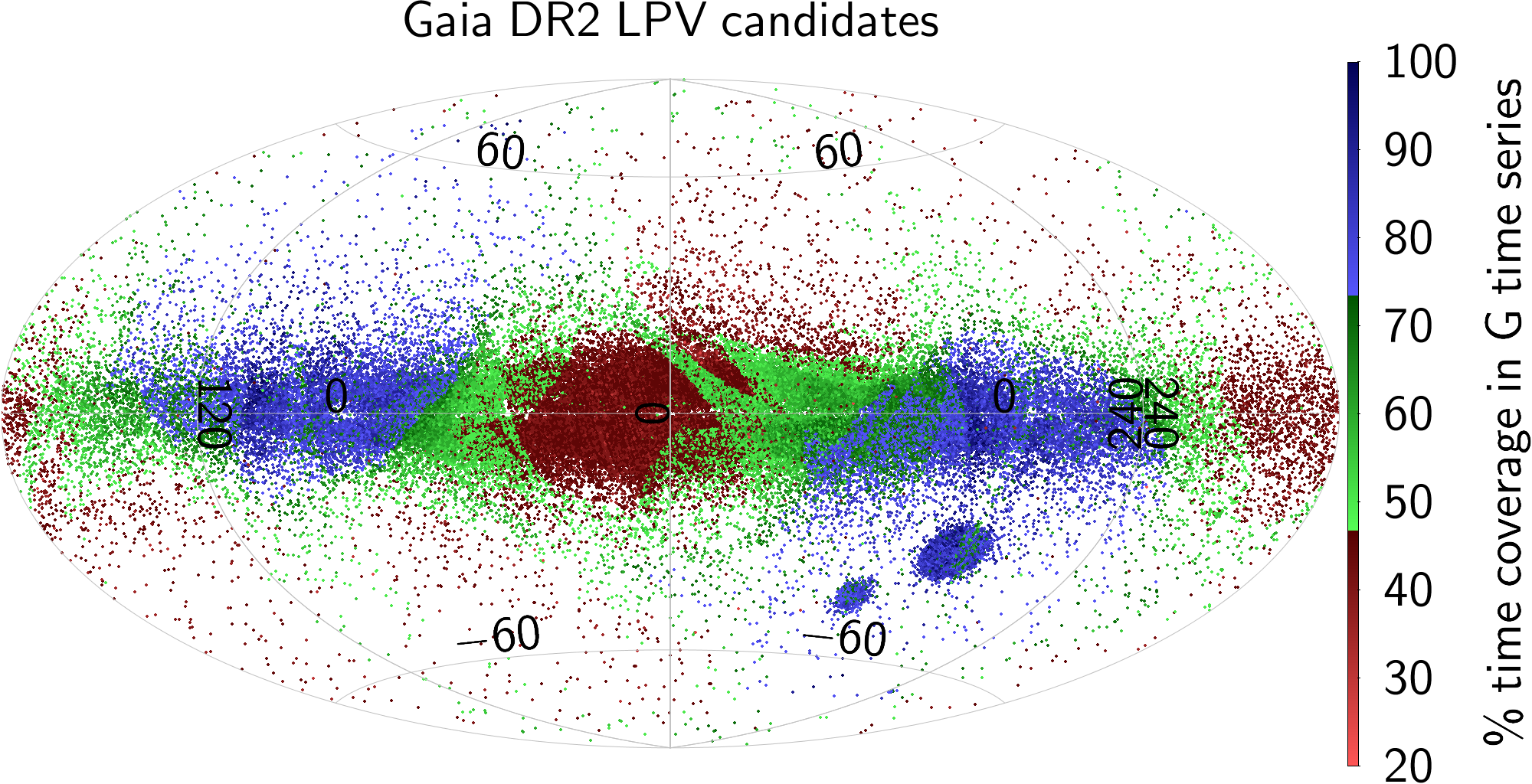}
\caption{Sky distribution of the time coverage percentage (see Sect.~\ref{Sect:overview_lcs}) of \gmag-band cleaned measurements within the time-series duration of each LPV candidate, according to the colour scale shown on the right of the figure.
        }
\label{Fig:skyTimeCoverage}
\end{figure}

\begin{figure}
\centering
\includegraphics[width=0.94\hsize]{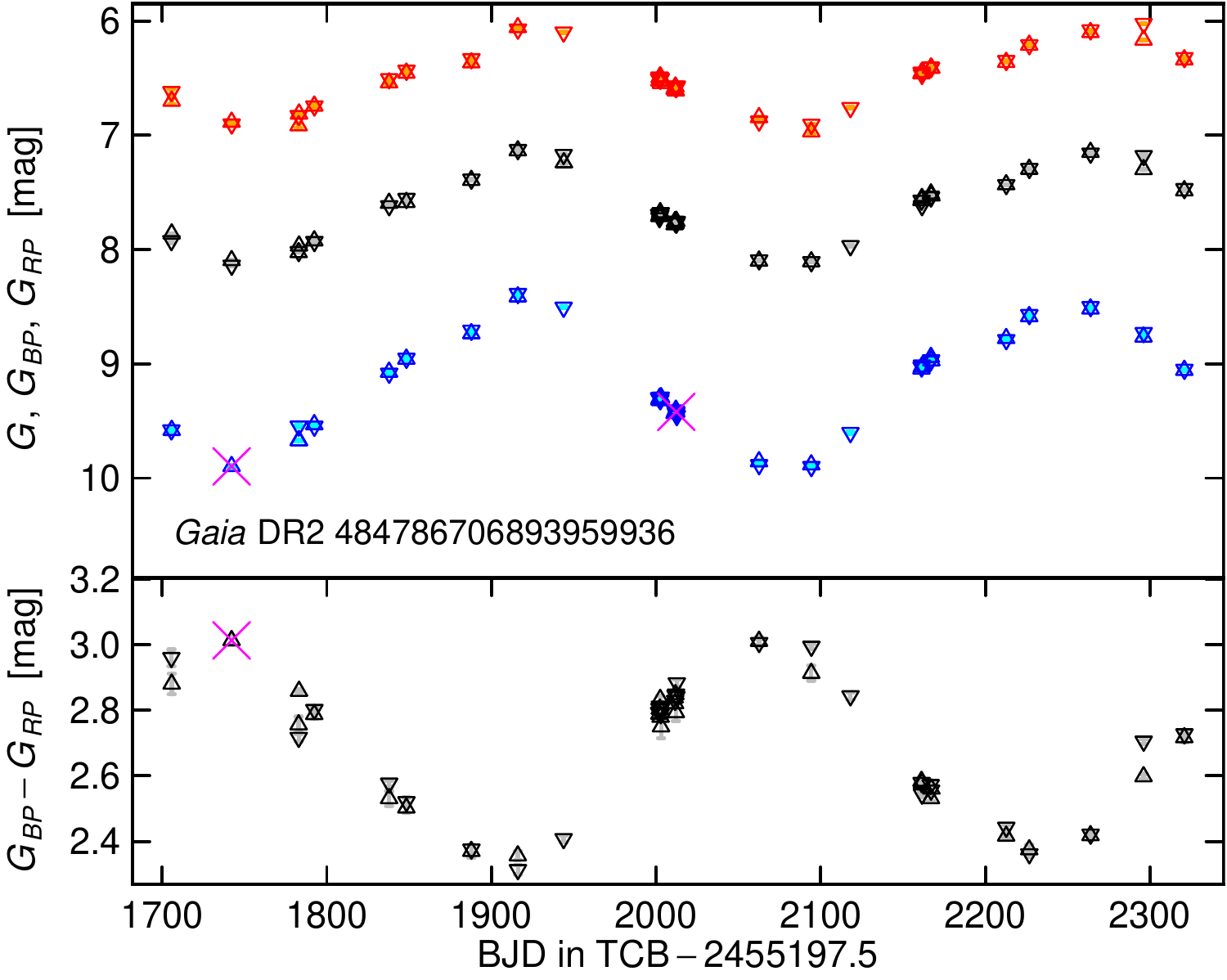}
\includegraphics[width=0.94\hsize]{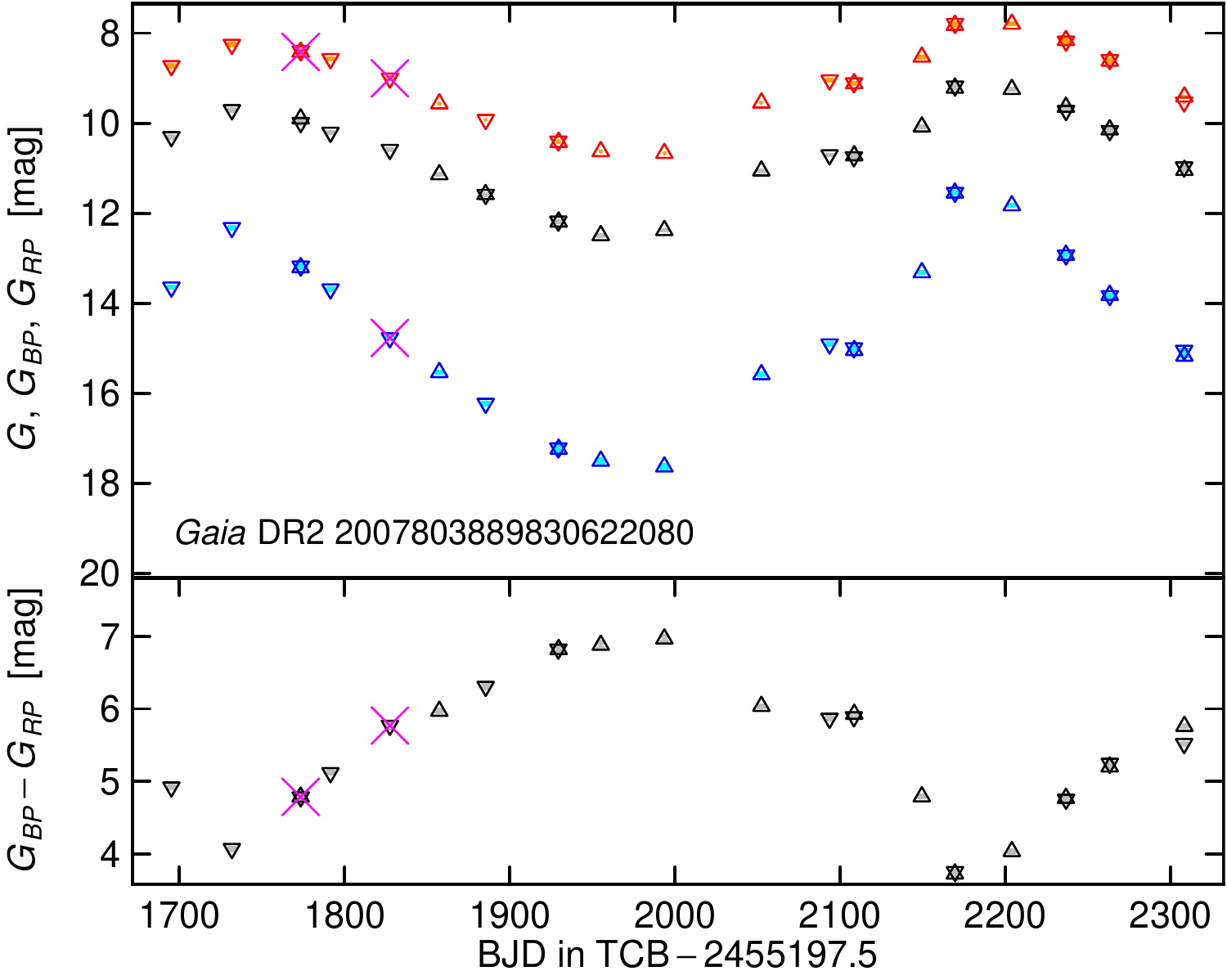}
\includegraphics[width=0.94\hsize]{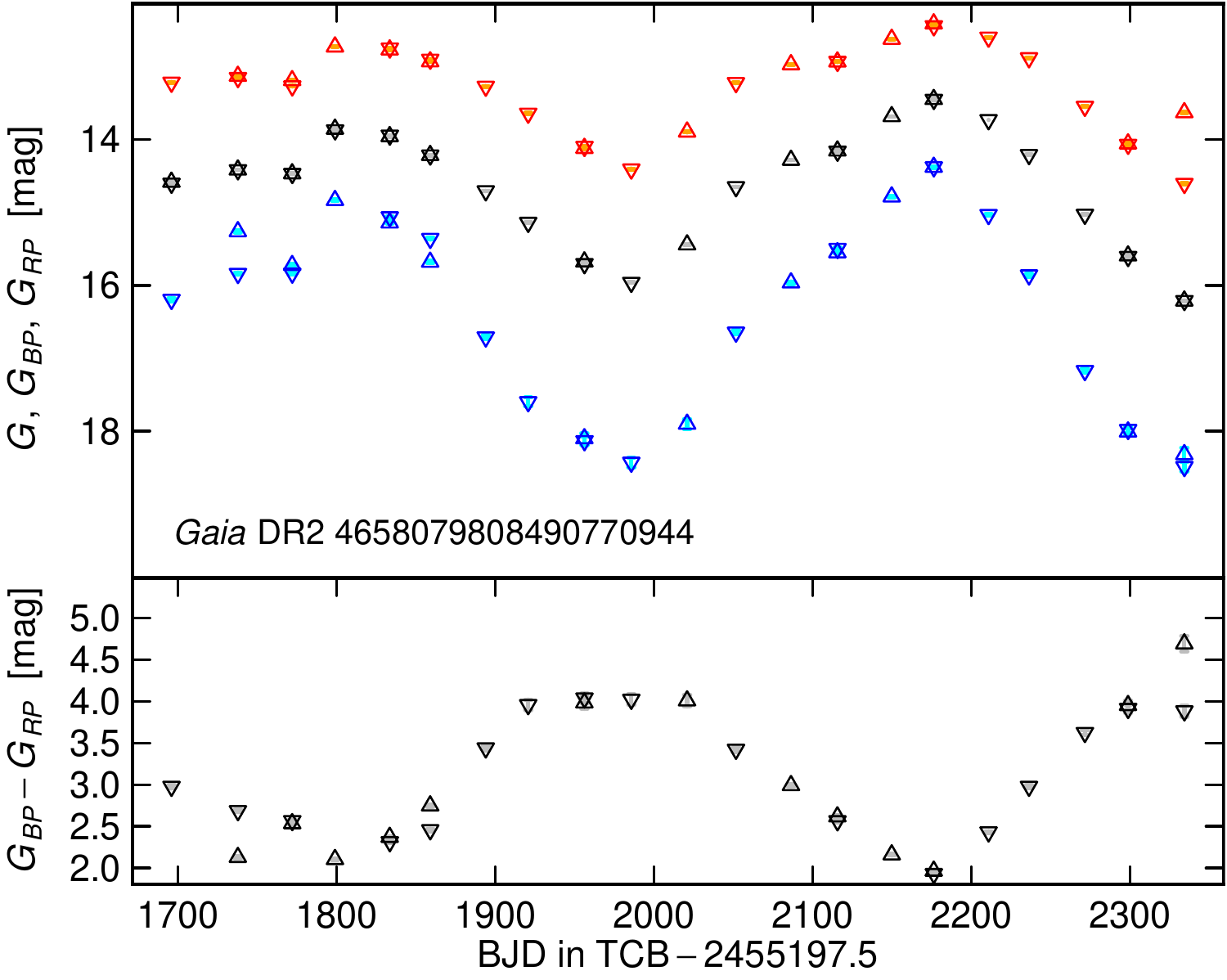}
\caption{Light-curve examples of three LPV candidates with good time coverage, from bright (top figure) to faint (bottom figure).
         \textbf{Upper panel in each sub-figure:} Black, blue, and red symbols identify \gmag, \gbp , and \grp light curves, respectively, with their measurement uncertainties shown at each point in grey, cyan, and orange, respectively (the uncertainties are often too small to be visible on the magnitude ranges of the panels).
         Upward and downward triangles indicate measurements in \Gaia preceding and following fields of view, respectively.
         Measurements that have been discarded in the variability processing pipeline are identified with magenta crosses (see text).
         \textbf{Lower panel in each sub-figure:} Same as upper panel, but for the colour $\gbp-\grp$ time-series.
        }
\label{Fig:LcsExamples1}
\end{figure}

\begin{figure}
\centering
\includegraphics[width=0.94\hsize]{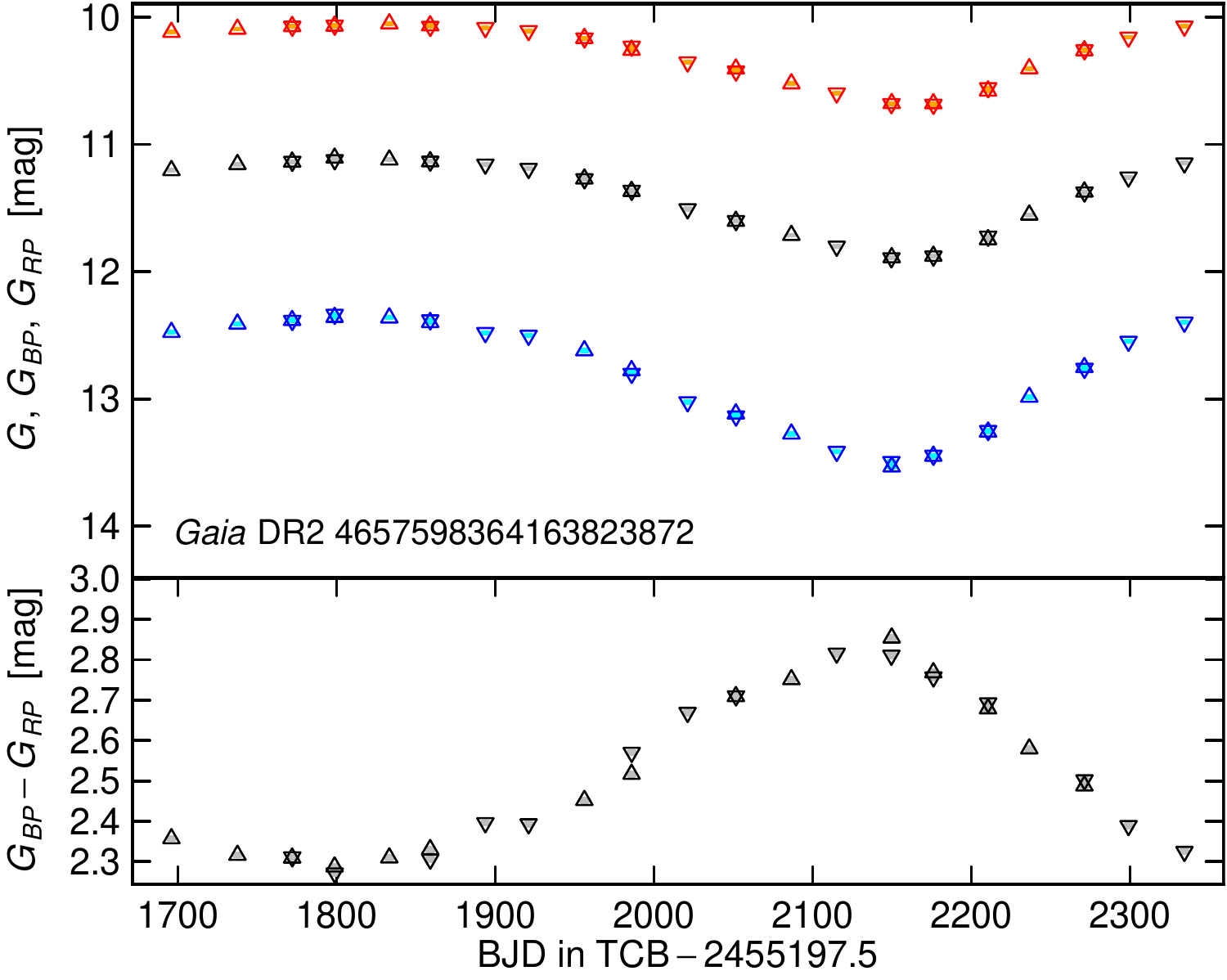}
\includegraphics[width=0.94\hsize]{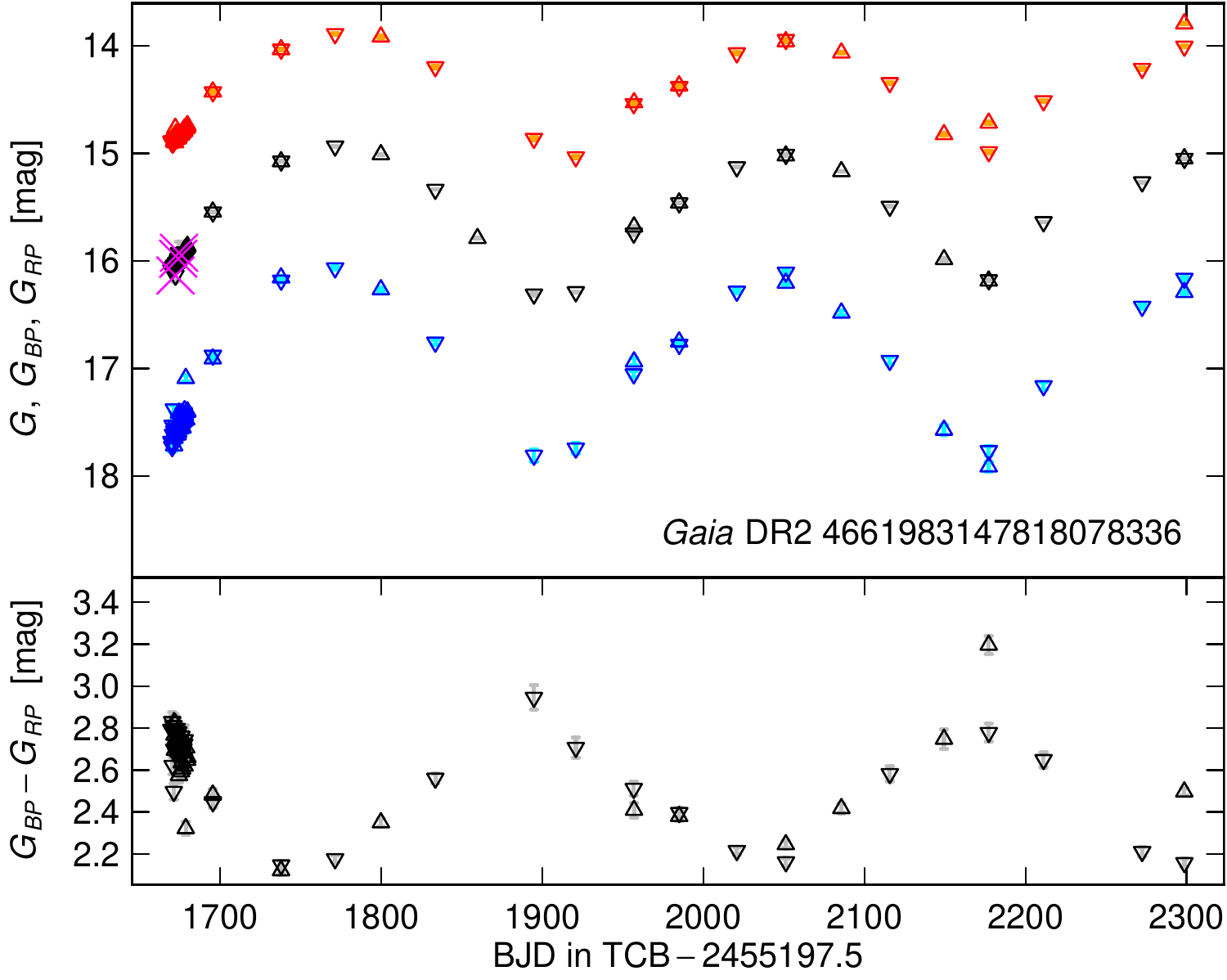}
\caption{Same as Fig.~\ref{Fig:LcsExamples1}, but for a source with a long period ($P=612$~d) compared to the observation duration (top panel) and a source in the LMC (bottom panel).
        }
\label{Fig:LcsExamples1a}
\end{figure}

\begin{figure}
\centering
\includegraphics[width=0.92\hsize]{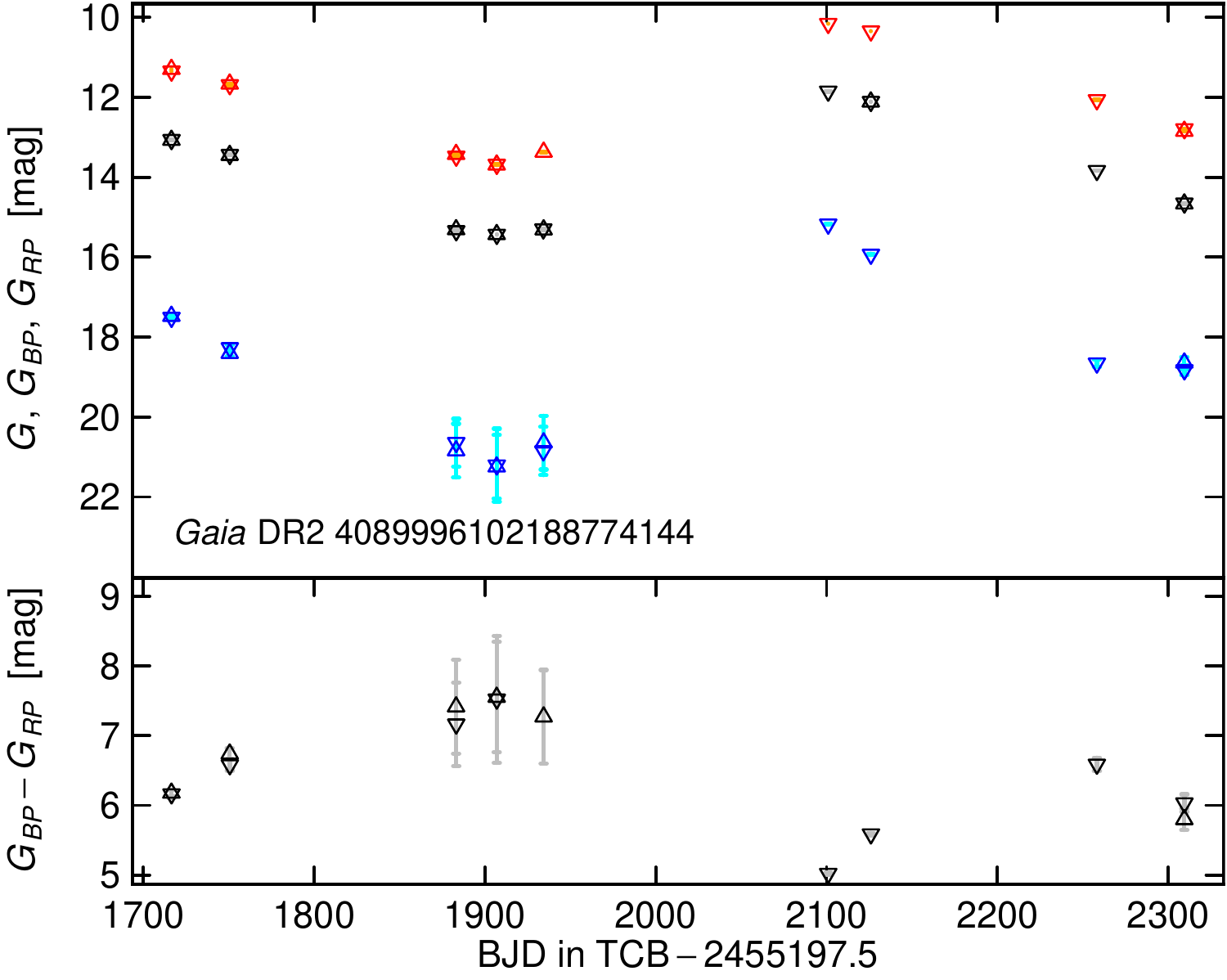}
\includegraphics[width=0.94\hsize]{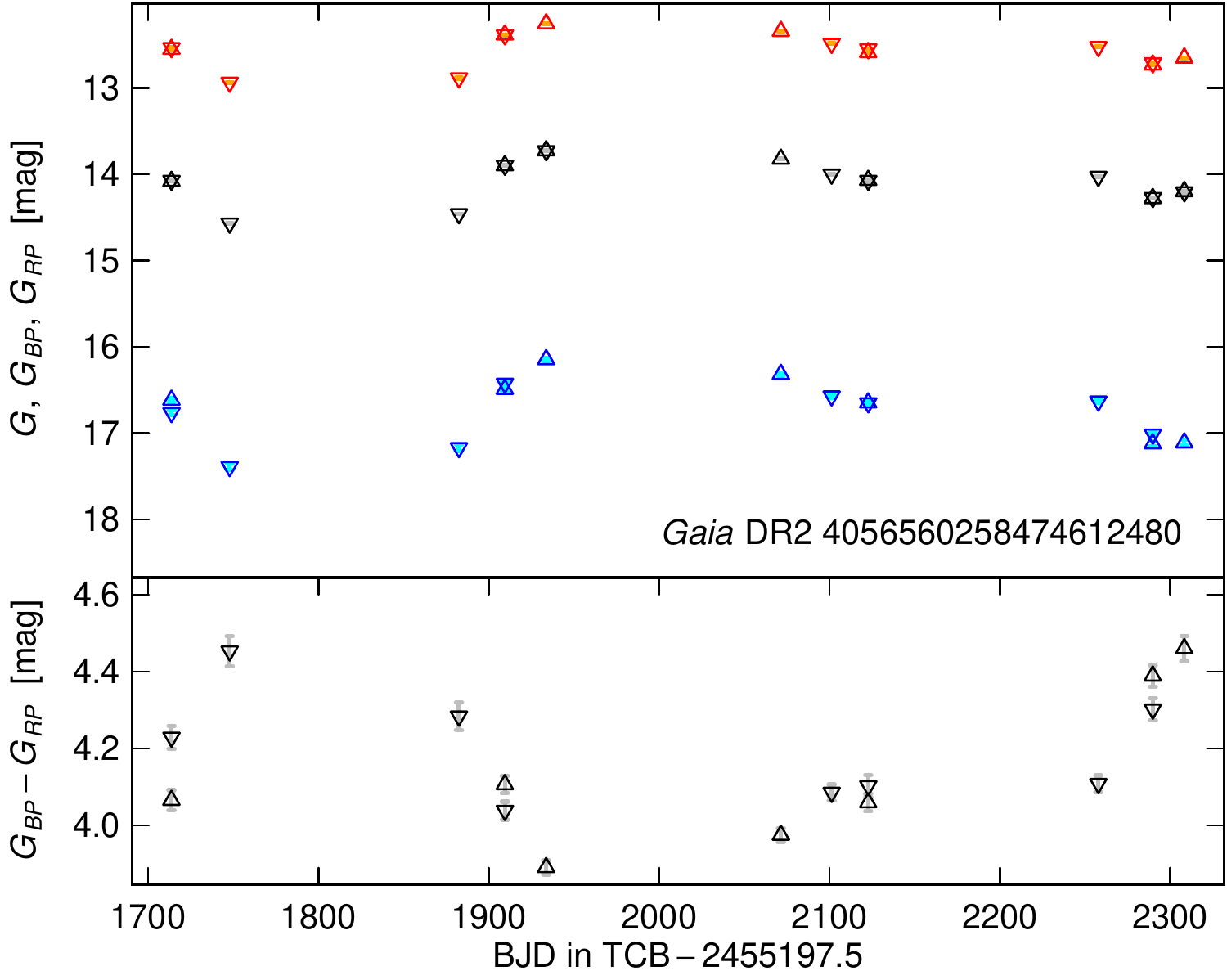}
\caption{Same as Fig.~\ref{Fig:LcsExamples1}, but for two cases with a smaller fraction of time coverage.
         The top example also illustrates the case where some \gbp measurements are fainter than 20~mag.
         The bottom example is a source in the direction of the bulge.
        }
\label{Fig:LcsExamples2}
\end{figure}

\begin{figure}
\centering
\includegraphics[width=\hsize]{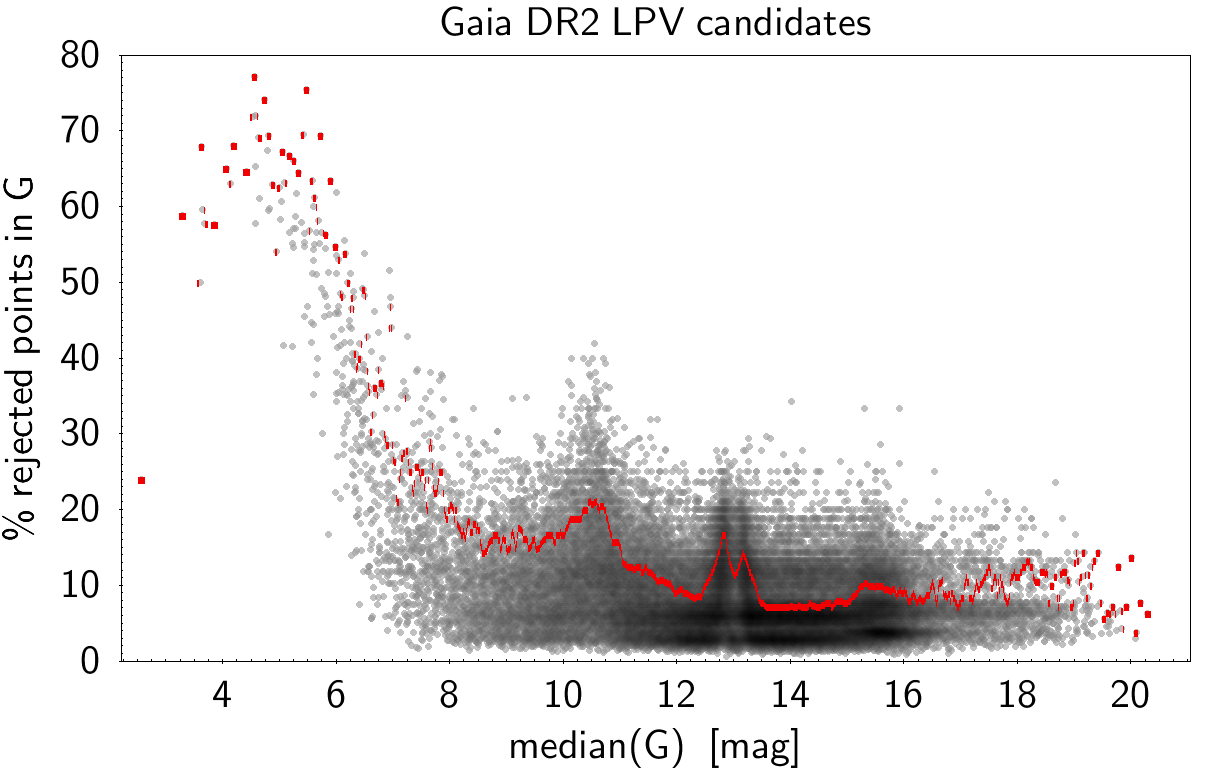}
\caption{Percentage of rejected points in \gmag-band versus median \gmag magnitude of all \Gaia DR2 LPV candidates.
         The red markers locate the 90\% quantile of the distribution at any given magnitude.
        }
\label{Fig:percRejectedPtsG}
\end{figure}

\begin{figure}
\centering
\includegraphics[width=0.9\hsize]{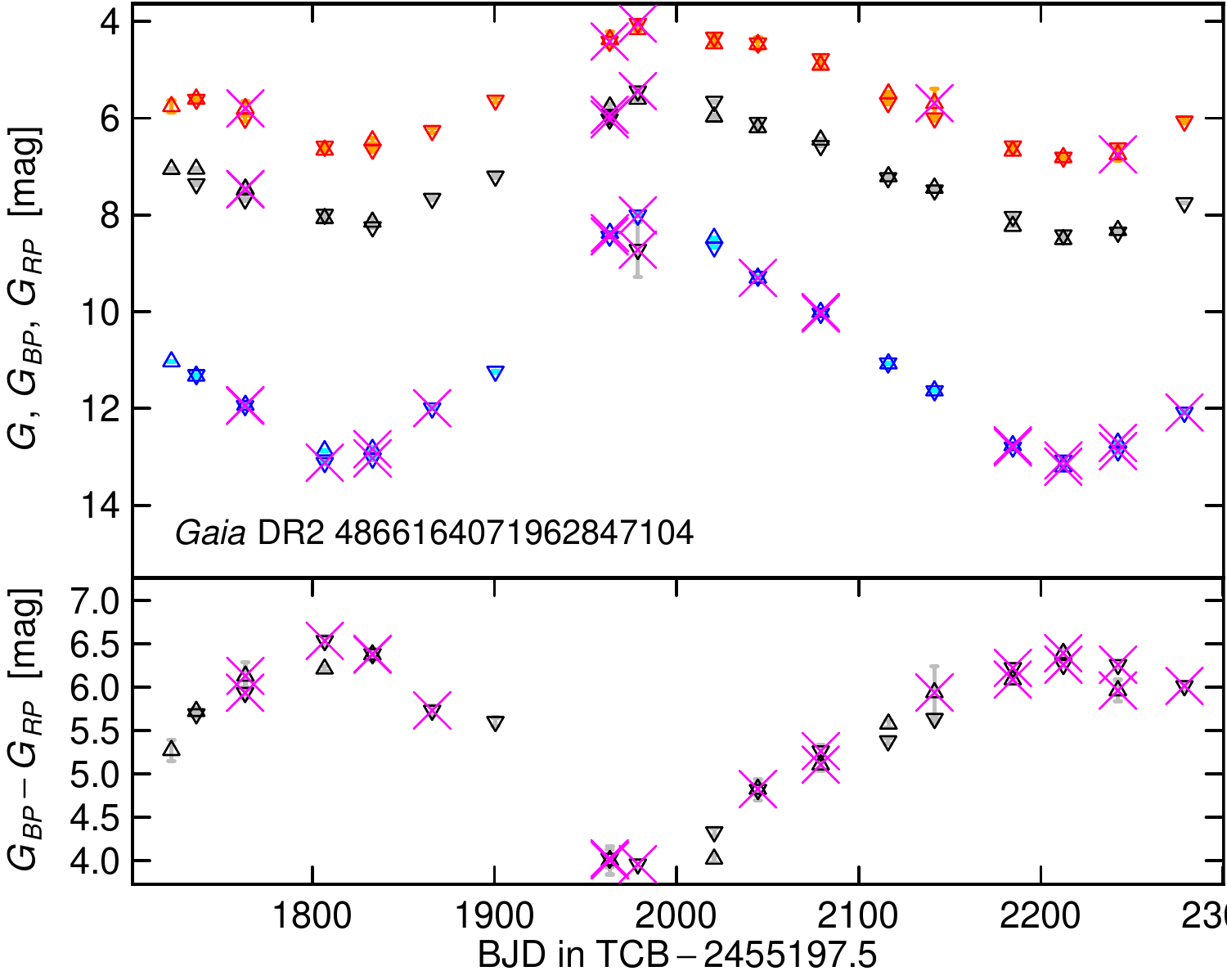}
\includegraphics[width=0.9\hsize]{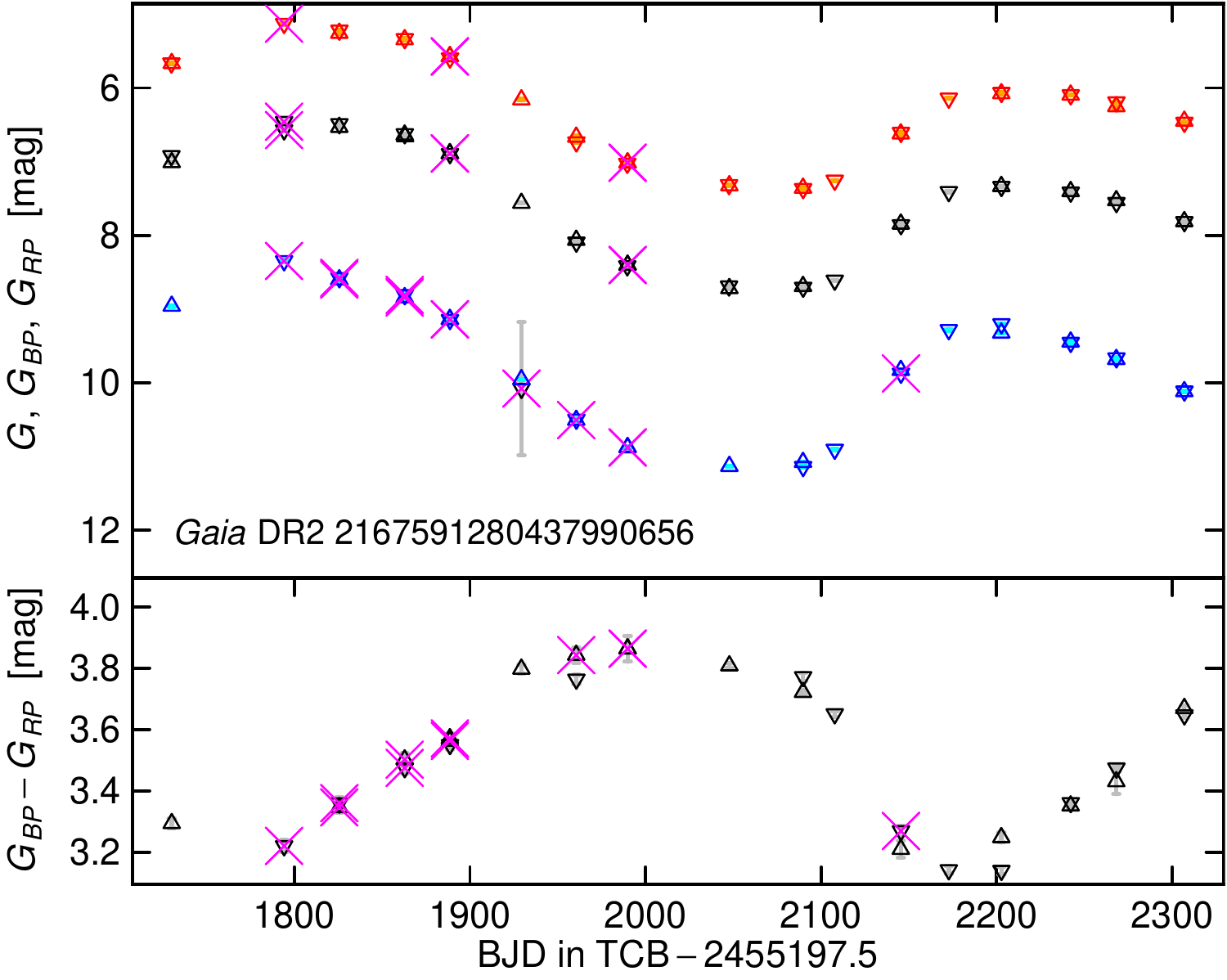}
\includegraphics[width=0.9\hsize]{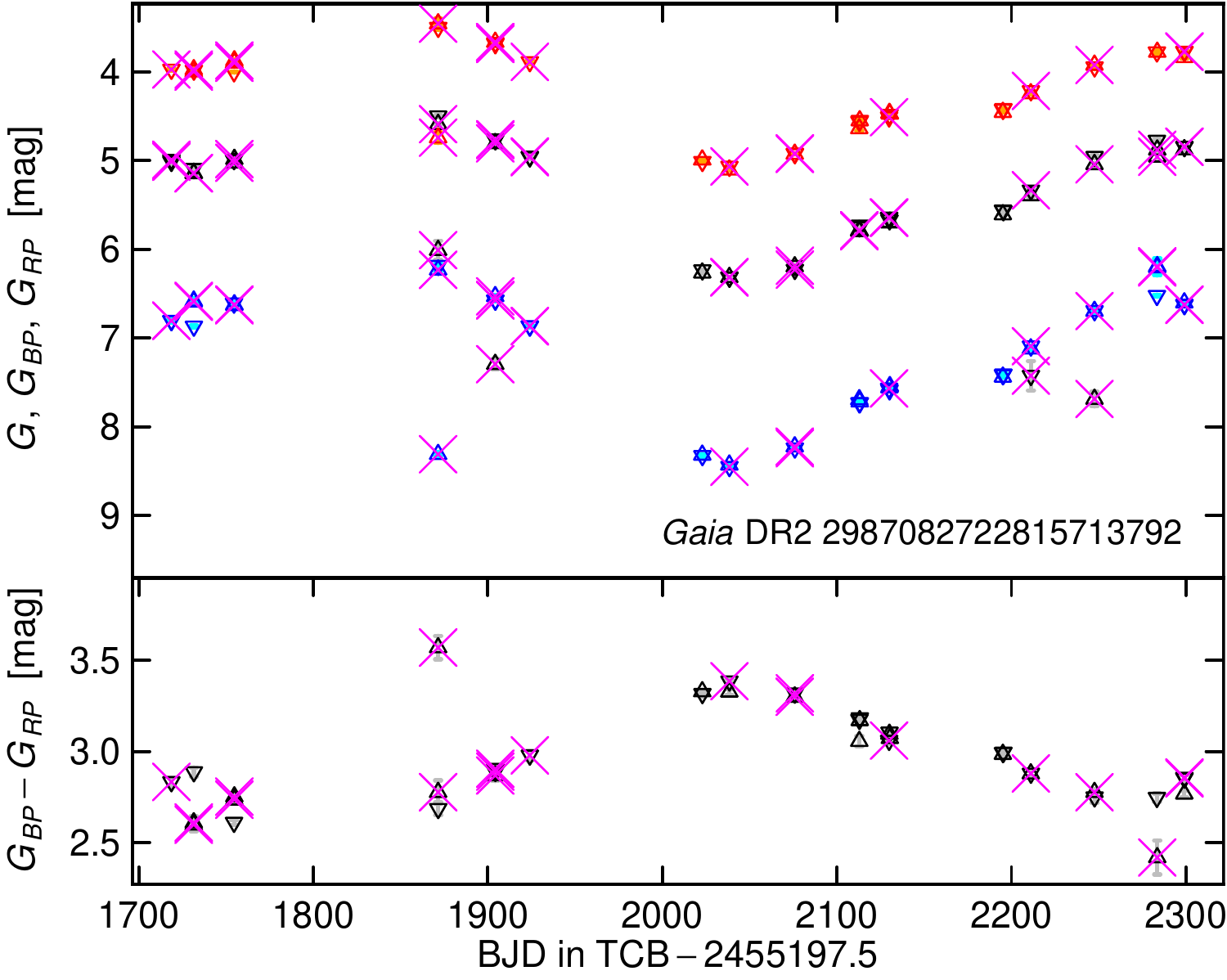}
\caption{Same as Fig.~\ref{Fig:LcsExamples1}, but for three cases with a large fraction of rejected measurements.
         Rejected measurements in these bright objects are mainly due to pairs of transits that occur within 105 min from one another. This time separation is smaller than the time separating successive observations from the two \Gaia fields of views.
        }
\label{Fig:LcsExamples3}
\end{figure}

\begin{figure}
\centering
\includegraphics[width=\hsize]{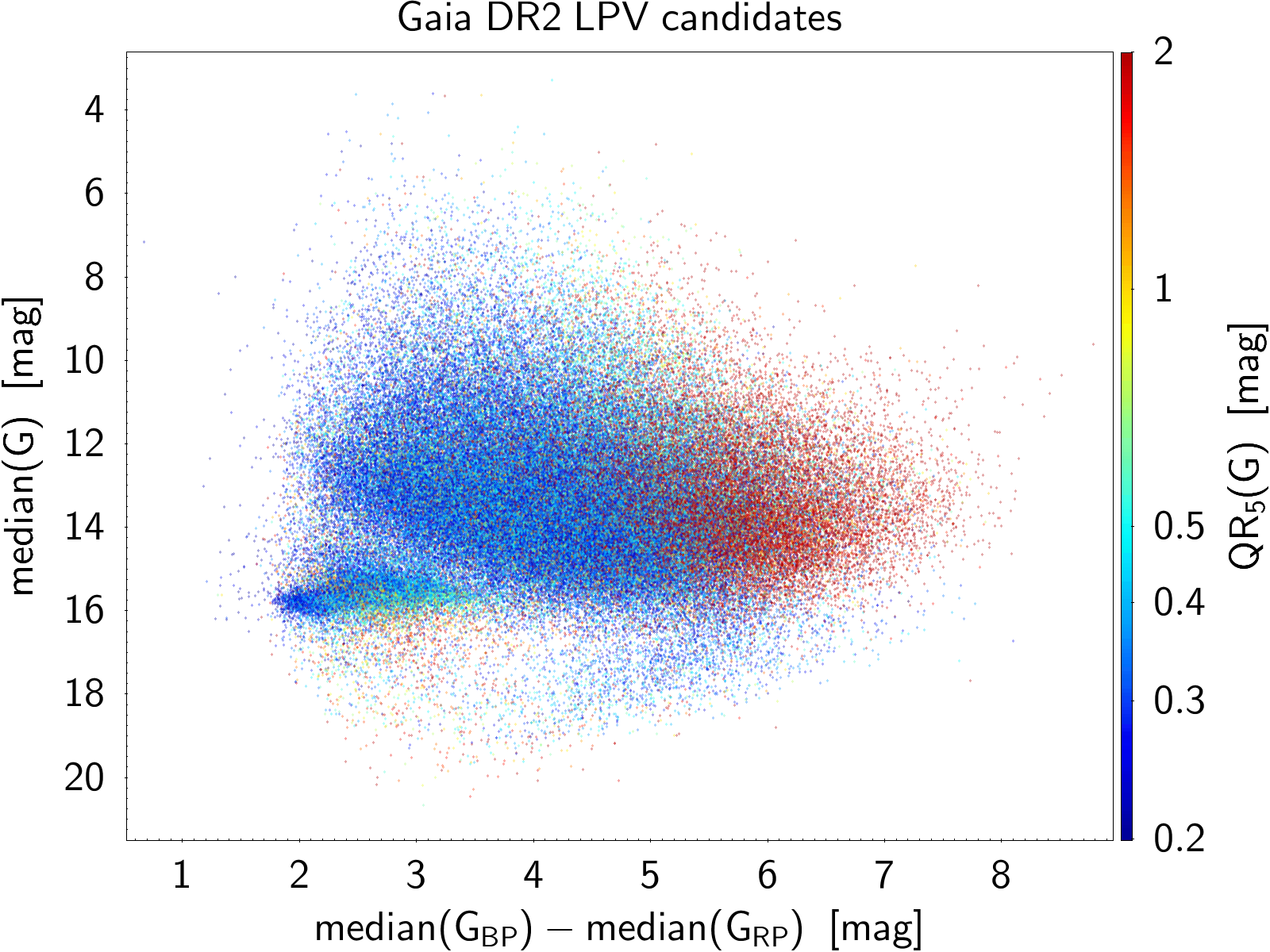}
\caption{Colour-magnitude diagram of all LPV candidates published in \Gaia DR2, with the 5--95\% quantile range $QR_5$ of the cleaned \gmag time-series shown in colour according to the colour scale drawn on the right of the figure.
The clump observed around $\gmag\simeq 15.6$~mag contains stars from the Magellanic Clouds, while the Sagittarius dwarf galaxy appears around $\gmag\simeq 14.7$~mag. 
        }
\label{Fig:CM_trimmedrange}
\end{figure}

\begin{figure}
\centering
\includegraphics[width=\hsize]{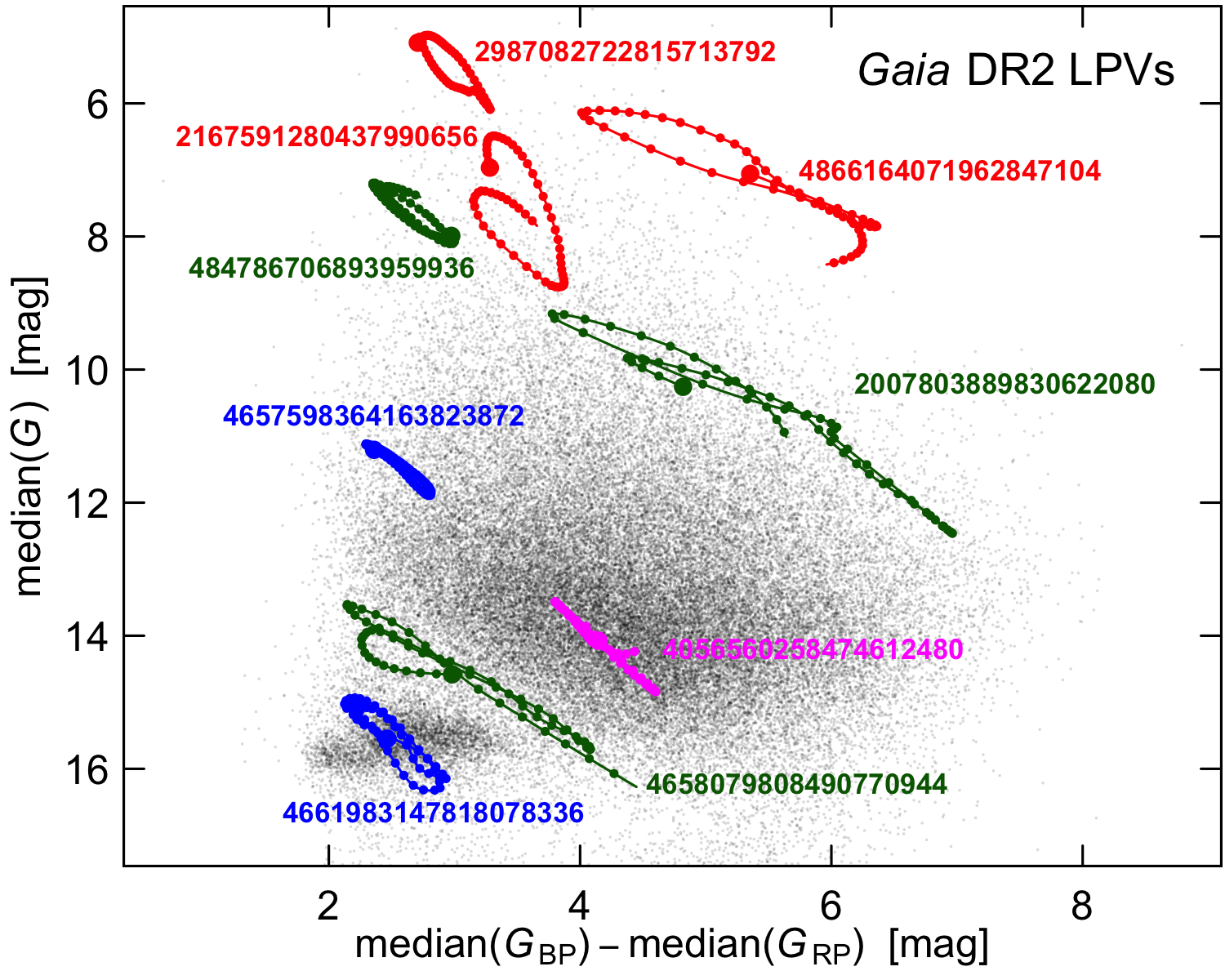}
\caption{Loops performed in the CMD by the LPVs shown as examples in Figs.~\ref{Fig:LcsExamples1} (loops drawn in green), ~\ref{Fig:LcsExamples1a} (in blue), \ref{Fig:LcsExamples2} (in magenta, only the source in the direction of the bulge is shown), and \ref{Fig:LcsExamples3} (in red).
         The loops are drawn from smoothed fits to the \gmag and colour time-series shown in these figures.
         The starting point is shown as a large filled circle for each source, and small filled circles mark every subsequent 10-day interval.
         The \Gaia DR2 source IDs are indicated next to each loop.
         The background grey points represent a random sample of 40\% of the \Gaia DR2 catalogue of LPV candidates, further plotted with a colour transparency of 15\% to avoid saturation in the figure.
        }
\label{Fig:CM_loops}
\end{figure}

\begin{figure}
\centering
\includegraphics[width=0.90\hsize]{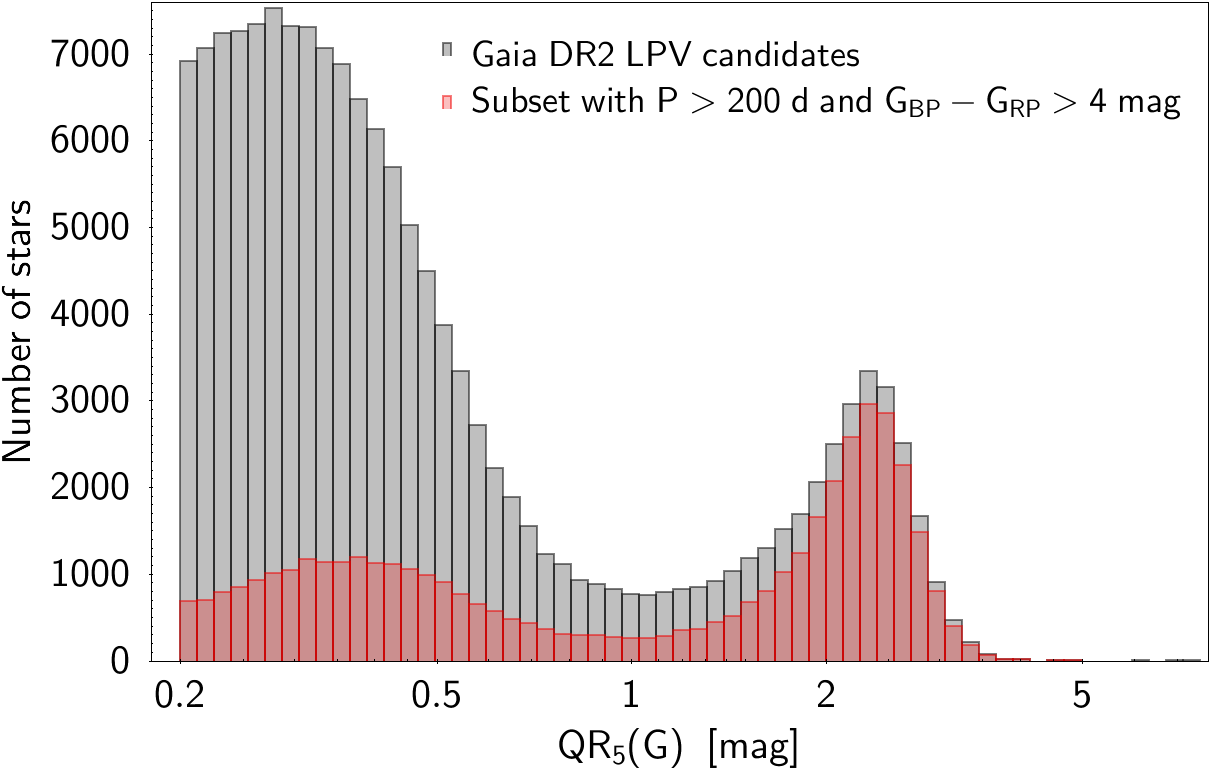}
\caption{Histogram, in grey, of DR2 LPV variability amplitudes, measured by the 5--95\% quantile range $QR_5$ of the cleaned \gmag time-series.
        In red we show the subset of LPV candidates with $\gbp-\grp$ colours larger than 4~mag and periods longer than 200~d.
        }
\label{Fig:histoAmplitudes}
\end{figure}

\begin{figure}
\centering
\includegraphics[width=0.90\hsize]{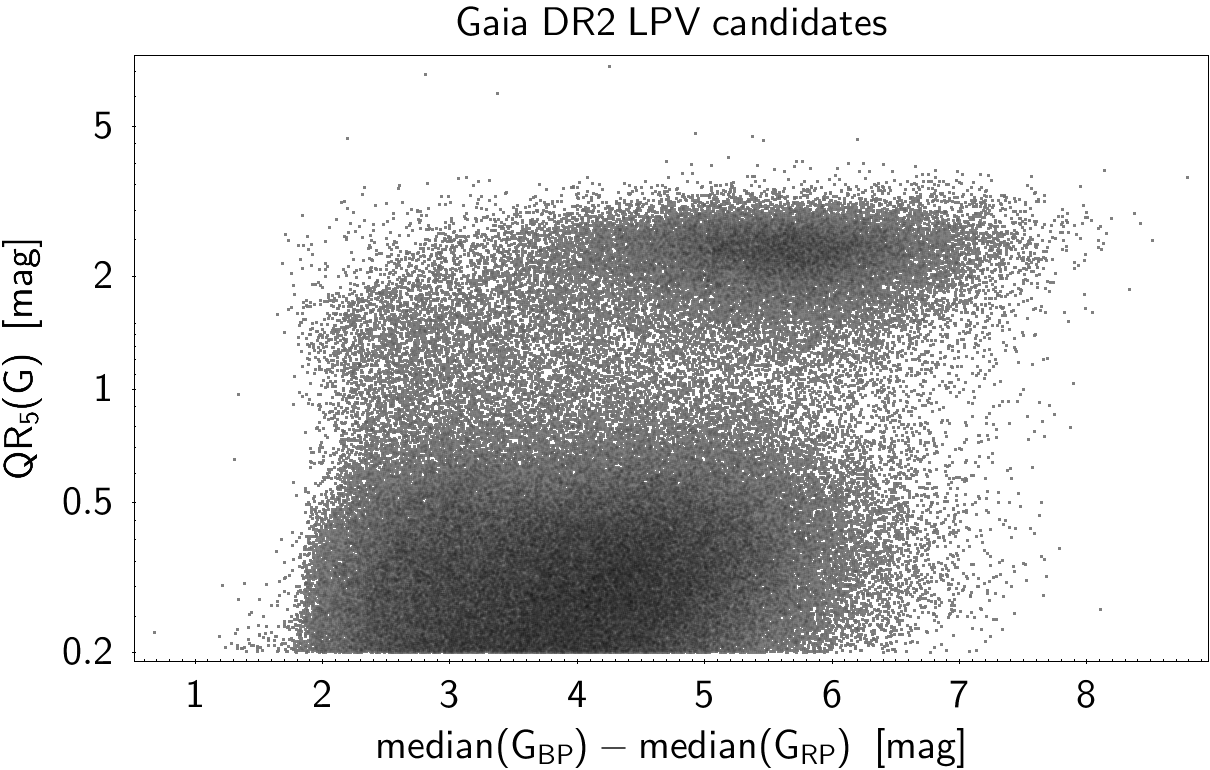}
\caption{\gmag-band variability amplitude, measured by the 5--95\% quantile range $QR_5$ of the cleaned \gmag time-series, versus $\gbp-\grp$ colour for all LPV candidates published in \Gaia DR2.
         The median values of \gbp and \grp measurements are used to compute the colours.
         The population of stars around $QR_5\simeq 2.5$~mag represents Mira-like candidates, while those with $QR_5\lesssim 1.0$~mag are SRV candidates.
        }
\label{Fig:colorAmplitude}
\end{figure}

\begin{figure}
\centering
\includegraphics[width=0.90\hsize]{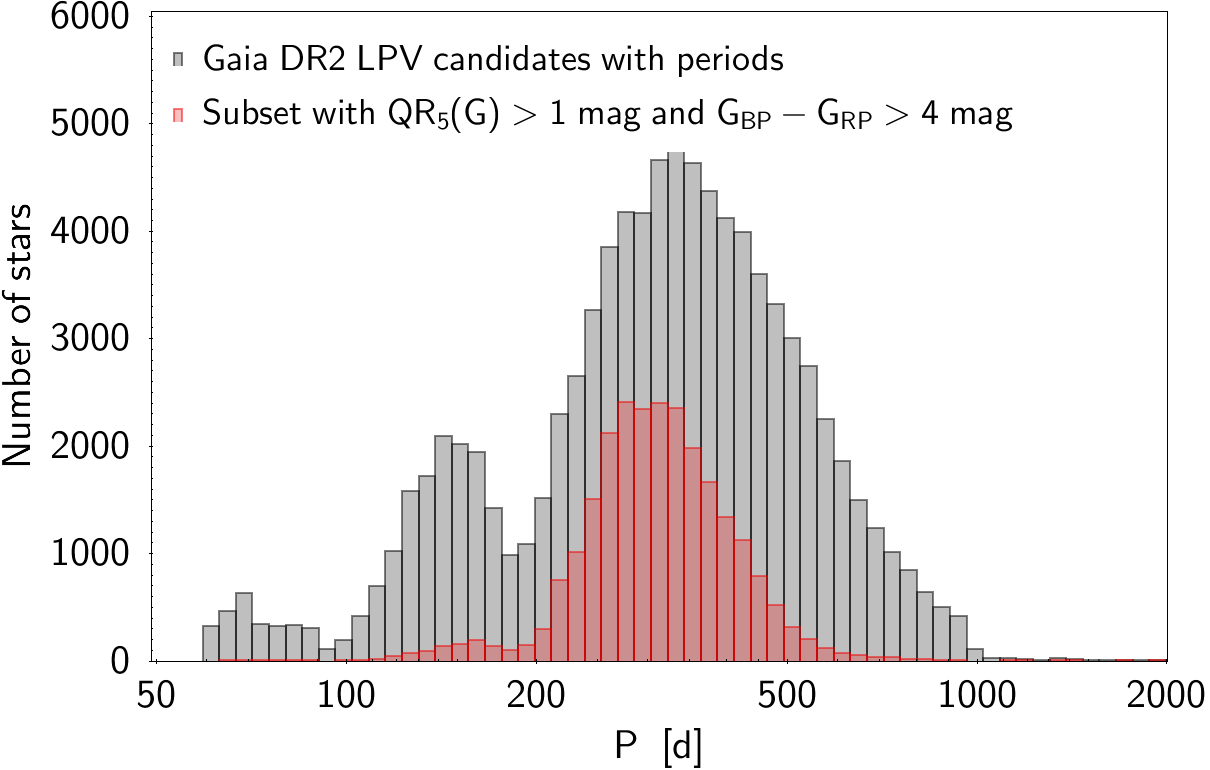}
\caption{Histogram, in grey, of the periods published in the DR2 catalogue of LPV candidates.
         The axis has been limited to $P<2000$~d for bettter visibility.
         There are 24 additional sources with periods above this limit.
         In red we show the subset containing Mira candidates with $\gbp-\grp$ colours larger than 4~mag and \gmag amplitudes larger than 1~mag as measured by the 5--95\% quantile range $QR_5$ of the \gmag light curve.
        }
\label{Fig:histoPeriods}
\end{figure}

\subsection{Light-curve overview}
\label{Sect:overview_lcs}

The histogram of the \gmag-band time-series durations of all LPV candidates is shown in Fig.~\ref{Fig:histoDuration}, with the subset of sources with published periods shown by the dashed blue histogram.
Only 27 sources have durations shorter than one year, while 90\% of the sources have observations over more than 1.5~years.
 The time coverage of the measurements over the observation duration is equally important.
To compute the time coverage, we divided the time-series into 20 equally spaced sub-time intervals, and computed the fraction of sub-time intervals that have non-zero measurements.
The histogram of the resulting time coverage is shown in Fig.~\ref{Fig:histoTimeCoverage}.
It shows a distribution of the number of stars with time coverages between 35\% and 90\% that is about equal.

The time-series duration and the time coverage of the measurements in the time-series both depend on the sky position of the star, according to the \Gaia scanning law.
This is shown in Figs.~\ref{Fig:skyDuration} and \ref{Fig:skyTimeCoverage}, which display the sky distributions of the time-series duration and time coverage, respectively.
Three cases with good time coverage are shown in Fig.~\ref{Fig:LcsExamples1}.
Two other cases are shown in Fig.~\ref{Fig:LcsExamples1a}, one with a very long period comparable to the observation duration, and the second for a source in the LMC, which illustrates a case with a good coverage during the \Gaia commissioning phase using the ecliptic pole scanning law%
\footnote{
\label{Footnote:EPSL}
During \Gaia commissioning, a special scanning law was used during one month, the ecliptic pole scanning law (EPSL).
With this scanning law, the ecliptic poles where scanned eight times a day (6~h rotation period of the spacecraft combined with two fields of view), which led to very dense light curves for sources near the ecliptic poles.
The LMC is not far from the south ecliptic pole.
}.
Two cases with smaller fractions of time coverage are displayed in Fig.~\ref{Fig:LcsExamples2}.
The first of these two illustrates a case with measurements in \gbp that exceed 20~mag, and the second shows a reddened source in the direction of the bulge.

The percentage of rejected points in the \gmag band is shown in Fig.~\ref{Fig:percRejectedPtsG}. Two main filters are responsible for the rejection: the first filter rejects pairs of transits that are too close in time (within 105 min) to be real observations of the same source, and the second filter rejects observations with too large errors or with outlying observations that show too extreme magnitudes compared to the median magnitude of all observations. 
Brighter stars ($\gmag\lesssim 11$~mag) have an increased fraction of rejected measurements, which is mainly due to the first filter, which removes duplicate observations.

Fig.~\ref{Fig:LcsExamples3} shows three examples of time-series with a large fraction of rejected measurements.
In these examples, the fraction of rejected measurements in \gmag is, from top to bottom in the figure,
17\% (\texttt{\Gaia} \texttt{DR2} \texttt{4866164071962847104}),
20\%  (\texttt{\Gaia} \texttt{DR2} \texttt{2167591280437990656}),
and 60\% (\texttt{\Gaia} \texttt{DR2} \texttt{2987082722815713792}), respectively.
Visually, however, the rejected measurements are seen to still have good enough magnitude estimates relative to the large variability amplitudes of LPV candidates.
The rejection criteria were quite strict and should be waived in future releases.
Additional light curves are presented in Sects.~\ref{Sect:individualLPVs} to \ref{Sect:ASAS}.

\subsection{Colour-magnitude diagram overview}
\label{Sect:overview_CMD}

The colour-magnitude diagram (CMD) of all DR2 LPV candidates is shown in Fig.~\ref{Fig:CM_trimmedrange}.
The colours extend from $\gbp-\grp \simeq 2$~mag to $\sim$8~mag for the reddest LPV candidates.
This spread in $\gbp-\grp$ largely originates from extinction due to interstellar and/or circumstellar dust.
The spread in \gmag magnitudes, on the other hand, is largely due to distance effects.
The LPV candidates from the Magellanic Clouds at $15 \lesssim \gmag\mathrm{~[mag]} \lesssim 16$ and $1.8 \lesssim (\gbp-\grp) \mathrm{~[mag]} \lesssim 3.5$ clearly stand out.

A tail of faint outliers is also visible in the CMD (Fig.~\ref{Fig:CM_trimmedrange}) below the main bulk of the distribution, at $\gmag \gtrsim 16$~mag and $\gbp-\grp \lesssim 6.5$~mag.
For these values of $\gmag$ and $\gbp-\grp$, $\gbp$ exceeds the magnitude threshold above which \Gaia measurements are disregarded, leading to unreliable \gbp light curves.
However, the \gmag and \grp light curves still allow a reliable classification as LPVs, and we therefore kept these sources in the catalogue despite their unreliable $\gbp-\grp$ colours.

The large-amplitude pulsations of LPVs are known to induce quite extended excursions in the CMD \citep{SpanoMowlaviEyer_etal09}.
Figure~\ref{Fig:CM_loops} shows the paths travelled in that diagram by the LPVs whose light curves were shown as examples in Figs.~\ref{Fig:LcsExamples1}, ~\ref{Fig:LcsExamples1a}, and \ref{Fig:LcsExamples3}.
To construct the loops, we first smoothed the \gmag-band and $\gbp-\grp$ time-series shown in these figures using the \texttt{LOESS} local polynomial regression fitting algorithm.
We then plot in the CMD (Fig.~\ref{Fig:CM_loops}) a continuous representation of the smoothed time-series.
The paths on the loops are marked by 50~d tags (small filled circles), the start of the path (first measurement in the time-series) being indicated by a large filled circle.
Figure~\ref{Fig:CM_loops} shows that LPVs can have very large amplitude motions in the CMD.
Various loop patterns are visible, revealing variability patterns that can be more  complex than a mono-periodic behaviour.
In addition to multi-periodicity, these variables can also have phase and amplitude changes from one cycle to the next.
Therefore, great caution must be taken when considering individual measurements of LPVs, which cannot be representative of the global properties of these stars. \citet{DR2-DPACP-Eyer} compared the loops of LPVs in the CMD to the loops of other types of variable stars.

\begin{figure}
\centering
\includegraphics[width=\hsize]{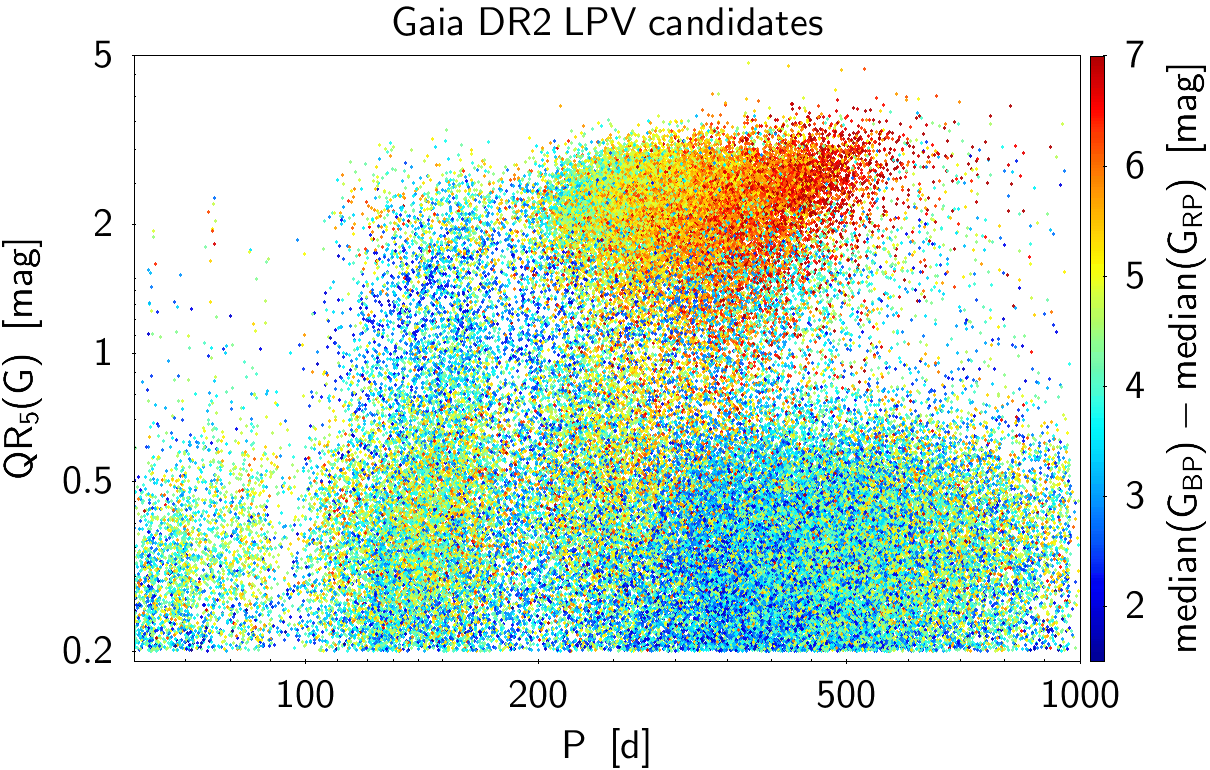}
\caption{Period-amplitude diagram of LPV candidates published in \Gaia DR2.
         The amplitude is measured by the 5--95\% quantile range $QR_5$ of the cleaned \gmag time-series.
         The colour of each point is related to the $\mathrm{median}(\gbp) - \mathrm{median}(\grp)$ colour of the stars according to the colour scale shown on the right of the figure.
         The population of stars around $QR_5\simeq 2.5$~mag and with $P>200$~d represents Mira-like candidates.
        }
\label{Fig:PA}
\end{figure}

\begin{figure}
\centering
\includegraphics[width=\hsize]{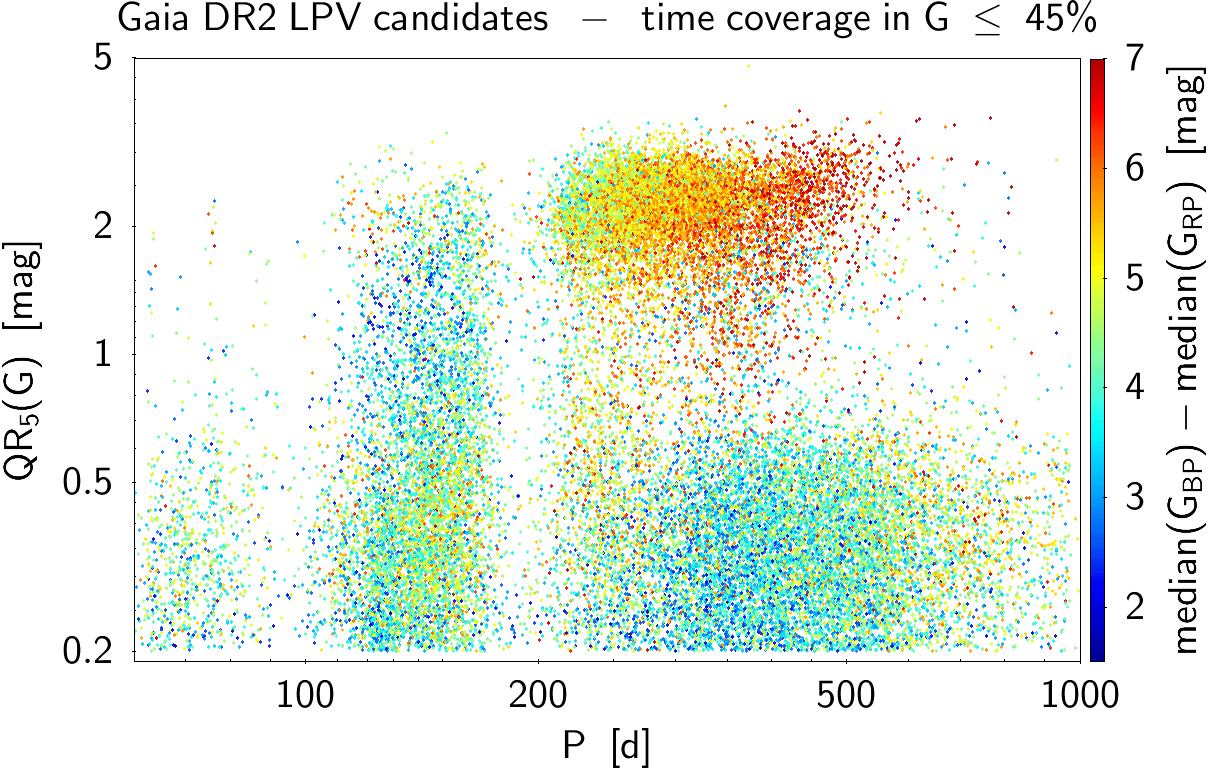}
\caption{Same as Fig.~\ref{Fig:PA}, but restricted to sources that have a light-curve time coverage of \gmag measurements of at most 45\%.
        }
\label{Fig:PA_timeCoverageLE0p45}
\end{figure}

\subsection{Variability amplitude overview}
\label{Sect:overview_amplitudes}

Figure~\ref{Fig:histoAmplitudes} shows the distribution of the \gmag-band variability amplitude, defined in this paper as the 5-quantile range (i.e. the 5--95\% inter-quantile range) of the \gmag-band light curve, and noted $QR_5(\gmag)$.
A group of large-amplitude LPV candidates is clearly observed at $QR_5(\gmag) \gtrsim 1$~mag, peaking at $\sim 2.3$~mag.
They constitute 20\% of the catalogue, and contain Mira-like variables.
They are among the reddest of the LPV candidates, as shown in the CMD in Fig.~\ref{Fig:CM_trimmedrange}, where the amplitudes are shown colour-coded.
This is in agreement with the expectation of large amplitudes for Miras \citep[e.g.][]{2003MNRAS.343L..79K}.
The distribution of the \gmag-band amplitude versus $\gbp-\grp$ colour is further shown in Fig.~\ref{Fig:colorAmplitude}, where the group of Mira candidates clearly stand out at $QR_5(\gmag) \gtrsim 1$~mag and $\gbp-\grp \gtrsim 4$~mag. 
An additional criterion for identifying Mira-like variables is provided in the next paragraph using the period.
The majority of stars with amplitudes smaller than one magnitude, on the other hand, are expected to be SRV candidates.

\subsection{Period overview}
\label{Sect:overview_periods}

\begin{figure}
\centering
\includegraphics[width=\hsize]{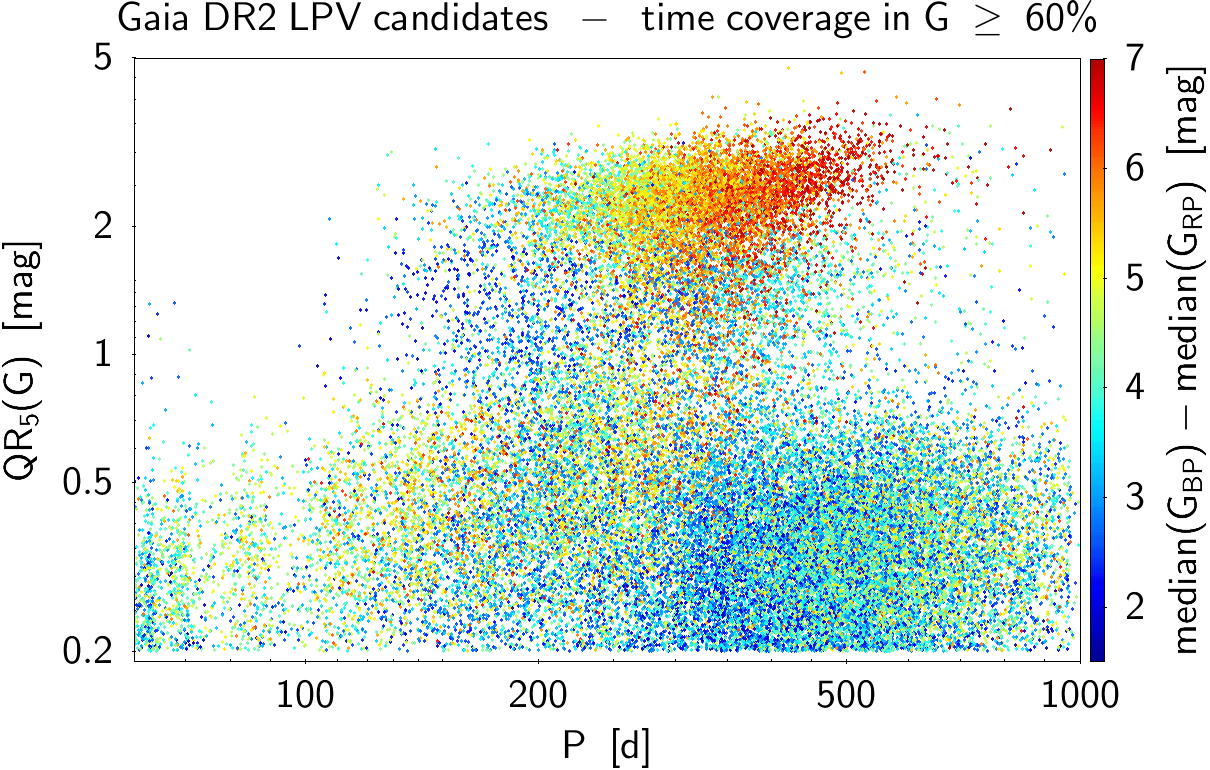}
\caption{Same as Fig.~\ref{Fig:PA}, but restricted to sources that have a light-curve time coverage of \gmag measurements of at least 60\%.
        }
\label{Fig:PA_timeCoverageLE0p60}
\end{figure}

The periods are provided in DR2 for 69\% (89\,617 stars) of all DR2 LPV candidates.
Their distribution is shown in black in Fig.~\ref{Fig:histoPeriods}. 
Using the criteria $QR_5(\gmag)>1$~mag and $\gbp-\grp > 4$~mag for Mira-like behaviour extracted from Fig.~\ref{Fig:colorAmplitude}, we highlight this group in red in Fig.~\ref{Fig:histoPeriods}. 
This shows that most of these stars have periods between 200 and 500 days, as expected for Miras. 
This distinct group of variables is also visible in the period-amplitude diagram shown in Fig.~\ref{Fig:PA}.

Adopting the definition of $\gbp-\grp>4$, $QR_5(\gmag)>1$~mag and $P>200$~d for Miras, DR2 includes at least 23\,405 Mira candidates, that is, more than one-fourth of the LPV candidates with published periods.
A period-amplitude trend is discernible in Fig.~\ref{Fig:PA} for these stars, their amplitude increasing with period, as expected for Miras \citep{2009AcA....59..239S}.
Additional Miras can be present among the other 62\,144 LPV candidates for which periods have not been published in DR2.

We also note in Fig.~\ref{Fig:PA} very red stars below the $P$--$QR_5(\gmag)$ trend that have $QR_5$ amplitudes smaller (in some cases much smaller) than expected from the period-amplitude trend.
This may be due to the relatively short DR2 observation duration compared to the Mira periods, combined with the scarcity of DR2 measurements (keeping in mind that a mean of only 13 five-day clumped observation groups per light curve is available in DR2, see Sect.~\ref{Sect:catalog}).
They are visible in Fig.~\ref{Fig:histoAmplitudes} as the low-amplitude tail of the $QR_5(\gmag)$ distribution of Mira candidates (red distribution).
However, these stars may also be related to a group of LPVs studied from the ground that show a pulsational behaviour between the classical Mira and SRV groups \citep{2005A&A...431..623L}.

\begin{table*}
\caption{Comparison sample of bright LPVs monitored from ground and published in the \Gaia DR2 catalogue of LPVs.
         The atmospheric chemistry (second column; M: O rich, C: C rich) is taken from the General Catalogue of Variable Stars \citep[GCVS;][]{2017ARep...61...80S}.
         Columns 3 to 8 list \Gaia  LPV attributes for the \Gaia crossmatches, with $QR_5(G)$ being the 5--95\% quantile range of the \gmag light curve, and $m_\mathrm{bol} = \gmag + BC$ with $BC$ being the bolometric correction.
         The integrated apparent bolometric magnitudes given in the ninth column are taken from \citet{1996A&A...308..489K}.
         The periods given in the last column come from the GCVS.
         }
\centering
\begin{tabular}{l l | c c c c c c | c c}
\hline\hline
       & GCVS       & \multicolumn{6}{| c |}{\Gaia} &  SED      & GCVS \\
Name   & VarType & \Gaia DR2 source ID & \gmag  & $\gbp-\grp$ & $QR_5(G)$ & $m_{\rm bol}$ & Period & $m_{\rm bol}$ & Period \\
       &         &    & [mag]  & [mag]       & [mag]     & [mag]     &  [d]   &  [mag]    &  [d]   \\
\hline
R Lep  & C Mira       & 2987082722815713792 & 5.61 & 3.01 & 1.49 & 4.05 & 501 & 2.72  & 427 \\
R Cas  & M Mira       & 1944073004731198336 & 4.83 & 5.02 & 3.93 & 2.63 & 621 & 1.26  & 434 \\
R Aqr  & M Mira/Symb. & 2419576358847019136 & 5.61 & 3.07 & 3.64 &  -   &  -  & 2.43  & 390 \\
g Her  & M SRV        & 1381118928134364544 & 2.57 & 3.09 & 0.85 & 1.00 &  62 & 0.73  &  89 \\
TT Mon & M Mira       & 3055512615681140736 & 7.88 & 4.73 & 3.00 & 5.68 & 342 & 4.94  & 318 \\
RX Tau & M Mira       & 3292447407137694208 & 7.97 & 5.04 & 1.83 & 4.46 & 320 & 4.63  & 337 \\
RR Ori & M Mira       & 3349083629043128320 & 8.79 & 3.97 & 2.61 & 5.94 & 255 & 6.04  & 252 \\
RU Vir & C Mira       & 3704116483406003072 & 9.15 & 3.76 & 1.44 & 6.96 & 425 & 4.58  & 434 \\
R Vir  & M Mira       & 3709971554622524800 & 6.30 & 2.86 & 2.14 & 4.46 & 141 & 4.84  & 146 \\
U Cet  & M Mira       & 5170512944979310208 & 6.84 & 3.22 & 3.12 & 4.64 & 235 & 5.54  & 235 \\
X Oct  & M Mira       & 5191703179748307456 & 6.47 & 3.63 & 1.87 & 4.27 & 206 & 4.75  & 200 \\
T Hya  & M Mira       & 5749870429386271488 & 7.28 & 3.89 & 1.60 & 4.53 & 291 & 5.23  & 291 \\
T Aqr  & M Mira       & 6913517223245165696 & 8.00 & 3.41 & 2.00 & 6.00 & 203 & 6.13  & 201 \\
\hline
\end{tabular}
\label{Tab:sampleBrightLPVs}
\end{table*}

The distribution of periods below 200 days is more complicated.
The apparent peak at about 150\,d in Fig.~\ref{Fig:histoPeriods}, also visible in Fig.~\ref{Fig:PA}, is an artefact due to the sparsity of detected periods below $\sim$120~d on the short-period side, and between $\sim$175~d and $\sim$210~d on the long-period side.
The sparsity of detected periods in these period ranges is in turn due to the short time-coverage of measurements in the DR2 light curves in specific regions of the sky owing to the \Gaia scanning law and limited DR2 observation time-span.
This is strikingly visible when plotting the period-amplitude diagram for the sample of LPVs with a \gmag-band light-curve time coverage shorter than 50\%, as shown in Fig.~\ref{Fig:PA_timeCoverageLE0p45}.
In contrast, the period-amplitude diagram of LPVs with a time coverage of at least 60\% shows no sparsity of periods between $\sim$175~d and $\sim$210~d (Fig.~\ref{Fig:PA_timeCoverageLE0p60}).
This conclusion is also reached in Sect.~\ref{Sect:OGLE_periods} when comparing \Gaia periods to OGLE-III periods for stars in the Magellanic Clouds (where \Gaia time coverage is good), and towards the Galactic bulge (where \Gaia time coverage is less good).
We refer in particular to Figs.~\ref{Fig:periodsGaiaVsOgle_Clouds} and \ref{Fig:periodsGaiaVsOgle_Bulge}, respectively.
The peak at about 150\,d in Fig.~\ref{Fig:histoPeriods} is thus not real, and a wide distribution of LPV periods extending down to periods below 100 days is expected.

Stars with $100 \lesssim P~\mathrm{[d]} \lesssim 200$ may include first-overtone red giant pulsators.
A small number of large-amplitude stars in that period regime is also visible. Their pulsation characteristics resemble those of the group of metal-poor Miras \citep{1991A&A...252..583H}.
These and further subgroups will be identified and discussed elsewhere.

At periods shorter than 100\,d (there are 2889 sources), aliases in the frequencygram are expected to be more present.
A comparison with well-known bright LPVs (Table~\ref{Tab:sampleBrightLPVs} in Sect.~\ref{Sect:individualLPVs_wellStudiedCases}), however, with ASAS\_SN LPVs (Fig.~\ref{Fig:periodsGaiaAsas} in Sect.~\ref{Sect:ASAS}), and with OGLE-III LPVs (Fig.~\ref{Fig:periodsGaiaVsOgle_Clouds} in Sect.~\ref{Sect:OGLE}) indicates a reliable period detection at all periods down to 60~d.
In reality, many more sources with periods shorter than 100~d are expected in the period distribution than what are found in DR2.

Long-period variables are known to obey period-luminosity relations that depend on the pulsation mode.
To see these relations, the intrinsic luminosities must be recovered, typically using Wesenheit extinction-free magnitudes \citep{Madore82,Wood15}.
This is the object of a separate study (Lebzelter et al., in prep.).

\section{Validation 1: Selected individual LPVs}
\label{Sect:individualLPVs}

\begin{figure}
\centering
\includegraphics[width=0.95\hsize]{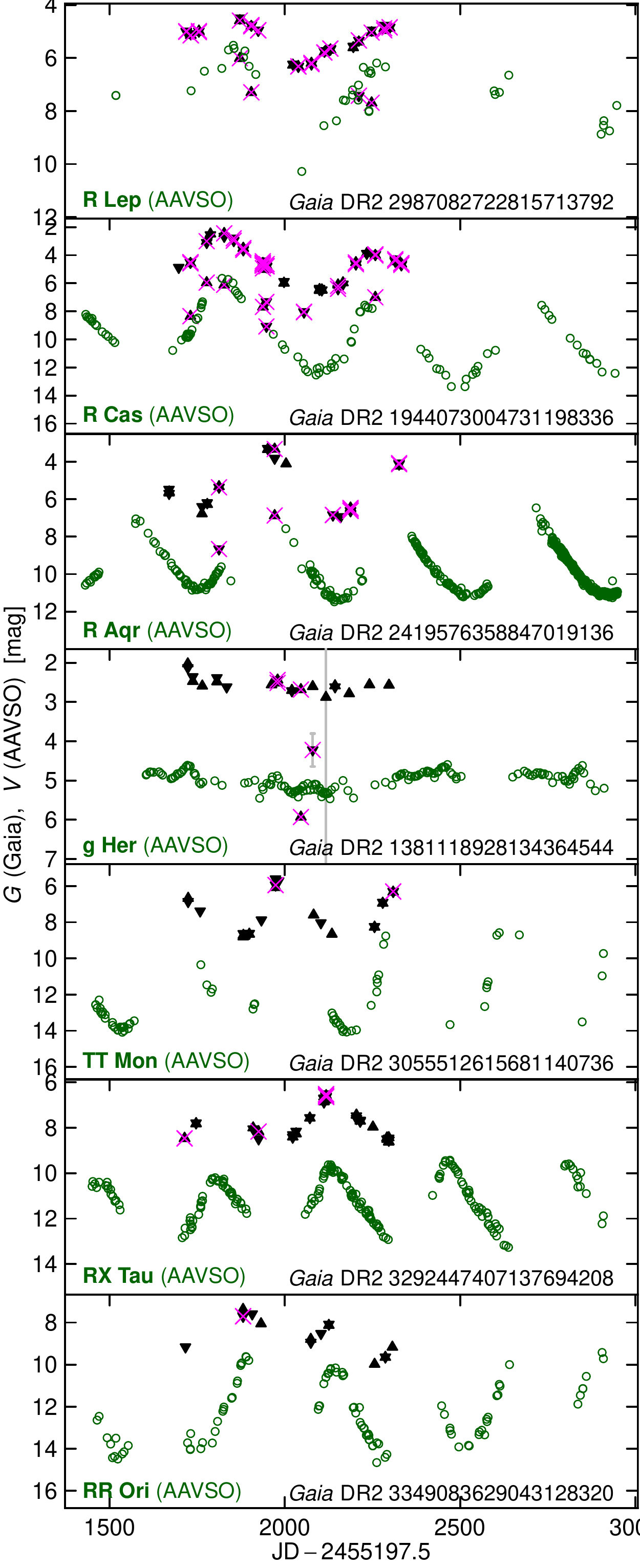}
\caption{Light-curve examples of the seven first LPVs listed in Table~\ref{Tab:sampleBrightLPVs}.
         The AAVSO $V$ magnitudes are shown as green open circles, and the \Gaia \gmag magnitudes as black filled triangles (downward-pointing and upward-pointing triangles for measurements in the \Gaia preceding and following fields of view).
        The magenta cross markers identify \Gaia measurements that have been disregarded in the \Gaia time-series processing pipeline (see text).
         The uncertainties in the \Gaia measurements are drawn as grey vertical bars, but they are within the size of the filled triangles in most cases.
         In the \Gaia light curve of \texttt{g Her}, an outlier measurement is present outside the Y-axis range at $G>7$~mag, the uncertainty of which is visible as a vertical grey line. 
        }
\label{Fig:LcsAAVSO_1}
\end{figure}

\begin{figure}
\centering
\includegraphics[width=0.95\hsize]{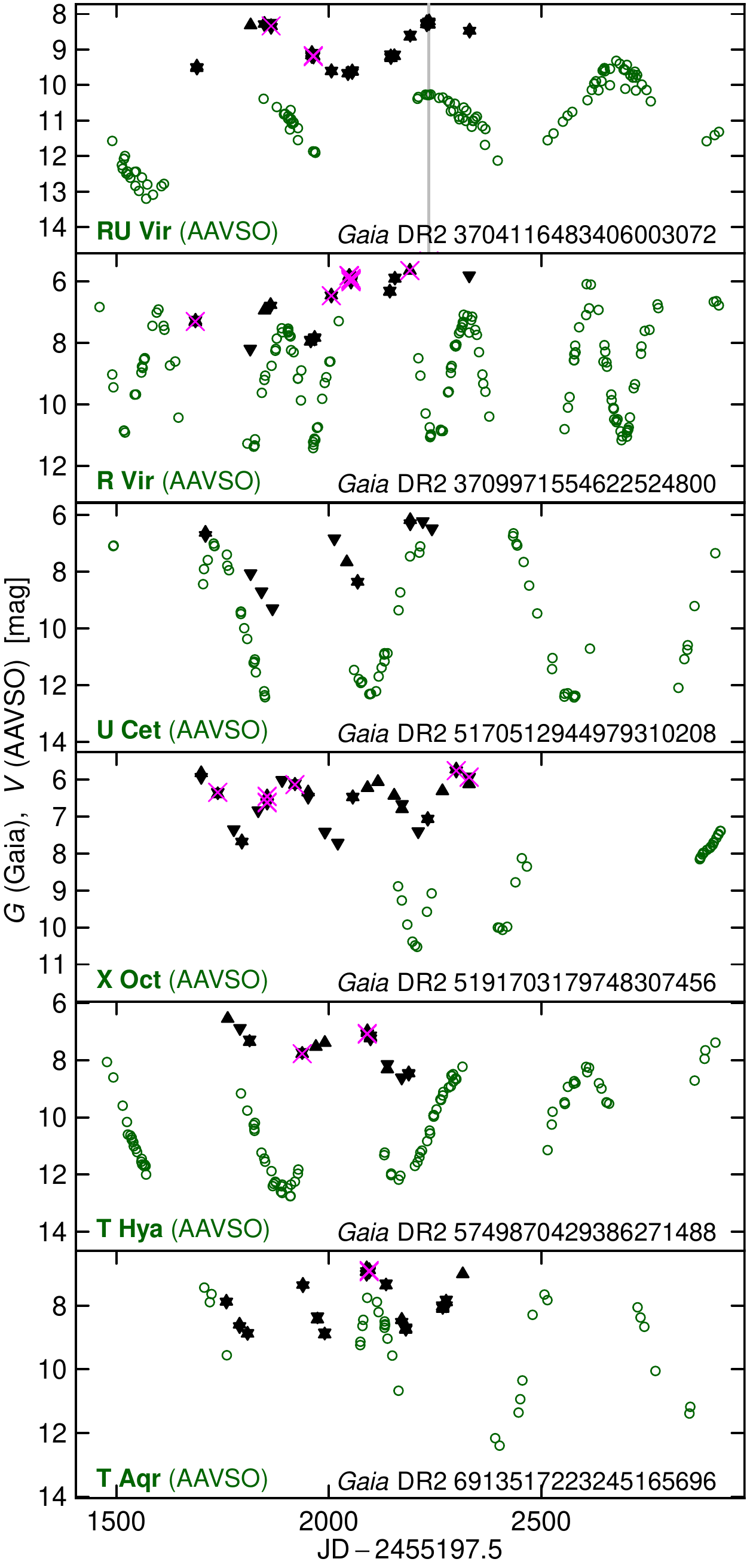}
\caption{Same as Fig.~\ref{Fig:LcsAAVSO_1}, but for the last six LPVs in Table~\ref{Tab:sampleBrightLPVs}.
         In the \Gaia light curve of \texttt{RU Vir}, an outlier measurements is present outside the Y-axis range at $G>14.5$~mag, the uncertainty of which is visible as a vertical grey line.
        }
\label{Fig:LcsAAVSO_2}
\end{figure}

\begin{figure}
\centering
\includegraphics[width=\hsize]{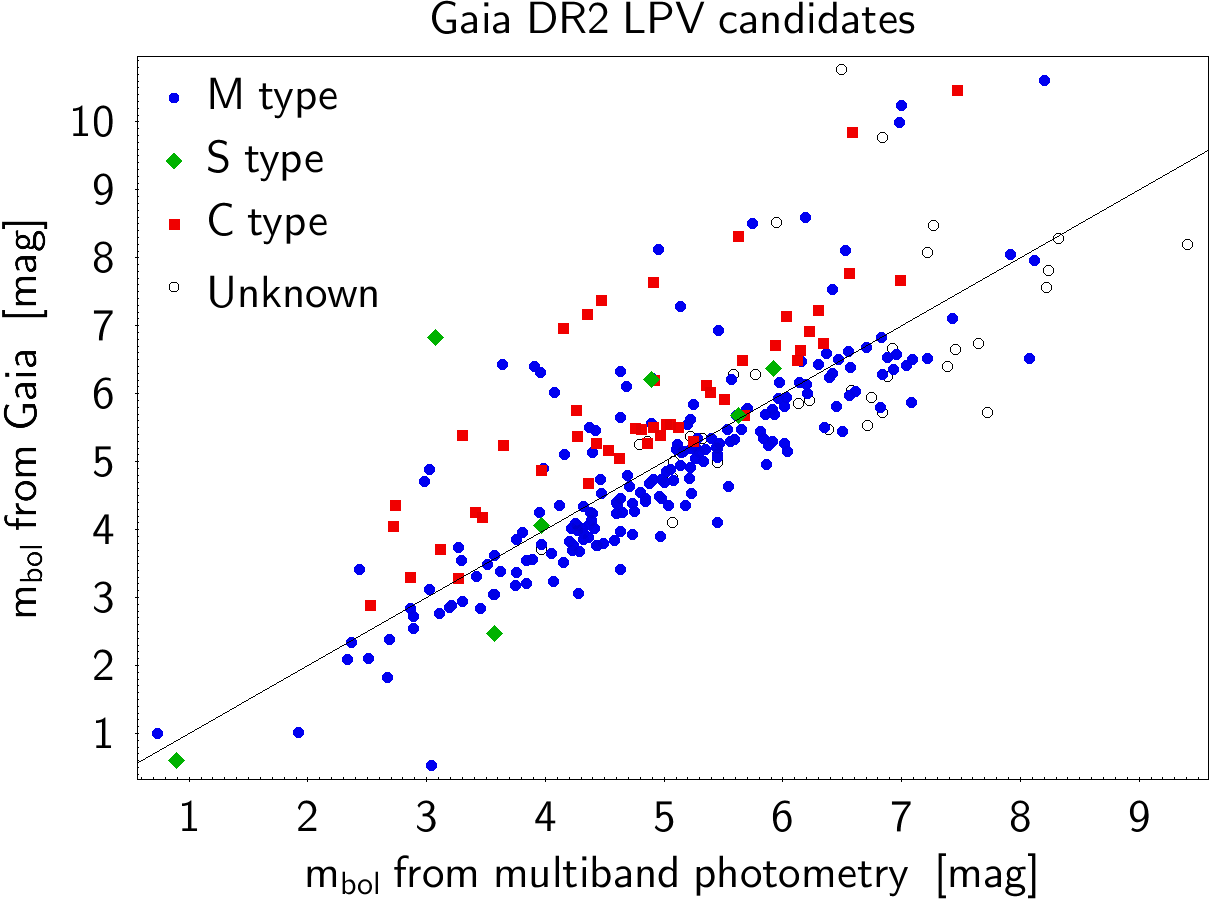}
\caption{Comparison of the apparent bolometric magnitude $m_{\rm bol}$ derived from \Gaia data using the bolometric corrections published in DR2 with the bolometric magnitudes derived by \citet{1996A&A...308..489K} from SED fits to multi-band photometry.
         The LPV sample consists of all LPVs from \citet{2010A&A...524A..87K} for which a \Gaia crossmatch is published in the DR2 LPV catalogue.
         M-, S-, and C-type stars are plotted as blue filled circles, green filled triangles, and red filled squares, respectively.
         Stars of unknown type are plotted as black open circles.
         A diagonal line is drawn for reference.
        }
\label{Fig:apparentMbol}
\end{figure}

\begin{figure}
\centering
\includegraphics[width=\hsize]{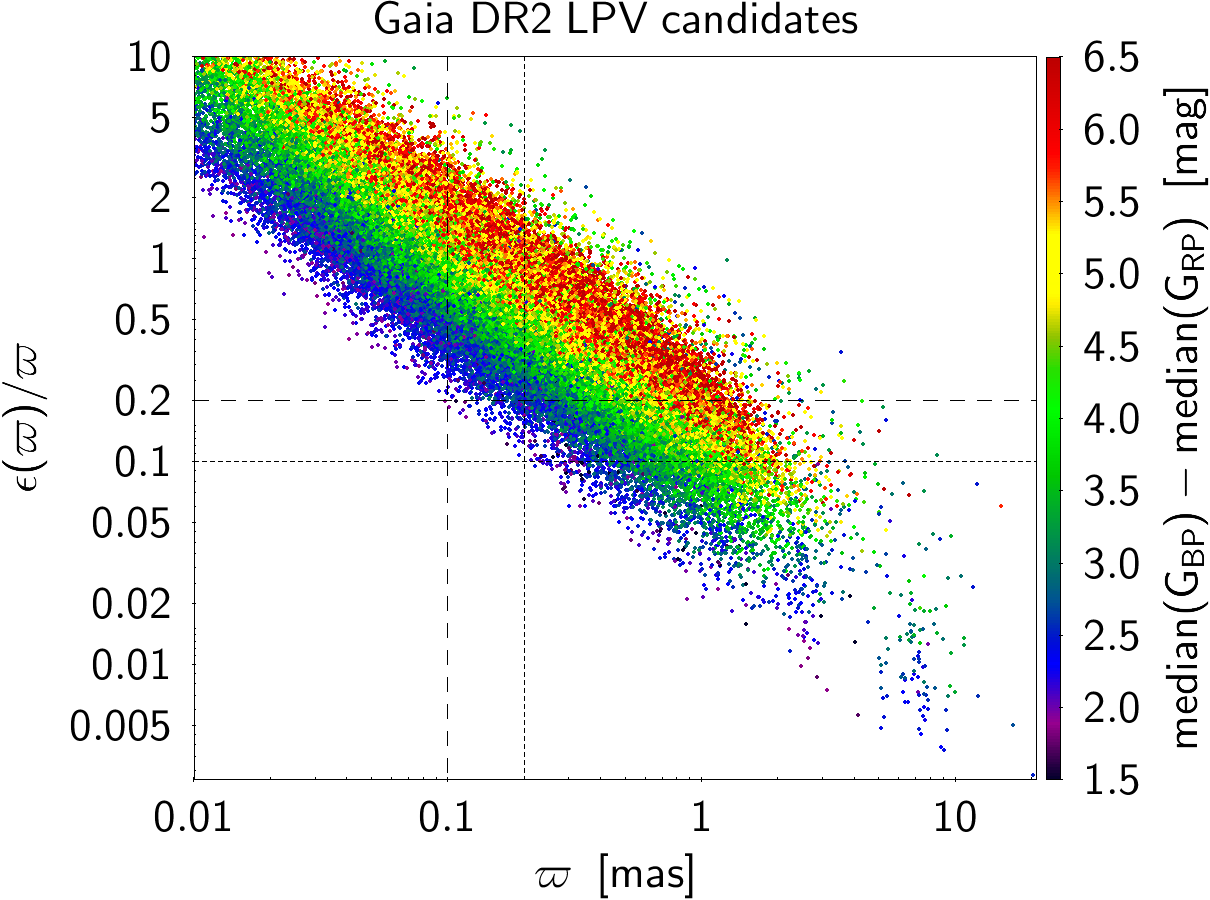}
\caption{Relative parallax uncertainty $\epsilon(\varpi) / \varpi$ versus parallax $\varpi$, in mas, of LPV candidates with $\varpi>0.01$~mas.
         The colour of each point is related to the $\mathrm{median}(\gbp) - \mathrm{median}(\grp)$ colour of the corresponding star according to the colour scale shown on the right of the figure.
         Dashed lines are drawn at $\epsilon(\varpi) / \varpi = 0.2$ and $\varpi=0.1$~mas, and dotted lines are drawn at $\epsilon(\varpi) / \varpi = 0.1$ and $\varpi=0.2$~mas as eye guide lines.
        }
\label{Fig:parallax}
\end{figure}

\begin{figure}
\centering
\includegraphics[width=\hsize]{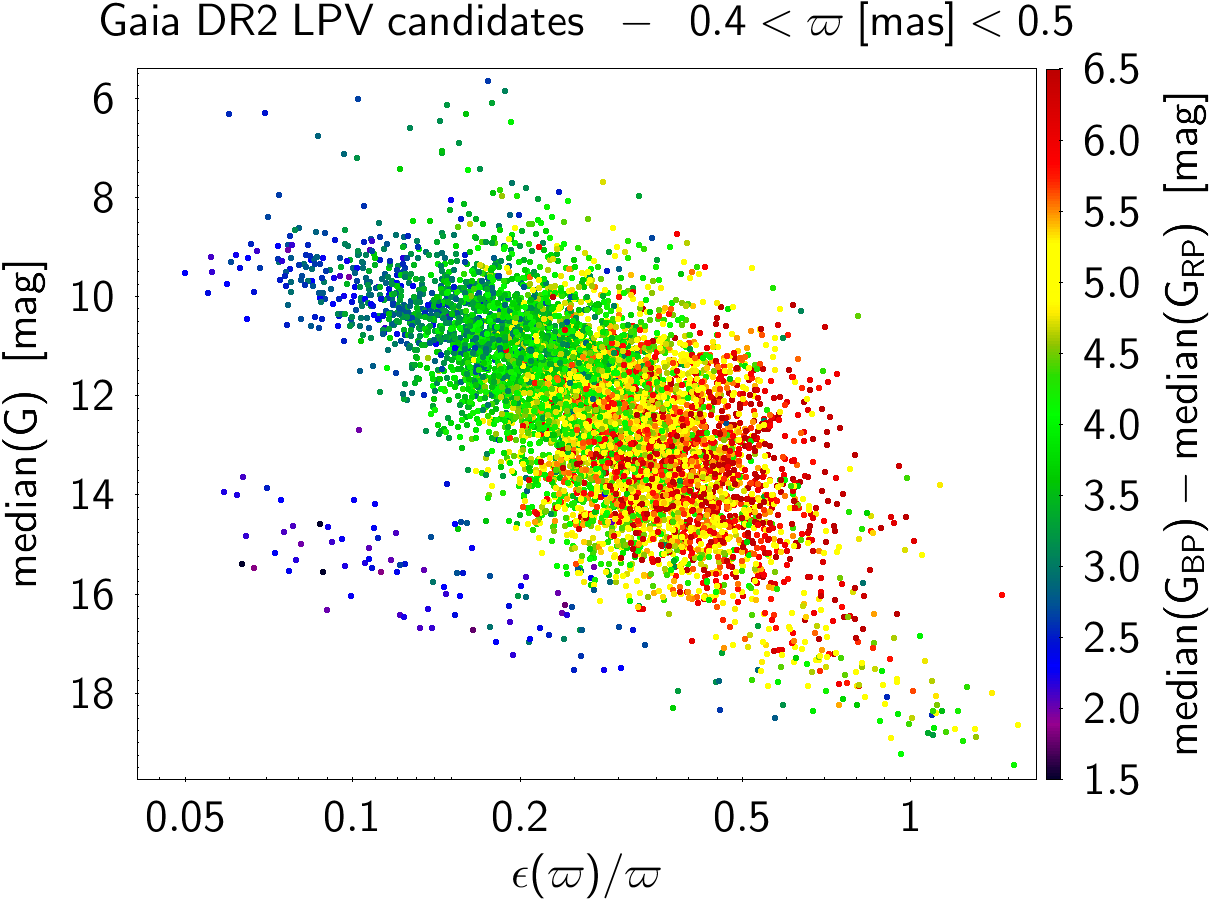}
\caption{Median \gmag magnitude versus relative parallax uncertainty of LPV candidates having $0.4 < \varpi \mathrm{[mas]} < 0.5$.
         The colour of each point is related to the $\mathrm{median}(\gbp) - \mathrm{median}(\grp)$ colour of the corresponding star according to the colour scale shown on the right of the figure.
        }
\label{Fig:magnitudeVsparallaxUncertainty}
\end{figure}

\begin{figure}
\centering
\includegraphics[width=0.95\hsize]{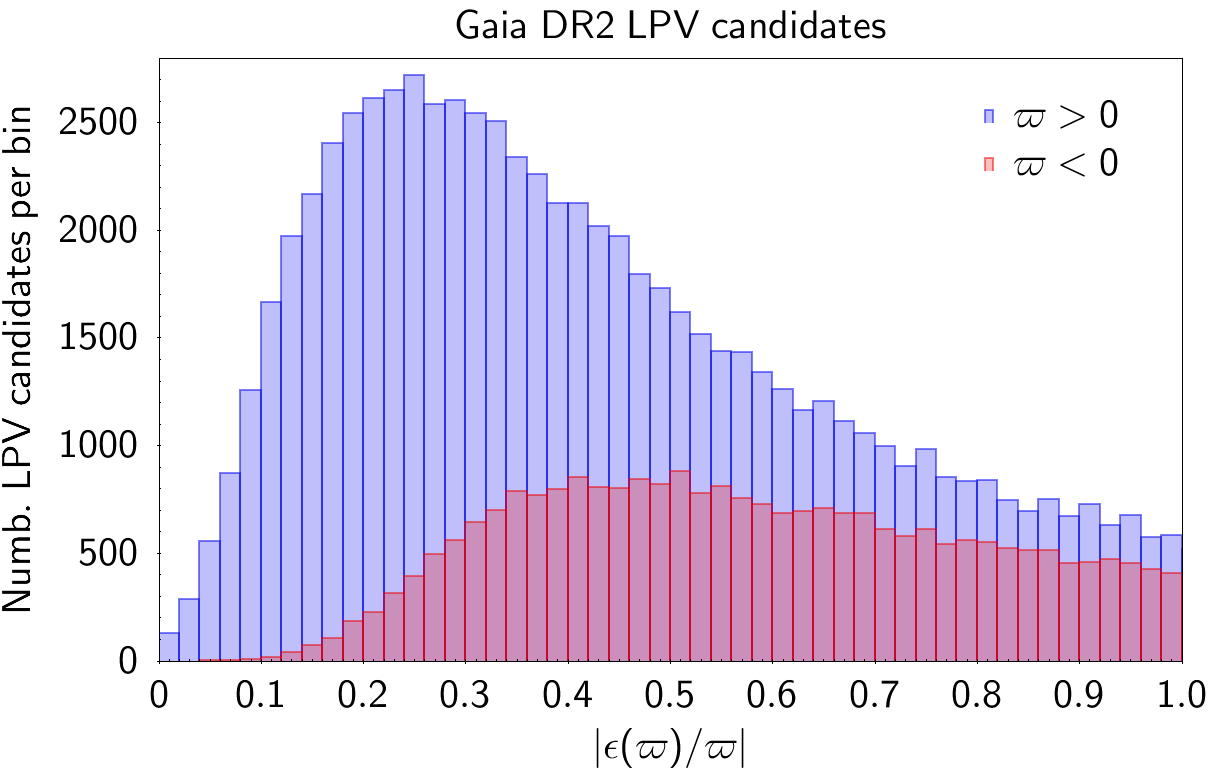}
\caption{Comparison of the relative parallax uncertainty distributions of LPV candidates with positive (blue histogram) and negative (red histogram) parallaxes.
        }
\label{Fig:histoParallaxRelativeError_agis2p2}
\end{figure}

\begin{figure}
\centering
\includegraphics[width=0.95\hsize]{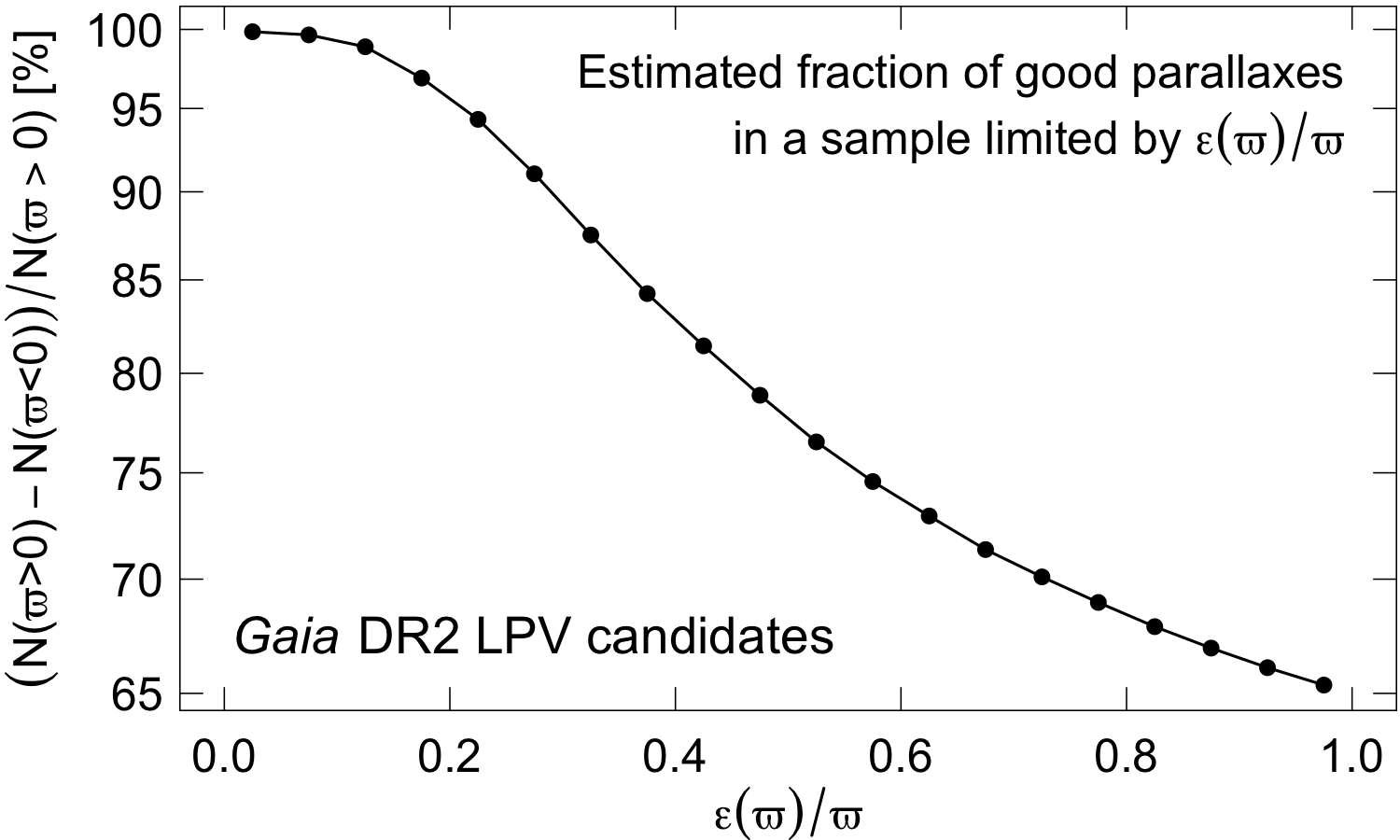}
\caption{Estimated percentage of good parallaxes in a sample of LPVs limited by an upper limit on the relative parallax uncertainty.
         The figure plots, as a function of this relative parallax uncertainty limit, $[N(\varpi>0)-N(\varpi<0)] / N(\varpi>0)$, where $N(\varpi>0)$ is the number of LPVs in the sample with positive parallaxes, and $N(\varpi>0)$ is the number of LPVs with negative parallaxes.
        }
\label{Fig:fracGoodParallaxes}
\end{figure}

The purpose of this section is to validate the LPV-specific attributes published in \Gaia DR2 on a small sample of well-known individual LPVs.
We first check the status of some of the best-studied LPVs in the literature  in the DR2 catalogue (Sect.~\ref{Sect:individualLPVs_wellStudiedCases}).
We then validate the periods and bolometric magnitudes for a sample of LPVs that have AAVSO light curves that overlap in time with \Gaia light curves (Sect.~\ref{Sect:individualLPVs_AAVSO}), and discuss the BCs in Sect.~\ref{Sect:individualLPVs_BCs}.

\subsection{Well-studied LPVs}
\label{Sect:individualLPVs_wellStudiedCases}

Of the 20 LPVs that have been studied most frequently, based on information in the SIMBAD database, only 4 stars are present in the DR2 LPV catalogue:
\texttt{R Lep},
\texttt{R Cas},
\texttt{R Aqr}, and
\texttt{g Her}.
They are listed at the top of Table~\ref{Tab:sampleBrightLPVs} with some of their ground-based and \Gaia-based properties.
Their light curves, both from Gaia and from the AAVSO, are shown in the top four panels of Fig.~\ref{Fig:LcsAAVSO_1}.
The \Gaia light curves follow the AAVSO light curves well; the \gmag magnitudes are brighter than the $V$ magnitudes by about three magnitudes, as expected for red stars \citep[see Fig.~11 in][]{2010A&A...523A..48J}.
Moreover, in three of the four stars, a majority of the \gmag measurements have been disregarded, while a good fraction of them seem visually good in comparison with the contemporaneous AAVSO light curves.
This is due to the presence of duplicate measurements in the \Gaia time-series (see Sect.~\ref{Sect:overview_lcs}), which are removed at the start of our variability processing pipeline (see Fig.~\ref{Fig:percRejectedPtsG}).
The number of duplicate measurements increases, in the mean, with brightness.
For the C star \texttt{R Lep}, for example, 30 of the 50 measurements recorded in its \Gaia \gmag-band light curve were excluded (see top panel of Fig.~\ref{Fig:LcsAAVSO_1}; note that all 50 measurements are available in DR2),
but the presence of still 20 measurements in the cleaned light curve ensured that the star was published in the DR2 catalogue of LPV candidates. Table~\ref{Tab:sampleBrightLPVs} also shows that the \Gaia period obtained for \texttt{R Lep} from its cleaned light curve agrees reasonably well with the GCVS period, despite the small number of measurements and limited time coverage.
Only the derived \Gaia bolometric magnitude is not compatible with the bolometric magnitude derived in the literature from multi-band photometry.
This discrepancy is expected for C~stars, since the BCs used for DR2 are valid only for M-type stars (see also Sect.~\ref{Sect:individualLPVs_BCs}).
The \Gaia light curve of \texttt{g Her} is also worth comparison with the AAVSO light curve in Fig.~\ref{Fig:LcsAAVSO_1}, this star being the only SRV in Table~\ref{Tab:sampleBrightLPVs}.
The light curves display the short time-scale variability that is typical for SRVs.

The 16 LPVs missing in our initial sample of 20 most-cited LPVs are
\texttt{o~Cet},
\texttt{R~Leo},
\texttt{$\,\chi\!\!$~Cyg},
\texttt{W~Hya},
\texttt{R~Aql},
\texttt{IK~Tau},
\texttt{U~Her},
\texttt{U~Ori},
\texttt{R~Hya},
\texttt{RX~Boo},
\texttt{RW~LMi},
\texttt{T~Cep},
\texttt{R~And},
\texttt{S~CrB},
\texttt{TX~Cam}, and
\texttt{R~LMi}.
It is instructive to check the reason for their non-selection for publication in DR2.
Ten of them were rejected because of too many duplicate measurements (see previous paragraph), which after light-curve cleaning led to too few measurements.
Three of the remaining stars show a \gmag--(\gbp-\grp) correlation factor smaller than 0.5, one star had only one \gbp measurement left after cleaning (the star is one of the reddest stars in the original sample of LPV candidates), which prevented us from computing the correlation factor, the Abbe value of one star is higher than 0.8, and the $QR_5(\gmag)$ variability amplitude of one star is smaller than 0.2~mag to start with. 
We refer to Appendix~\ref{Sect:SOS} for a description of the selection criteria.

\begin{figure*}
\sidecaption
\includegraphics[width=0.7\hsize]{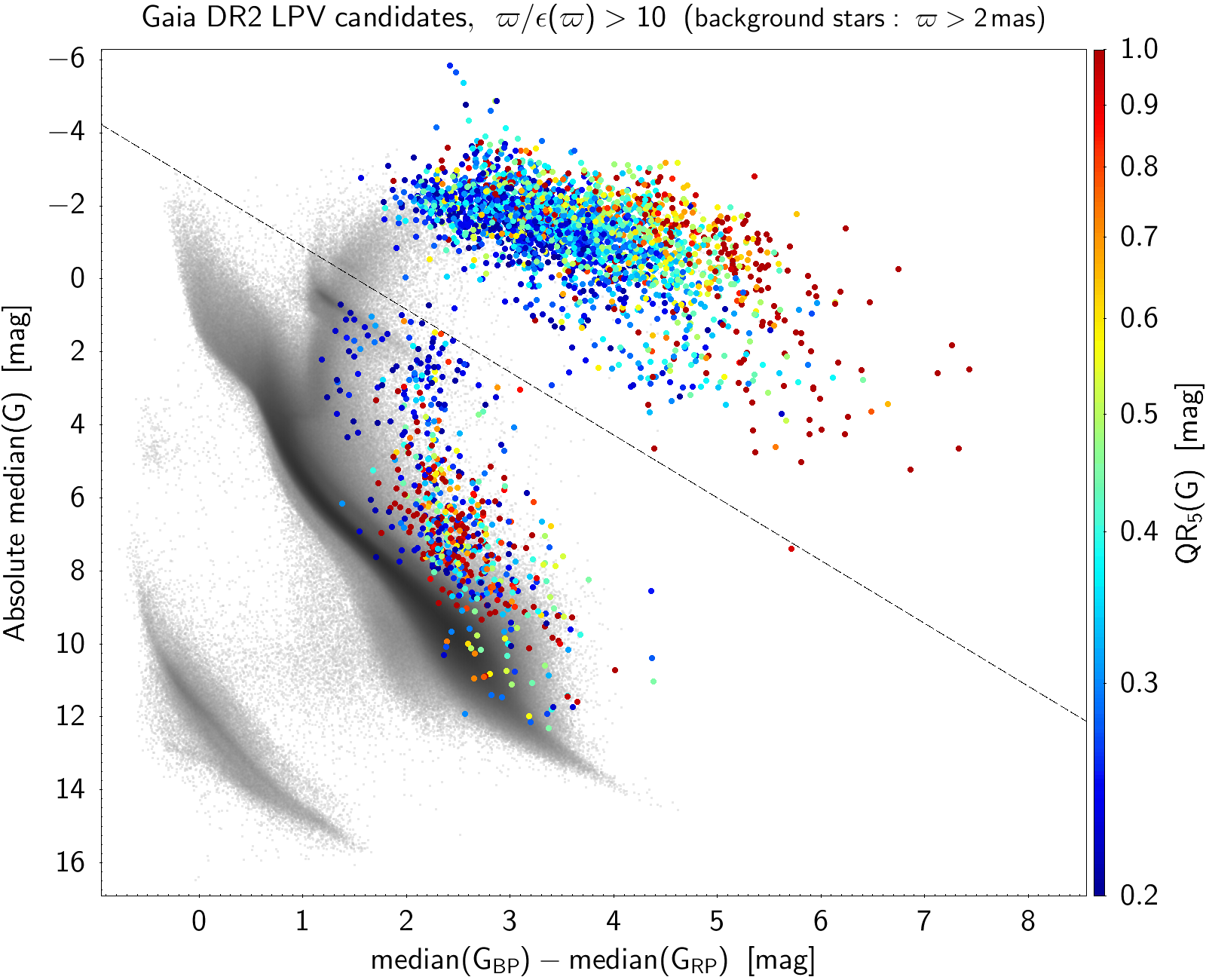}
\caption{Observational HR diagram.
         Colour points represent the sources from the DR2 LPV catalogue that have positive $\epsilon(\varpi)/\varpi <0.1$.
         The colour of the points is related to the \gmag variability amplitude, measured by the 5--95\% quantile range $QR_5$ of the cleaned \gmag time-series, according to the colour scale on the right of the figure (the upper limit of the colour scale has been truncated to $QR_5=1$~mag for better visibility, all objects with $QR_5>1$~mag being plotted in red).
         The background grey points represent the sample of stars from the whole \Gaia DR2 catalogue that have $\varpi > 2$~mas and $\epsilon(\varpi)/\varpi <0.1$.
         The dotted line separates LPV candidates from YSO candidates (see text).
        }
\label{Fig:observationalHR}
\end{figure*}

\subsection{Selected LPVs with AAVSO light curves}
\label{Sect:individualLPVs_AAVSO}

As a second approach, we considered a sample of LPVs that have good AAVSO light curves, are present in the \Gaia DR2 catalogue of LPVs, and have published integrated bolometric magnitudes derived from multi-band photometry  \citep{1996A&A...308..489K,2010A&A...524A..87K}.
The integrated bolometric magnitudes are based on spectral energy distribution (SED) fits to multi-band photometry from the visual to the far-infrared.
Our test sample of LPVs is listed in Table~\ref{Tab:sampleBrightLPVs} below \texttt{g Her}.
The sample includes only Miras because we lack AAVSO light curves for SRVs for which a bolometric magnitude is available. 
Furthermore, only objects that have AAVSO photometric time-series in the $V$ band are considered.

The \Gaia and AAVSO light curves of all the sources in our test sample are displayed in Figs.~\ref{Fig:LcsAAVSO_1} and \ref{Fig:LcsAAVSO_2}.
The \Gaia light curves follow  the AAVSO $V$-band light curves very well.

Table~\ref{Tab:sampleBrightLPVs} also shows a very good agreement between the \Gaia DR2 periods derived from the \gmag-band light curves and the periods published in the GCVS, despite the limited \Gaia observation time spans and number of measurements.
Additional validations of the \Gaia DR2 periods for LPVs are presented in Sect.~\ref{Sect:goodParallaxes} and below.

\subsection{Bolometric corrections}
\label{Sect:individualLPVs_BCs}

For the M-type stars listed in Table~\ref{Tab:sampleBrightLPVs} with amplitudes below 3 mag, the \Gaia $m_{\rm bol}$ value is within a few tenths of a magnitude of the SED fit value.
Considering that this difference is of the same order as the scatter around the BC relations \citep[e.g.][]{2010A&A...524A..87K}, the agreement is satisfactory. 
The two C~stars listed in Table~\ref{Tab:sampleBrightLPVs} show a much larger deviation between their \Gaia and SED $m_{\rm bol}$ values. 
This confirms the expected need for a separate BC relation for C-stars, which will be included in the \Gaia LPV analysis for DR3.
Similarly, the BC for Miras with very large amplitudes will be improved in DR3.

To confirm these conclusions, the \Gaia and SED $m_{\rm bol}$ values are compared in Fig.~\ref{Fig:apparentMbol} for 284 LPVs extracted from \citet{1996A&A...308..489K} and \citet{2010A&A...524A..87K}.
They represent 40\% of the sample of our reference sample of 710 stars with SED $m_{\rm bol}$, that is, those that have a crossmatch in the \Gaia DR2 LPV catalogue.
The remaining stars that do not have a \Gaia DR2 LPV crossmatch are, in the mean, the brighter stars of the sample, in agreement with Sect.~\ref{Sect:individualLPVs_wellStudiedCases}.
Figure~\ref{Fig:apparentMbol} confirms the overall good agreement of \Gaia $m_{\rm bol}$ values compared to SED values for M-type stars, although the \Gaia values are slightly too bright in the mean.
For C~stars, the \Gaia $m_{\rm bol}$ values are too faint.
We observe a tendency whereby the quality of our BCs decreases for stars showing an infrared excess from circumstellar dust as identified from the SED fit, as expected from the BC implementation in DR2.

\section{Validation 2: Sample with good parallaxes}
\label{Sect:goodParallaxes}

\begin{figure*}
\centering
\includegraphics[width=0.49\hsize]{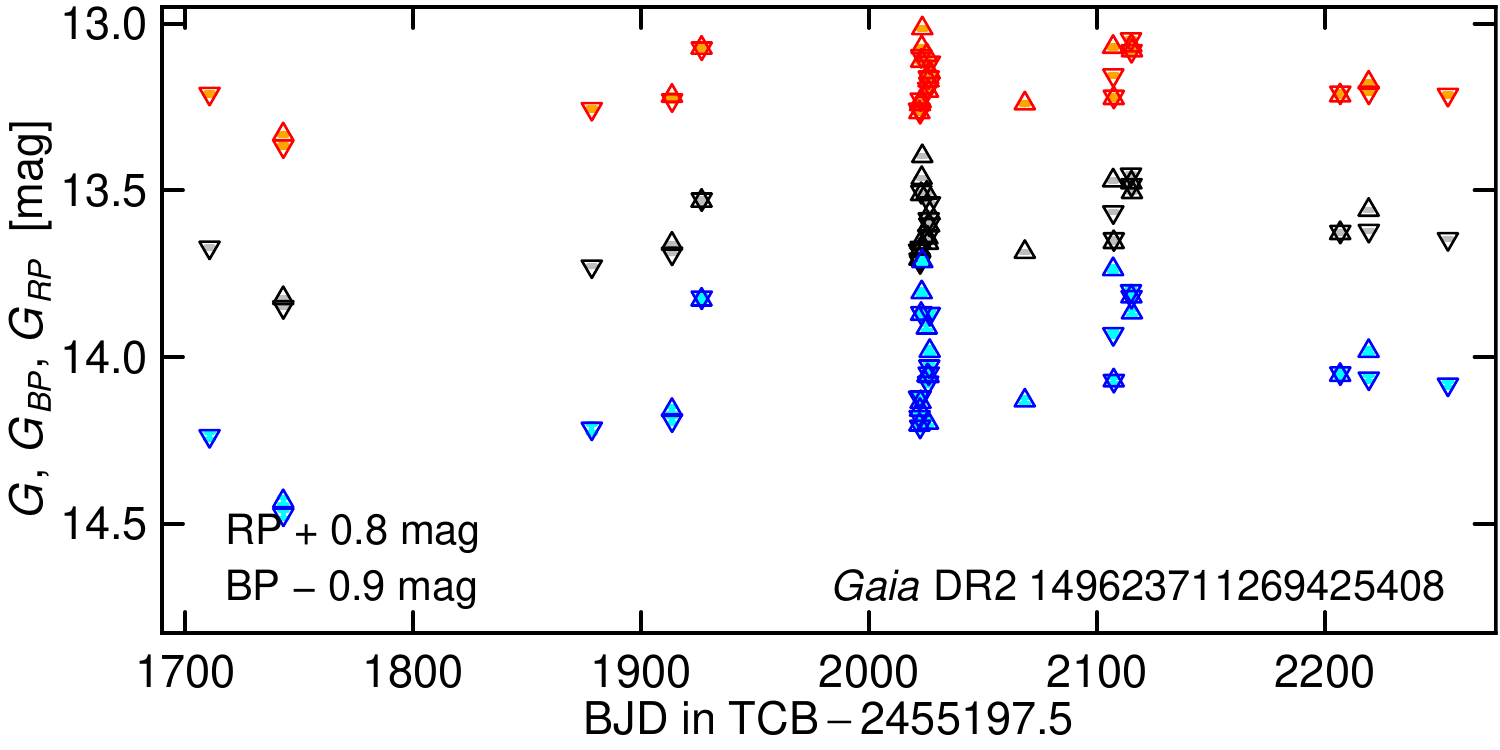}
\includegraphics[width=0.49\hsize]{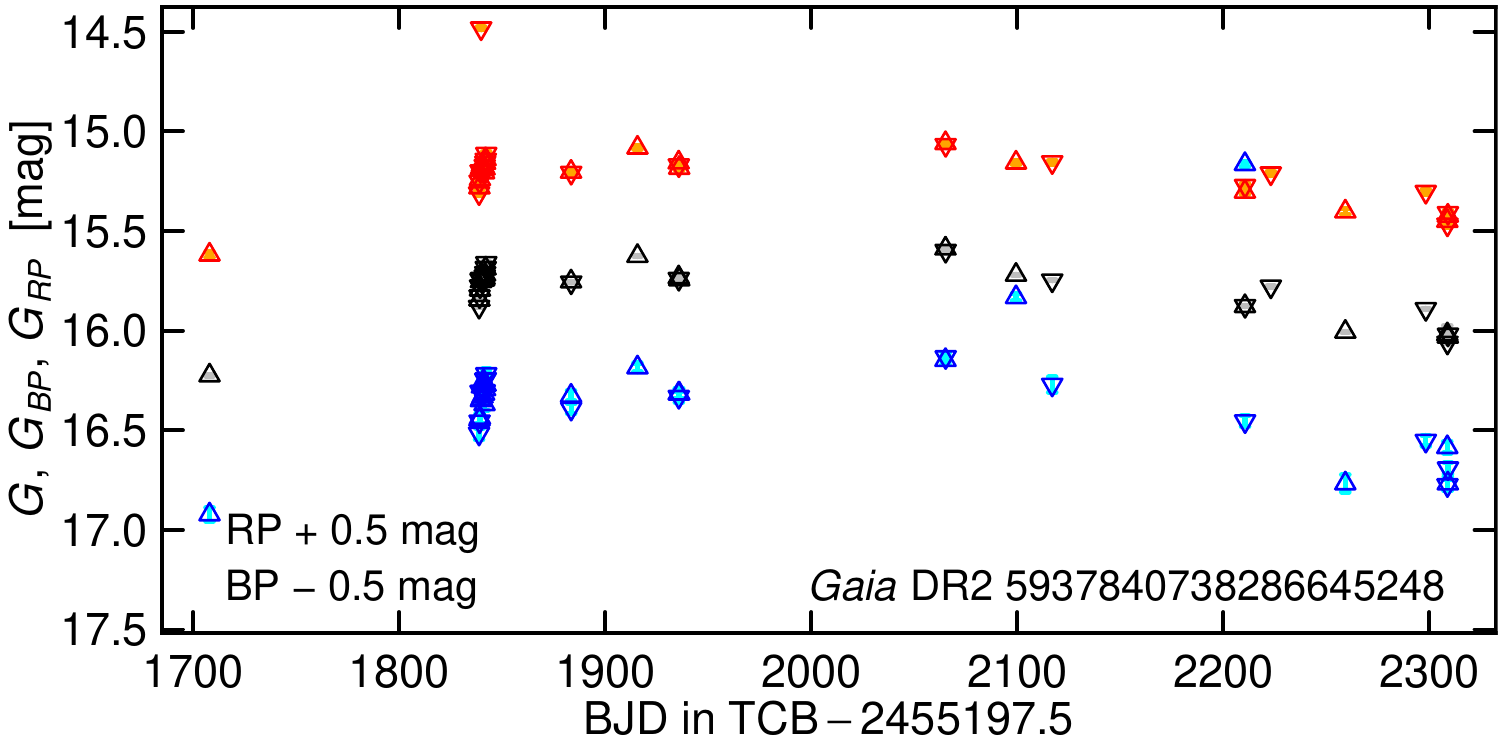}
\includegraphics[width=0.49\hsize]{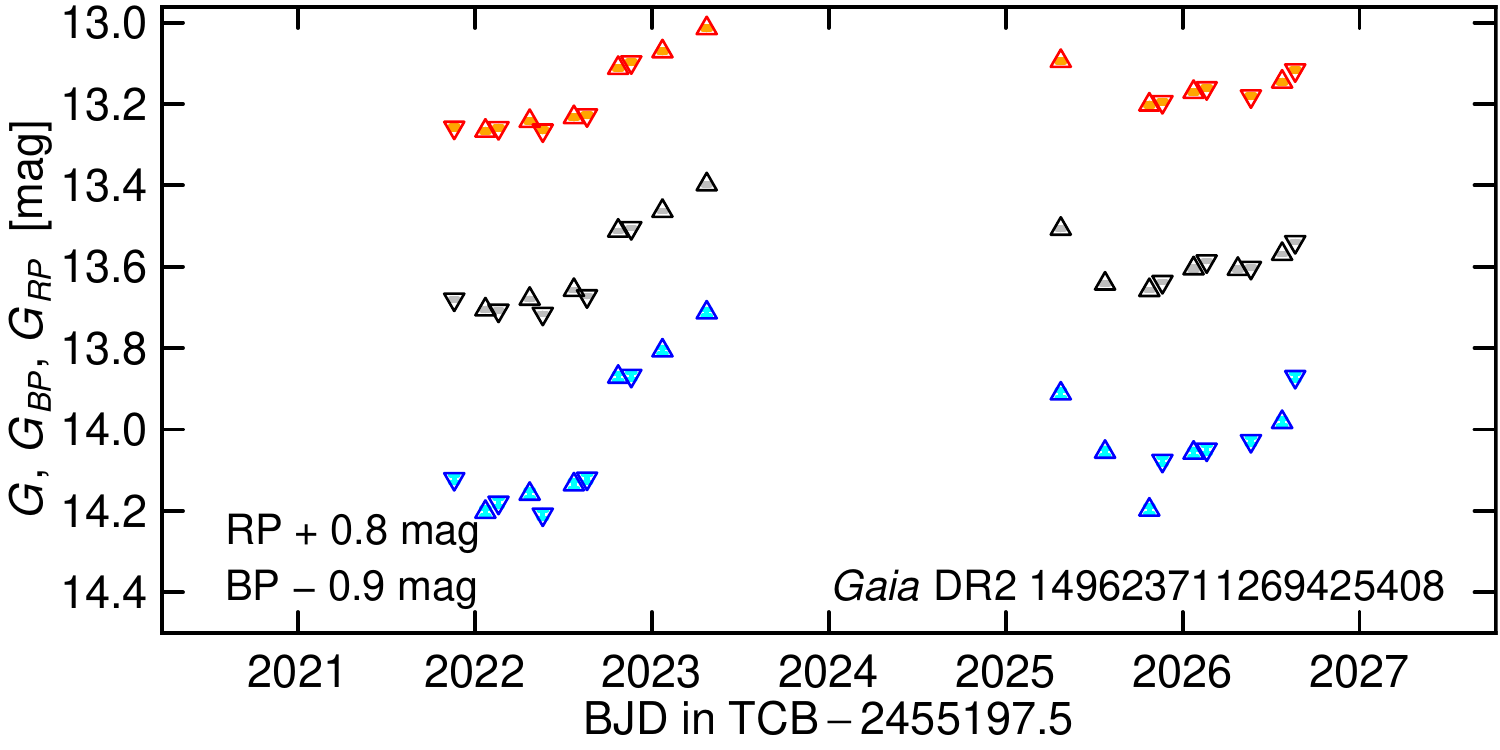}
\includegraphics[width=0.49\hsize]{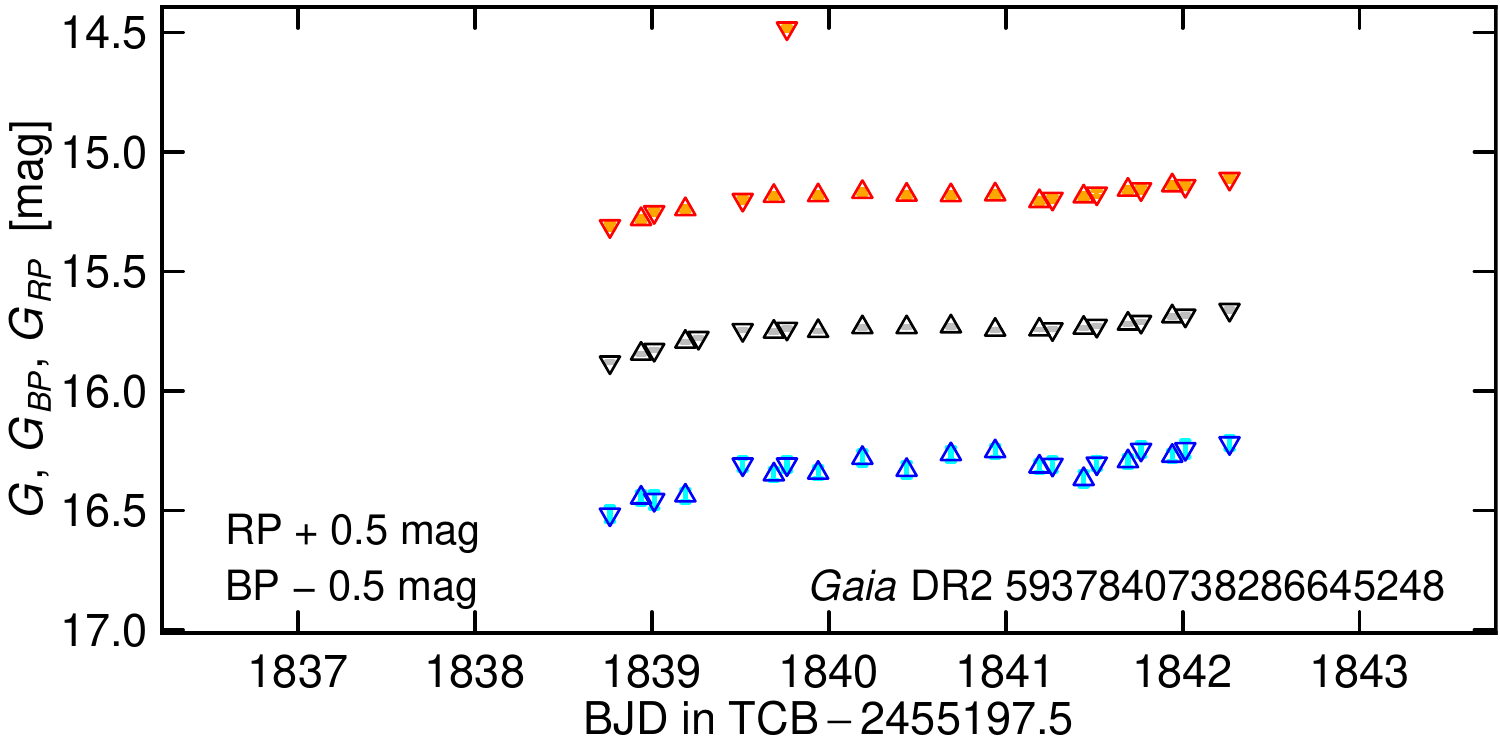}
\caption{\textbf{Top panels:} Light curve examples of two YSOs present in the \Gaia DR2 catalogue of LPV candidates.
         The meaning of the symbols is identical to that in Fig.~\ref{Fig:LcsExamples1}.
         The \gbp and \grp magnitudes are shifted by the amount indicated in each panel for better visibility of the light variations.
         \textbf{Bottom panels:} Same as the top panels, but zoomed on a 7-day time interval around a concentration of measurements in time.
        }
\label{Fig:LCs_YSOs}
\end{figure*}

\begin{figure*}
\centering
\includegraphics[width=\hsize]{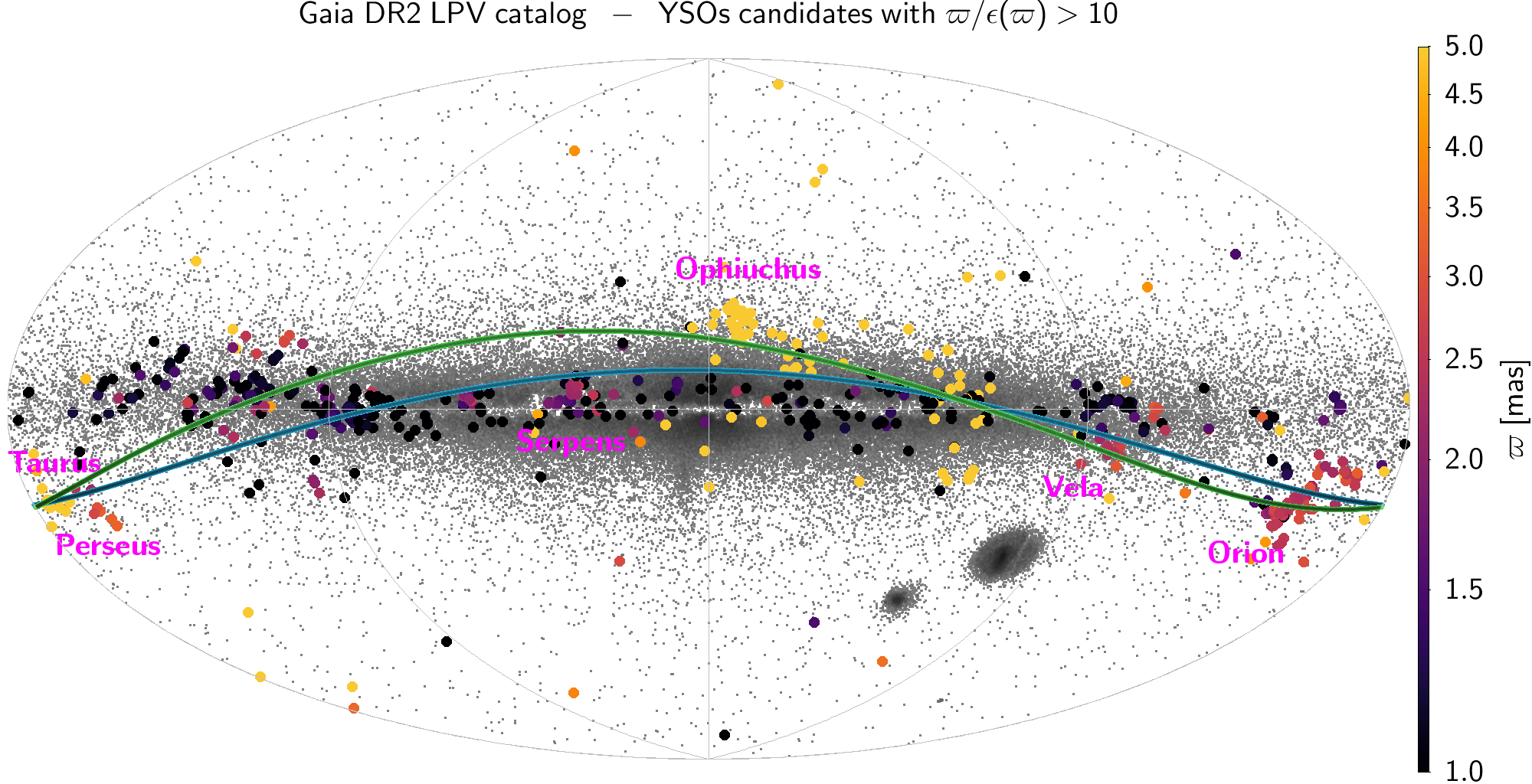}
\caption{Sky distribution, in Galactic coordinates, of YSO candidates (large filled circles) in the \Gaia DR2 catalogue of LPVs with $\varpi/\epsilon(\varpi) > 10$.
         The colour of the markers is related to the parallax according to the colour scale shown on the right of the figure (sources with parallaxes below the minimum or above the maximum values shown on the scale are coloured with the colour at the respective end of the scale). 
         The background small grey points represent all LPV candidates in the catalogue.
         The location of the Gould belt determined from regions of stellar formation and dark clouds \citep{TaylorDickman_scoville87} are indicated by the green and blue lines, respectively, and its main stellar associations are labelled in magenta.
        }
\label{Fig:sky_YSOs}
\end{figure*}

\begin{figure}
\centering
\includegraphics[width=\hsize]{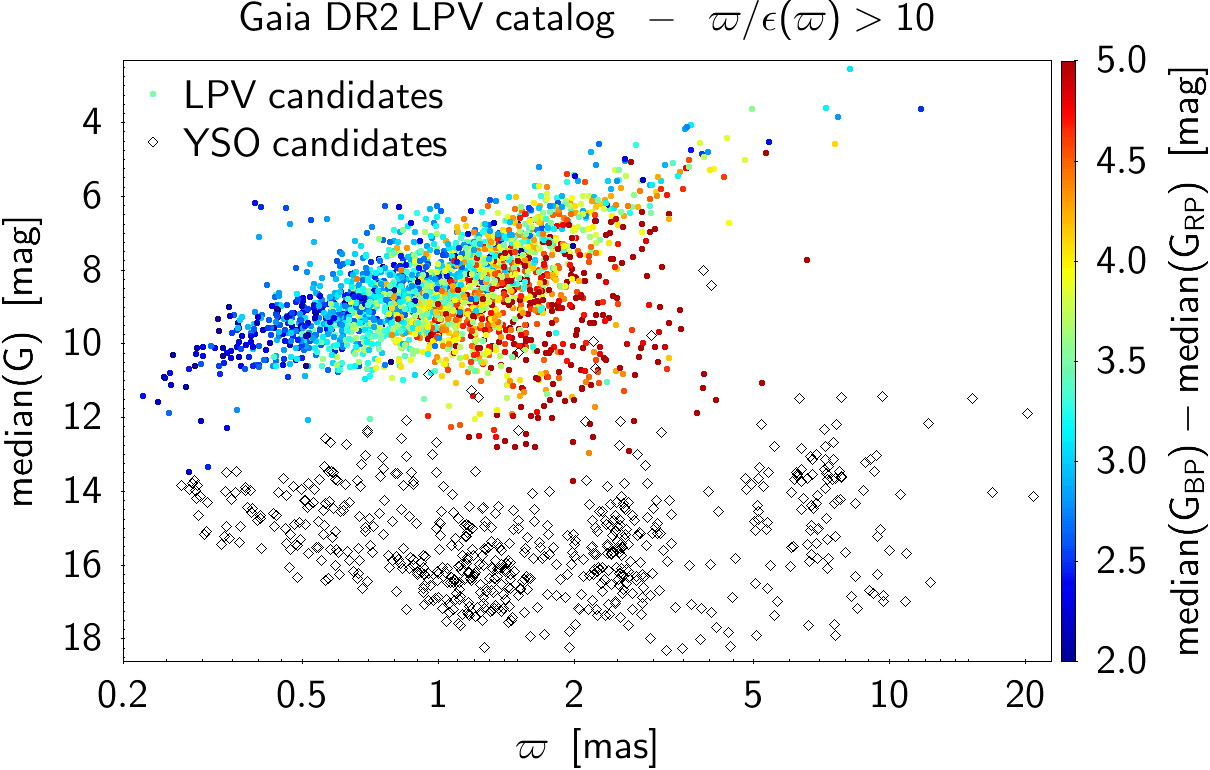}
\caption{\gmag magnitude vs. parallax in mas of LPV (filled circles) and YSO candidates (open diamonds) identified by their location in the observational HR diagram (see Fig.~\ref{Fig:observationalHR}).
          The sample displayed in the figure is restricted to sources with $\varpi/\varepsilon(\varpi) > 10$.
          The marker colours for the LPVs are related to their $\mathrm{median}(\gbp) - \mathrm{median}(\grp)$ colours according to the colour scale on the right of the figure.
        }
\label{Fig:gVersusParallaxYSOs}
\end{figure}

\begin{figure}
\centering
\includegraphics[width=\hsize]{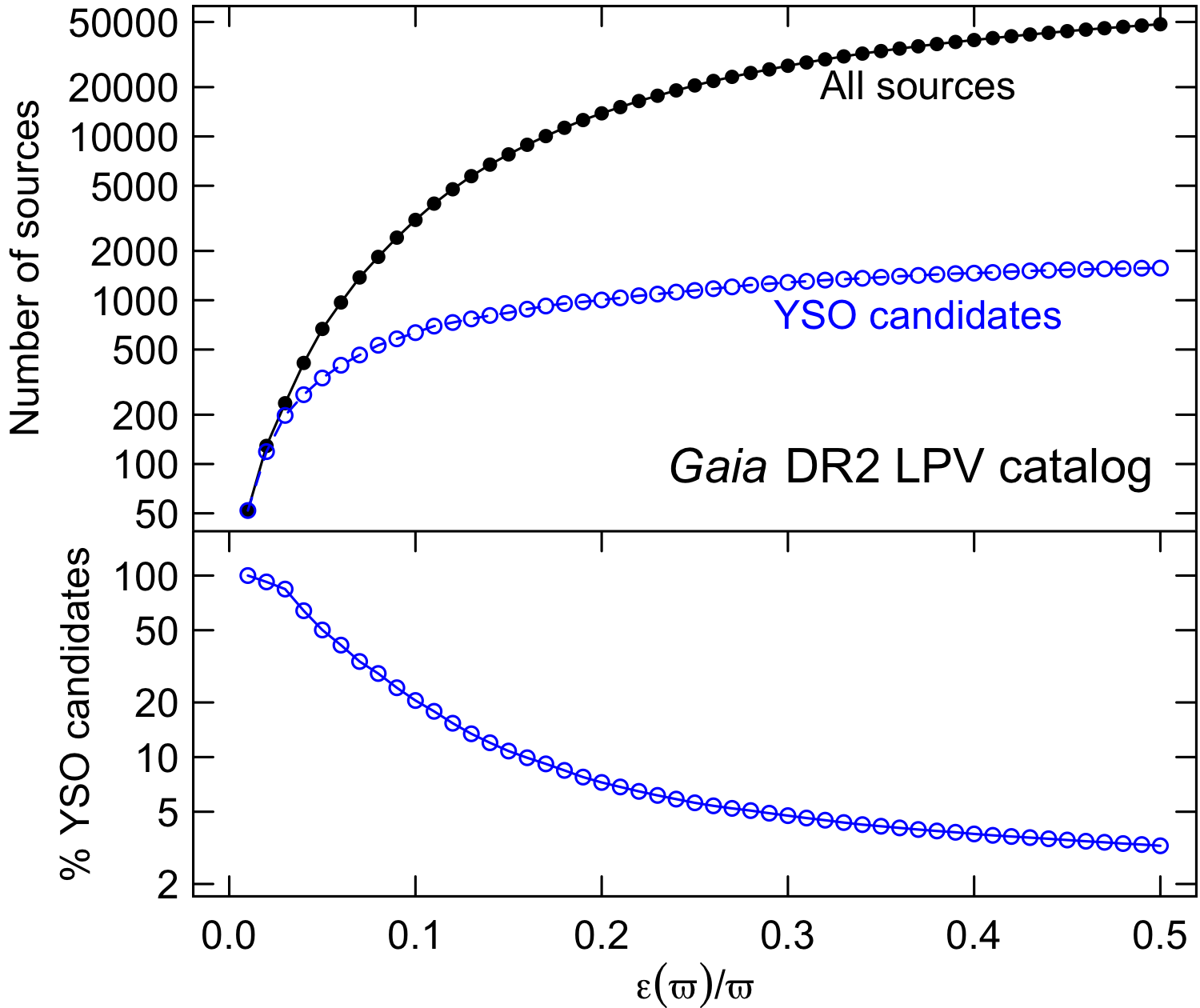}
\caption{\textbf{Top panel:} Cumulative distribution of the total number of sources (upper black solid line) and of the number of YSO contaminants (lower blue dashed line) in the \Gaia DR2 catalogue of LPVs as a function of relative parallax uncertainty.
         Young stellar objects are defined as objects lying in the observational diagram below the dashed line shown in Fig.~\ref{Fig:observationalHR}.
         \textbf{Bottom panel:} Percentage of the cumulative number of YSOs relative to the cumulative number of all sources as a function of relative parallax uncertainty.
        }
\label{Fig:YsoVsRelParalUncertainty}
\end{figure}

In this section, we check the sample of DR2 LPV candidates with good parallaxes ($\varpi/\epsilon(\varpi) > 10$).
Their positioning in the HR diagram, in particular, will allow us to identify potential contaminants.
The sample is defined in Sect.~\ref{Sect:goodParallaxes_selection} and analysed in Sects.~\ref{Sect:goodParallaxes_observationalHR} and \ref{Sect:goodParallaxes_YSOs}.

\subsection{Sample selection}
\label{Sect:goodParallaxes_selection}

We first wish to select a sample of LPV candidates based on parallax precision.
For LPVs, the precision reached on the parallax depends on the $\gbp-\grp$ colour of the star.
This is shown in Fig.~\ref{Fig:parallax}, where relative parallax uncertainties are plotted versus parallax, with each point colour-coded according to the $\gbp-\grp$ colour of the stars.
Very red LPV candidates have, in the mean, less precise parallaxes than less red LPV candidates at any given parallax. 
At $\varpi=0.2$~mas (distance of $\sim$5~kpc), for example, the relative precision uncertainty of a star with $\gbp-\grp=2.5$~mag can reach close to 10\%, while it can increase above 100\% at the same distance if $\gbp-\grp \gtrsim 5$~mag.

The colour dependence of parallax precision seen in Fig.~\ref{Fig:parallax} for LPVs reflects the magnitude dependence of parallax precision, according to which fainter stars have less precise parallaxes.
LPVs have SEDs that peak in the infrared.
As a consequence, the redder an LPV star, the fainter it appears in \gmag at a given distance. This leads to less precise parallaxes.
This is shown in Fig.~\ref{Fig:magnitudeVsparallaxUncertainty}, which plots the \gmag magnitude of all LPV candidates at a given parallax, taken equal to $(0.45 \pm 0.05)$~mas.
The anti-correlation between \gmag magnitude and relative parallax uncertainty is clearly visible in the figure, as is the $\gbp-\grp$ colour dependence (shown in colour in the figure), according to which the reddest stars are the faintest and have the least precise parallaxes.

The relative parallax uncertainty threshold used to define our sample with good parallaxes is estimated by comparing the distribution of $|\varepsilon(\varpi)/\varpi|$ for positive and negative parallaxes.
This is shown in Fig.~\ref{Fig:histoParallaxRelativeError_agis2p2}.
The positive-parallax distribution, plotted in blue, peaks around $\varepsilon(\varpi)/\varpi \simeq 0.25$.
The negative-parallax distribution, on the other hand, plotted in red, peaks around $\varepsilon(\varpi)/\varpi \simeq 0.45$, but with far fewer star per $\varepsilon(\varpi)/\varpi$ bin than the positive-parallax distribution.
The number of good parallaxes is estimated by the number of positive parallaxes that are in excess of the number of negative parallaxes in any given $\varepsilon(\varpi)/\varpi$ bin.
Therefore, to estimate the fraction of good parallaxes in a sample limited by $\varepsilon(\varpi)/\varpi$, we compute the number of positive-parallax sources that are in excess of the number of negative-parallax sources in the sample, and compare it to the number of positive-parallax sources.
This ratio is shown (in percentage) in Fig.~\ref{Fig:fracGoodParallaxes} as a function of the maximum $\varepsilon(\varpi)/\varpi$ defining the sample.
Based on this figure, we define our sample of good-parallax LPV candidates with the criterion $\varpi/\varepsilon(\varpi) > 10$.
This corresponds to a contamination from poor parallaxes at the percent level.

The sample of DR2 LPV candidates with $\varpi/\varepsilon(\varpi) > 10$ contains 3093 sources. According to Fig.~\ref{Fig:parallax}, this corresponds to a sphere around the Sun of $\sim$1~kpc radius for the reddest LPV candidates ($\gbp-\grp \gtrsim 5$~mag) to $\sim$5~kpc radius for the bluest LPV candidates ($\gbp-\grp \lesssim 2.5$~mag).

\paragraph{A note on \Gaia DR2 parallaxes of LPVs.}
We end this section by describing the problem of magnitude and colour variations in the LPV parallax determination.
The calibration contains colour- and magnitude-dependent terms when computing the parallax.
For the Hipparcos survey, \citet{PourbaixPlataisDetournay_etal03} showed that for LPVs, the use of a single chromaticity correction is not sufficient.
Because these stars show large colour variations within a cycle, chromaticity corrections must be applied to each epoch data.
In \Gaia DR2, parallaxes are determined assuming constant mean colour and magnitude for each source \citep{DR2-DPACP-14}.
Therefore, the parallaxes of some LPVs may be incorrectly evaluated in DR2.
In a future release, epoch photometry will be used instead of constant values.

\subsection{Observational HR diagram: checking LPV classification}
\label{Sect:goodParallaxes_observationalHR}

\begin{figure}
\centering
\includegraphics[width=\hsize]{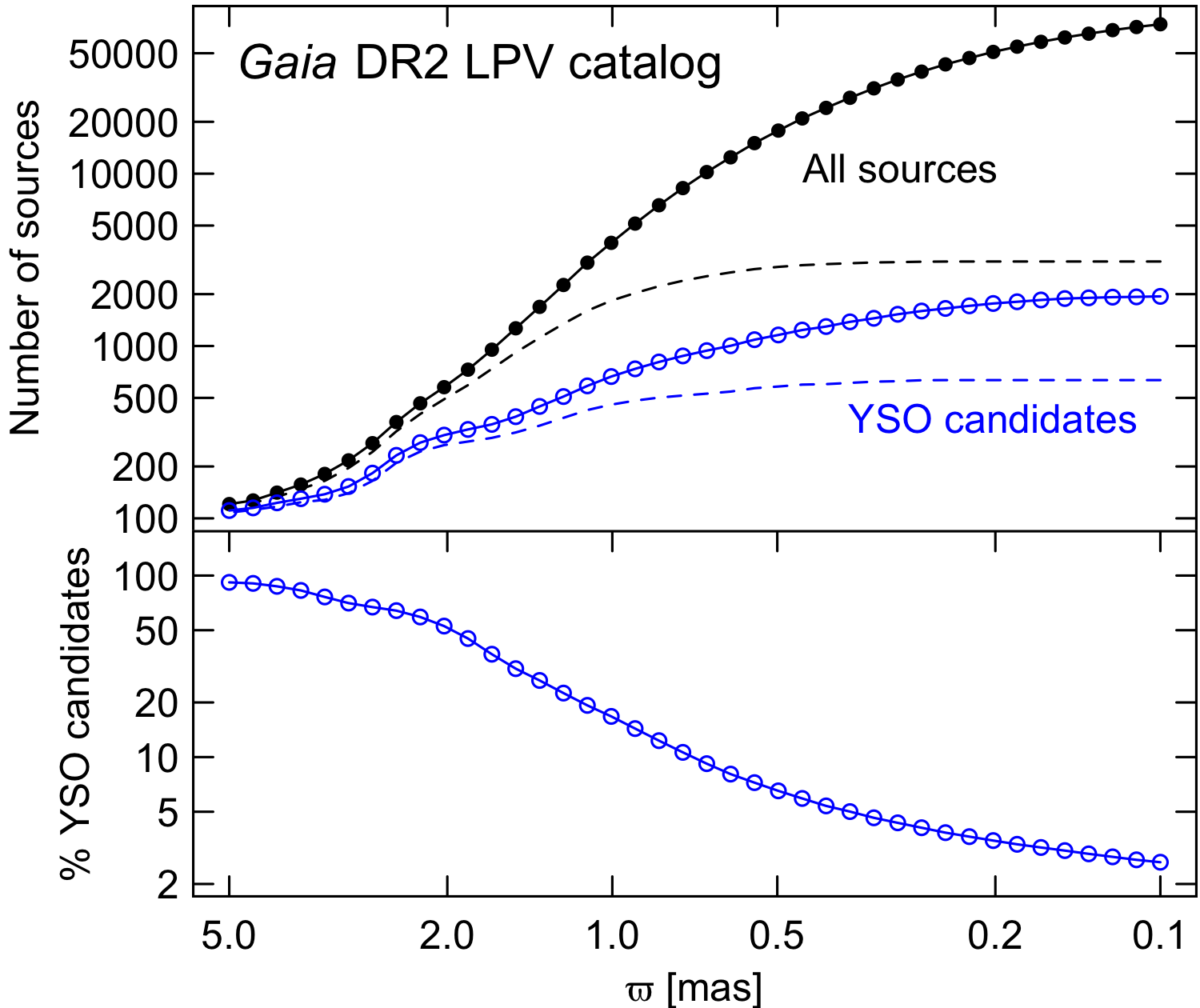}
\caption{Same as Fig.~\ref{Fig:YsoVsRelParalUncertainty}, but as a function of decreasing parallax, expressed in mas.
         The black and blue dashed lines in the upper panel are similar to the respective solid lines, but considering the sub-sample of sources with $\varpi/\epsilon(\varpi)>10$.
        }
\label{Fig:YsoVsParallax}
\end{figure}

Figure~\ref{Fig:observationalHR} shows the observational HR diagram of the sample of DR2 LPV candidates (in colour) with $\varepsilon(\varpi)/\varpi<0.1$ (positive parallaxes only)  defined in Sect.~\ref{Sect:goodParallaxes_selection}, in addition to all Gaia DR2 data with $\varpi > 2$~mas and the same relative precision of $\varepsilon(\varpi)/\varpi<0.1$ \citep[in grey; see][for more details on the selection of these background stars]{DR2-DPACP-Eyer}.
At $\epsilon(\varpi) / \varpi = 0.1$, the parallax uncertainty contributes to a $\sim$0.22~mag uncertainty on $M_\textrm{bol}$ (see Eq.~\ref{Eq:sigma_Mbol} in Appendix~\ref{Sect:SOS}).
The plot reveals a split of DR2 LPV candidate sample into two groups, with a clear separation indicated by the dotted line drawn in the diagram.
The slope of the line is taken equal to 1.72 to reproduce the ratio of the total-to-selective absorption in \gmag to $E$($\gbp-\grp$) that was visually determined from the slope of the red clump extension visible in the background of Fig.~\ref{Fig:observationalHR}.
The intercept of the line, on the other hand, has been visually determined to have a reasonable separation of the two populations.

The majority of the DR2 LPV candidates (80\%) plotted in Fig.~\ref{Fig:observationalHR} are found to lie above the dashed line.
A comparison with the distribution of the whole \Gaia sample drawn in the background of the figure confirms that these LPV candidates are located in the upper part of the giant branch.
As the stars evolve along the giant branch, they become redder and their variability amplitude increases.
The overall increase in $QR_5(\gmag)$ amplitude, shown colour-coded in Fig.~\ref{Fig:observationalHR}, with increasing $\gbp-\grp$ agrees with these expectations.
This validates the LPV classification of more than 90\% of the DR2 LPV candidates with good parallaxes.
 
A group of 634 stars (i.e. 20\%) in our sample of 3093 objects with good parallaxes ($\varpi/\epsilon(\varpi) > 10$) are located below the dashed line shown in Fig.~\ref{Fig:observationalHR}.
Their luminosities are too low to be associated with the upper part of the giant branch, suggesting a different evolutionary stage.
They are further discussed in Sect.~\ref{Sect:goodParallaxes_YSOs}.

\subsection{Contamination by young stellar objects}
\label{Sect:goodParallaxes_YSOs}

Sources located below the dashed line shown in Fig.~\ref{Fig:observationalHR} lie close to the main sequence in the observational HR diagram.
They are associated with mainly YSOs.
These objects are red and have variability amplitudes larger than 0.2~mag, like LPVs, but they show large-amplitude variability on very short timescales, unlike LPVs.
This short-timescale variability can be observed in the \Gaia light curves depending on their occurrence (they are not regular) and on the \Gaia scanning law.
The light curves of several YSO candidates are shown in Fig.~\ref{Fig:LCs_YSOs}.

The YSO classification is further confirmed by their distribution on the sky, shown in Fig.~\ref{Fig:sky_YSOs}.
YSO candidates that are closer than $\sim$600~pc from the Sun are seen to be distributed along the Gould belt (the parallax is colour-coded in Fig.~\ref{Fig:sky_YSOs}).
They clump in the main young stellar associations that are associated to the belt, including Taurus ($\sim$140~pc), Perseus ($\sim$400~pc), Serpens ($\sim$415~pc), Ophiuchus ($\sim$140~pc), and Orion ($\sim$500~pc) \citep{deZeeuwHoogerwerfdeBruijne_etal99,PerrotGarnier2003}.
We also note in Fig.~\ref{Fig:sky_YSOs} that a fraction of the YSO candidates have larger distances (they appear as black filled circles in the figure) and are mainly distributed throughout the Galactic plane.
They may be associated with distant star-forming regions, such as Cyg X and Cyg OB7.
Additional studies will be required to confirm or refute the YSO nature of these contaminants.
The \gmag -magnitude distribution of all the YSO and LPV candidates with $\varpi / \epsilon(\varpi) > 10$ is shown in Fig.~\ref{Fig:gVersusParallaxYSOs} as a function of parallax.

The percentage of YSO contamination as a function of relative parallax uncertainty is shown in Fig.~\ref{Fig:YsoVsRelParalUncertainty}.
A similar figure, but as a function of parallax, is shown in Fig.~\ref{Fig:YsoVsParallax}.
Since YSOs are intrinsically much fainter than LPVs, they contaminate the catalogue only up to a (relatively) short distance from the Sun.
Figure~\ref{Fig:YsoVsParallax} shows that about half of the LPV candidates contained in a sphere of radius $\sim$500~pc are YSO contaminants.
The contamination decreases below 10\%  at $\varpi=0.7$~mas ($\sim 1.4$~kpc), and below 5\% at $\varpi=0.37$~mas ($\sim$2.7~kpc).
Figure~\ref{Fig:YsoVsParallax} further suggests a maximum number of YSO contaminants in the DR2 catalogue of LPV candidates of about 2000.
When we consider only sources with relative parallax uncertainties better than 10\% (dashed lines in Fig.~\ref{Fig:YsoVsParallax}), the number of YSO candidates amounts to about ~600.
We thus expect a total number of YSOs between 600 and 2000 in the \Gaia DR2 catalogue of LPVs.
We did not further analyse these sources, for example, by crossmatching them with SIMBAD.

\section{Validation 3: All-sky comparison with ASAS\_SN}
\label{Sect:ASAS}

\begin{figure*}
\centering
\includegraphics[width=\hsize]{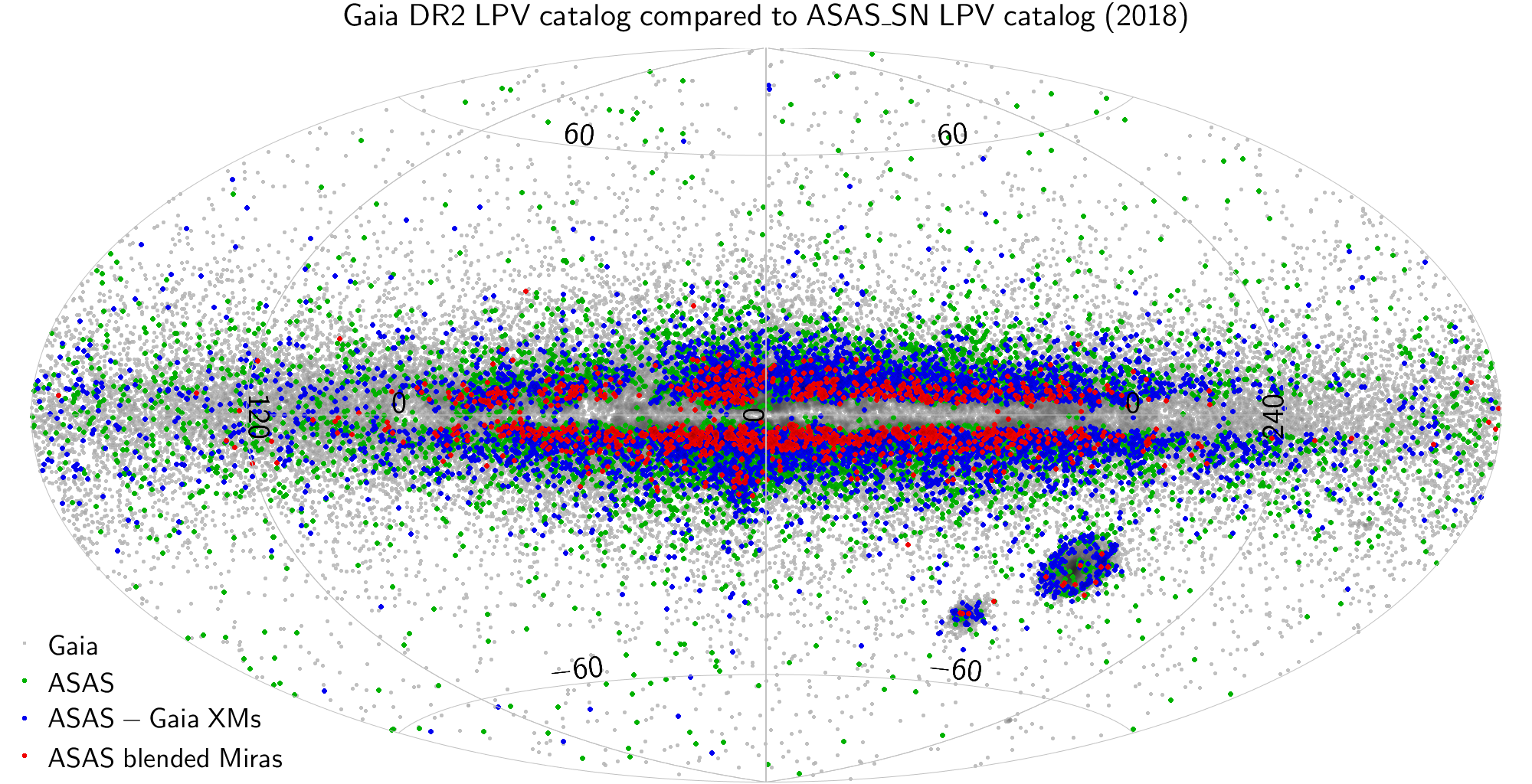}
\caption{Sky distribution in Galactic coordinates of \Gaia LPV candidates (background grey points), of ASAS\_SN LPV candidates (Mira and semi-regular variables; green points), of ASAS\_SN--\Gaia crossmatches (blue points), and of ASAS\_SN--\Gaia crossmatches in which the ASAS\_SN source is a potential blended source (red points, see text for a discussion on blended sources).
        }
\label{Fig:sky_ASAS}
\end{figure*}

\begin{figure}
\centering
\includegraphics[width=\hsize]{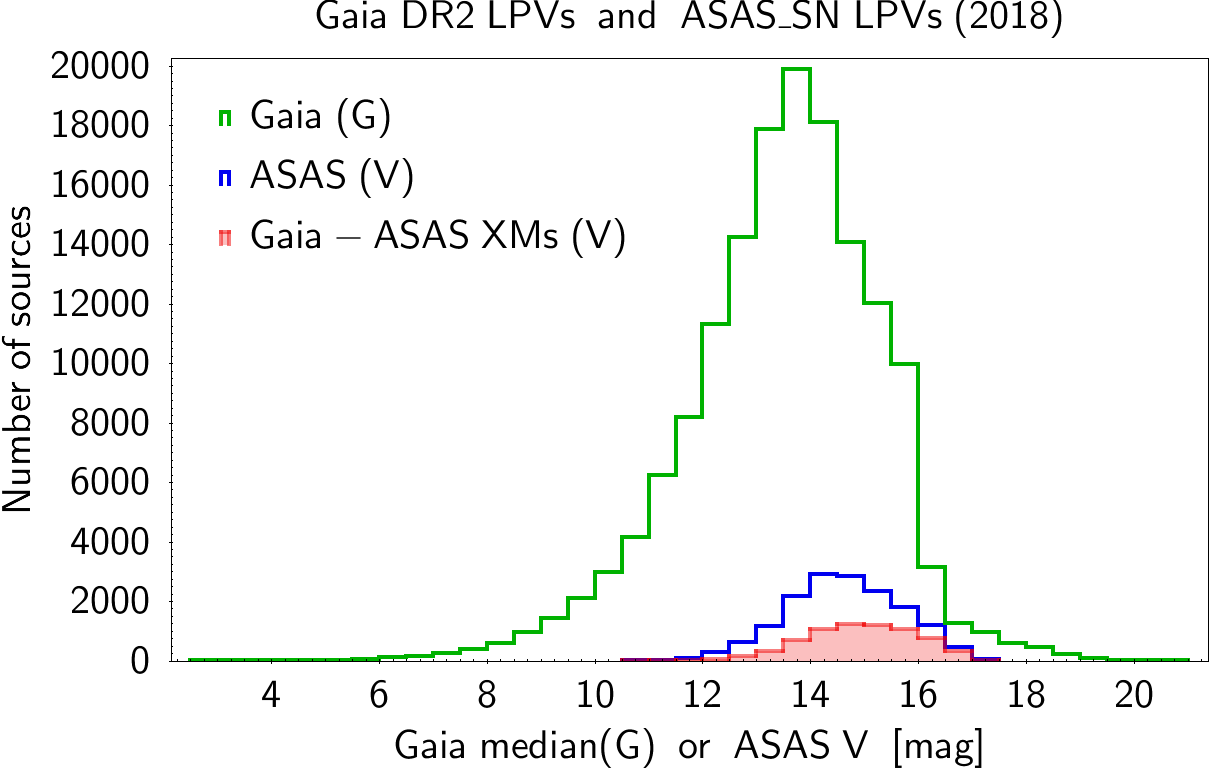}
\caption{Magnitude histograms of \Gaia sources (median \gmag magnitudes; in green), of the ASAS\_SN sources ($V$ magnitudes; in blue), and of \Gaia-ASAS\_SN crossmatches ($V$ magnitudes; filled red histogram).
        }
\label{Fig:histoMagnitudesGaiaAsas}
\end{figure}

\begin{figure}
\centering
\includegraphics[width=\hsize]{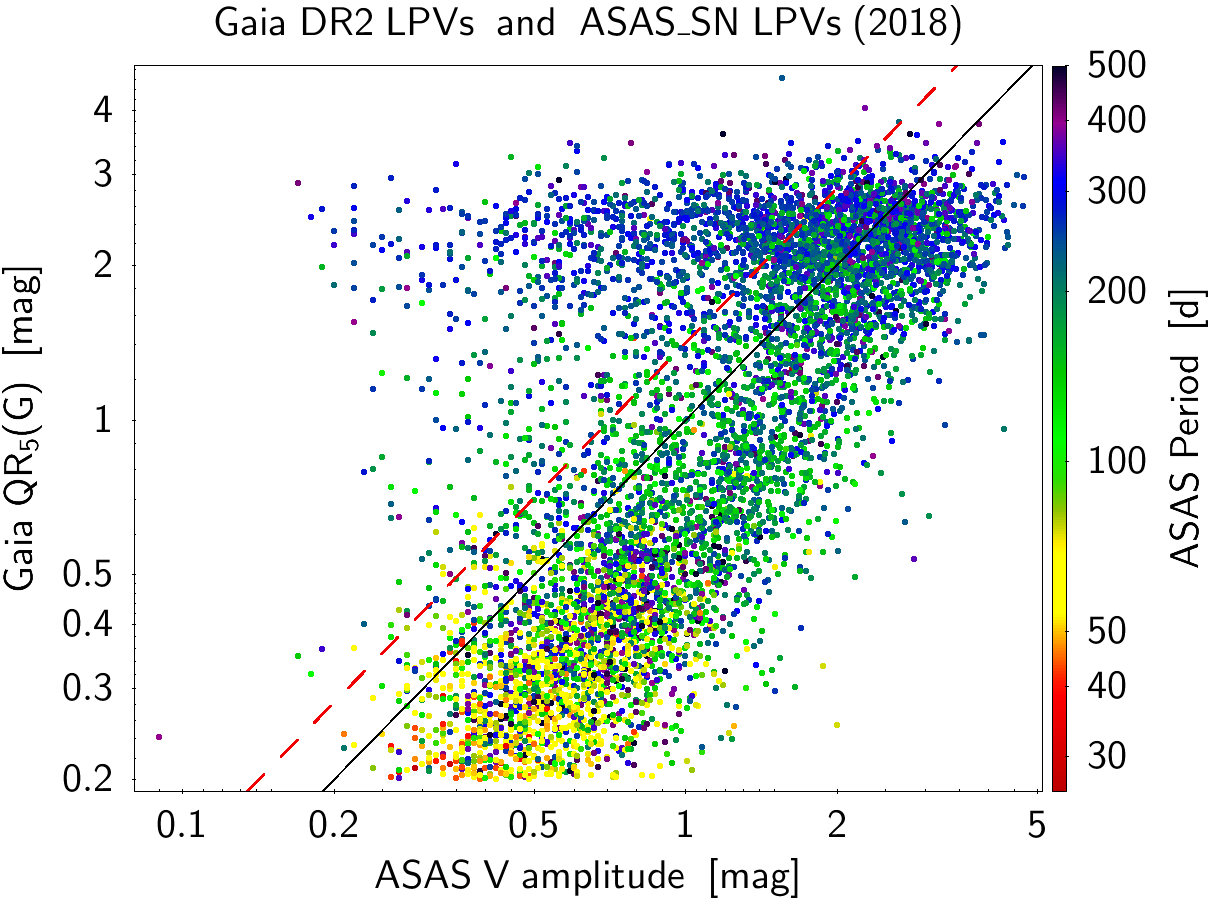}
\caption{Comparison of \Gaia amplitudes (5--95\% quantile range $QR_5$ of the cleaned \gmag time-series) to ASAS\_SN amplitudes ($V$ band) of the 6915 \Gaia--ASAS\_SN LPV crossmatches.
         The colour of each point is related to the periods published by ASAS\_SN periods according to the colour scale shown on the right of the figure.
         A diagonal line is drawn in black to guide the eye.
         Potential ASAS\_SN-blended sources are preferentially located above the dashed red line.
        }
\label{Fig:amplitudesGaiaAsas}
\end{figure}

\begin{figure}
\centering
\includegraphics[width=0.95\hsize]{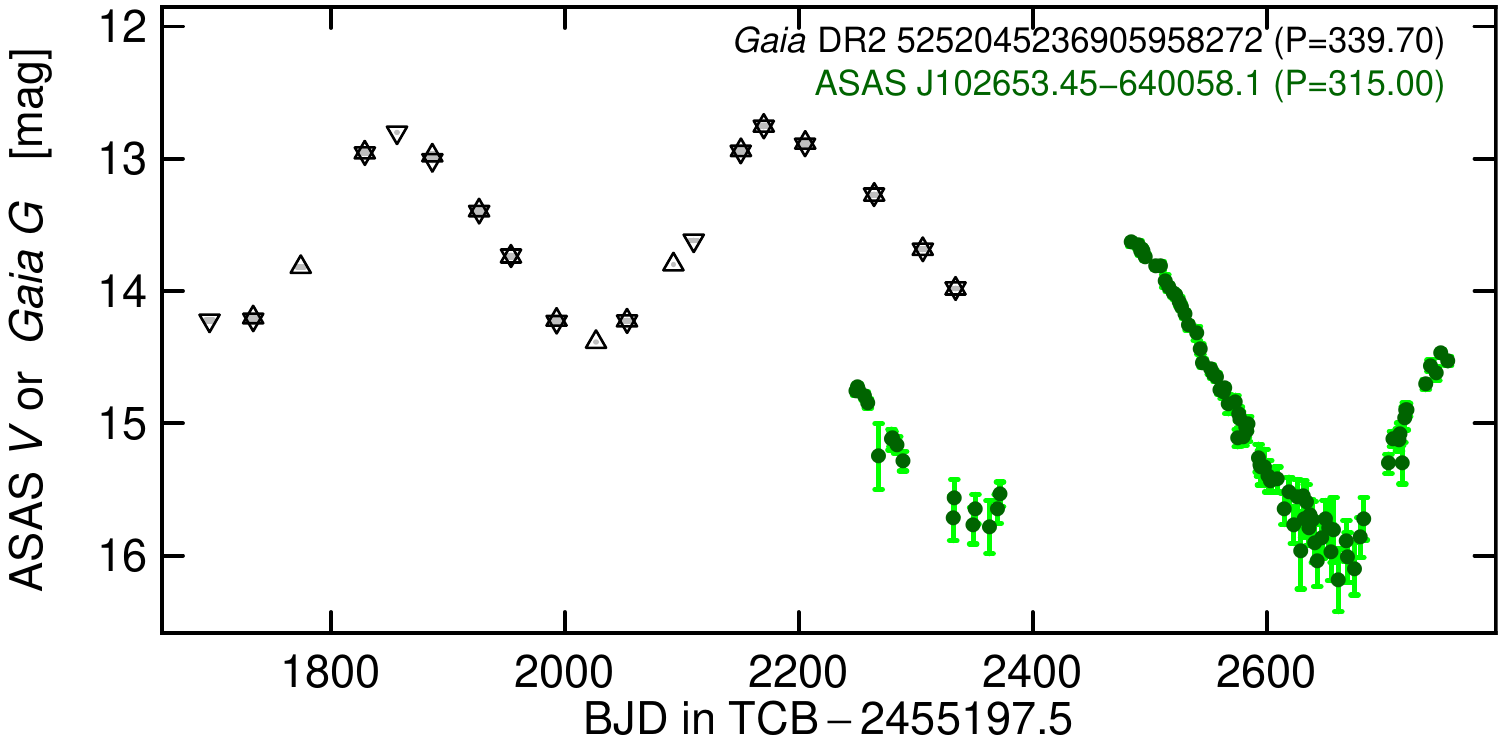}
\includegraphics[width=0.95\hsize]{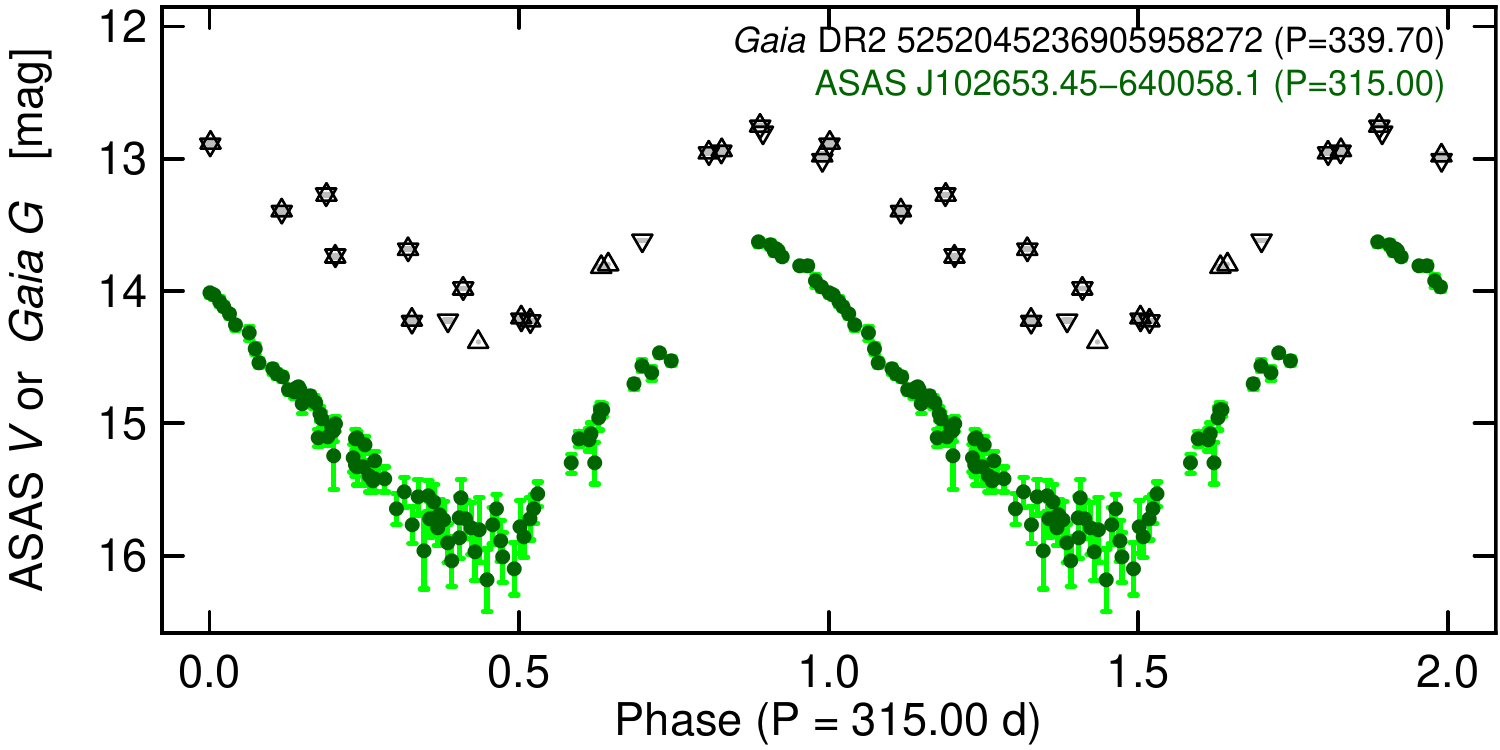}
\caption{\textbf{Top figure:} Light curves of ASAS\_SN Mira variable J102653.45-640058.1 ($V$ band, in green) and of its \Gaia crossmatch (\gmag\  band, in black).
         Upward- and downward-pointing triangles correspond to the first and second \Gaia fields of view, respectively.
         The origin of the barycentric Julian Day is set to the \Gaia reference time BJD$_0$ = 2455197.5.
         Measurement uncertainties are plotted in light green for ASAS\_SN and in grey for \Gaia (although \Gaia measurement uncertainties are within the size of the markers given the magnitude scale of the Y-axis).
         \textbf{Bottom figure:} Same as the top panel, but for the folded light curves.
         The folding period is taken from the ASAS\_SN catalogue.
        }
\label{Fig:lcASAS}
\end{figure}

\begin{figure}
\centering
\includegraphics[width=0.95\hsize]{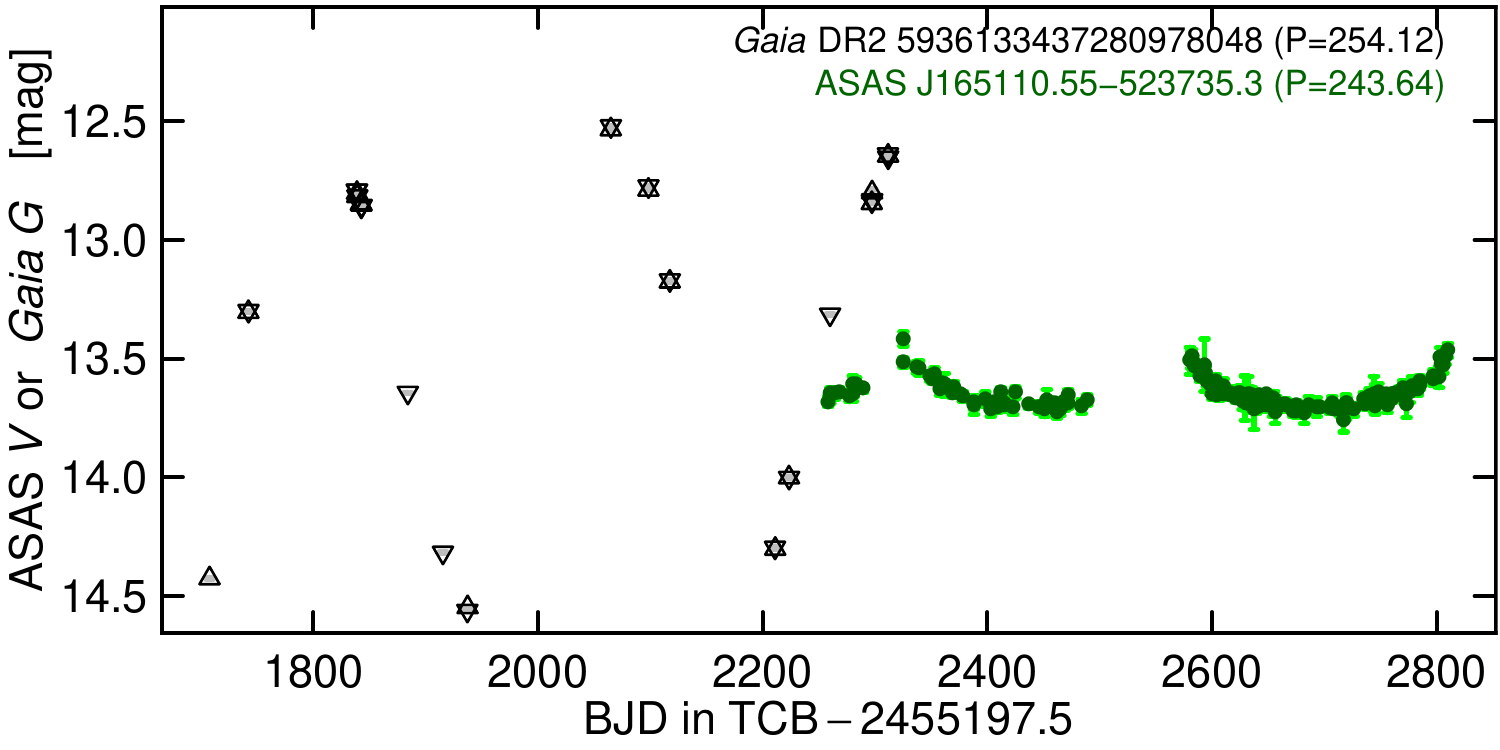}
\includegraphics[width=0.95\hsize]{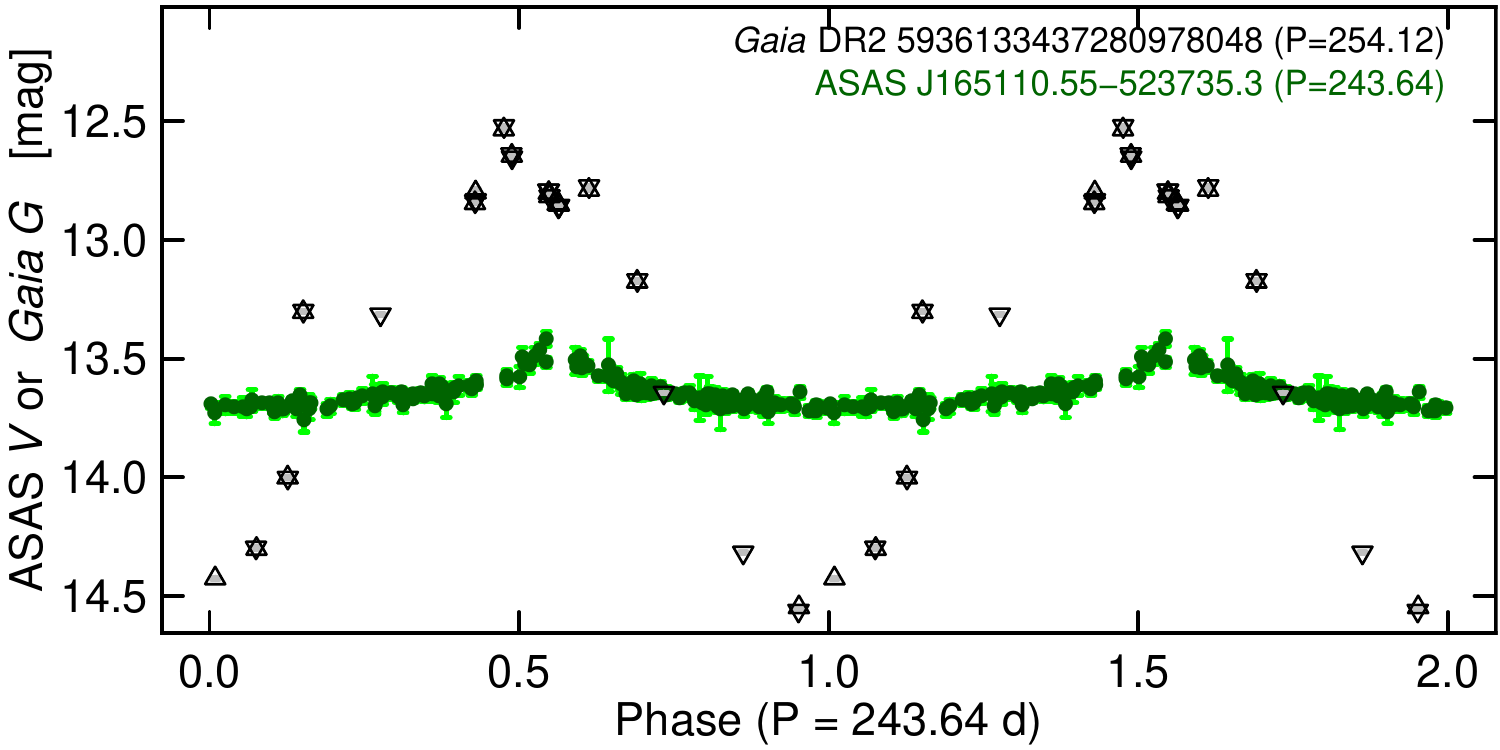}
\caption{Same as Fig.~\ref{Fig:lcASAS}, but for the blended ASAS\_SN source J165110.55-523735.
         Blending of a large-amplitude variable star with neighbouring constant star(s) leads to a reduced amplitude in the light curve of the blended source.
        }
\label{Fig:lcASAS_blended}
\end{figure}

\begin{figure}
\centering
\includegraphics[width=\hsize]{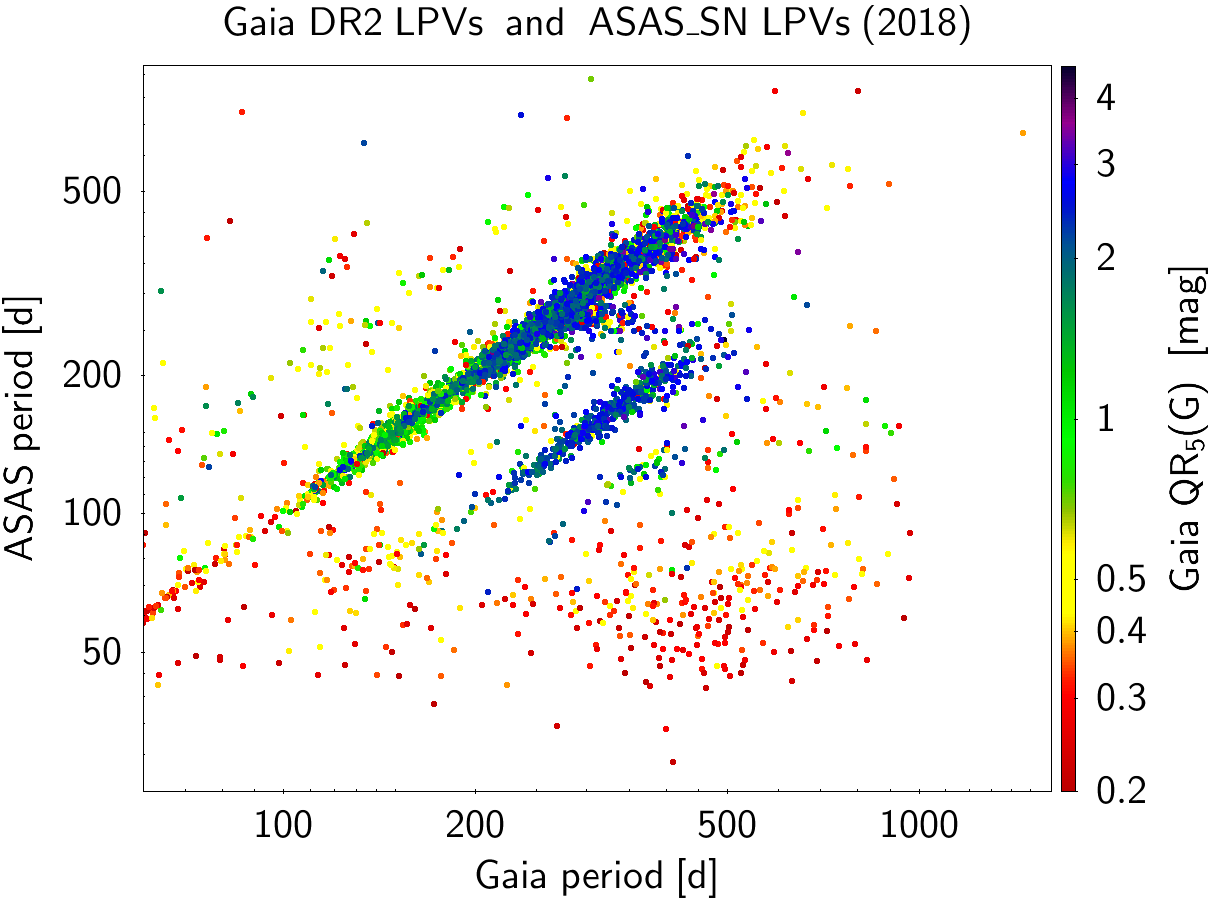}
\caption{Comparison of \Gaia (X-axis) and ASAS\_SN (Y-axis) periods of the 5491 LPV candidates in common in both catalogues whose \Gaia period is published in DR2.
         The colour of each point is related to the \Gaia \gmag-band 5--95\% quantile range $QR_5$ according to the colour scale shown on the right of the figure.
        }
\label{Fig:periodsGaiaAsas}
\end{figure}

\begin{figure}
\centering
\includegraphics[width=\hsize]{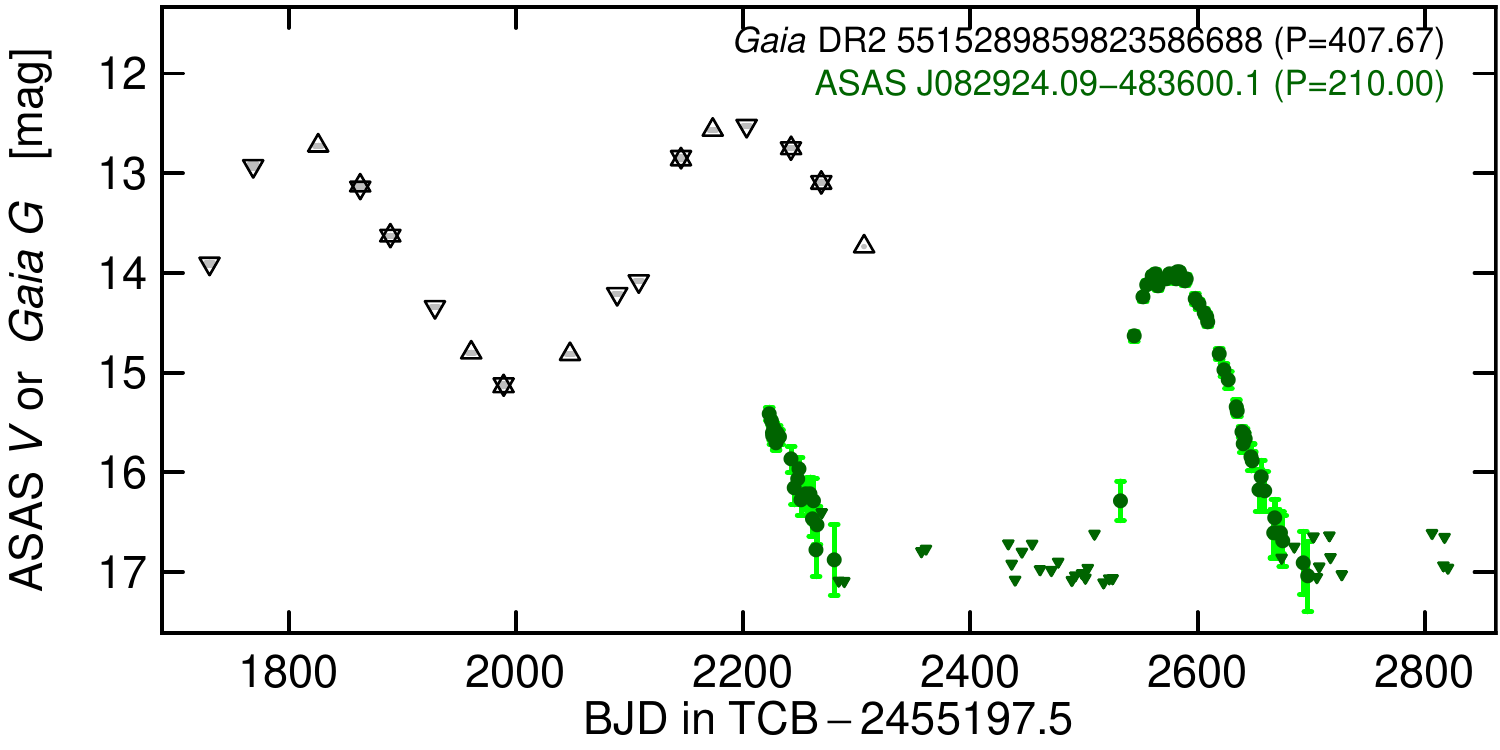}
\caption{Light curves of ASAS\_SN J102653.45-640058.1 ($V$ band) and of its \Gaia crossmatch (\gmag\  band).
         Symbols and colours have the same meanings as in Fig.~\ref{Fig:lcASAS}.
         Downward-pointing filled green triangles indicate upper limits to ASAS\_SN $V$ magnitudes.
        }
\label{Fig:lcASAS_faint}
\end{figure}

The ASAS\_SN catalogue of variable stars
\citep{Jayasinghe_etal18,2014ApJ...788...48S} offers an excellent source of comparison for our Gaia results.
A similar analysis can be conducted on other existing surveys, such as the North \citep{DrakeGrahamDjorgovski14} and South \citep{DrakeDjorgovskiCatelan_etal17} Catalina surveys, or the first ATLAS catalogue of variable stars \citep{HeinzeTonryDenneau_etal18}.
The ASAS\_SN $V$-band survey observed the whole sky over several years down to $V\simeq17$~mag. 
In total, 100 to 500 epochs per star were obtained, and these epochs partly overlap with the \Gaia observations.
The catalogue includes 1855 Miras and 14\,047 SRVs \citep{Jayasinghe_etal18}. 
Periods and amplitudes are listed, and the complete light curves can be accessed.

Given the larger number of measurements in the ASAS\_SN time-series, the comparison of \Gaia results with ASAS\_SN data allows us to validate the \Gaia periods of the bright \Gaia LPVs.
A total of 6915 ASAS\_SN sources was crossmatched with \Gaia sources, which represents a \Gaia recovery rate of 43\%.

In the next sections, we successively compare the sky and magnitude distributions between ASAS\_SN and \Gaia (Sect.~\ref{Sect:ASAS_sky}), the variability amplitudes reported in the two surveys (Sect.~\ref{Sect:ASAS_amplitudes}), and the periods (Sect.~\ref{Sect:ASAS_periods}).

\subsection{Sky and magnitude distributions}
\label{Sect:ASAS_sky}

The sky distribution of the ASAS\_SN sources is shown in Fig.~\ref{Fig:sky_ASAS}, in green for ASAS\_SN sources with no \Gaia crossmatch, in blue and red for ASAS\_SN sources with a \Gaia crossmatch (red symbols identify those ASAS\_SN--\Gaia crossmatches that are blended sources in ASAS\_SN, see next paragraph), and with \Gaia sources with no ASAS\_SN crossmatch plotted in grey in the background.
The magnitude distributions of the \Gaia (median values of the cleaned \gmag time-series; green distribution) and ASAS\_SN ($V$ magnitudes; blue distribution) are shown in Fig.~\ref{Fig:histoMagnitudesGaiaAsas}.

\subsection{Variability amplitudes}
\label{Sect:ASAS_amplitudes}

The variability amplitudes of the crossmatched LPVs in \Gaia and in ASAS\_SN are compared in Fig.~\ref{Fig:amplitudesGaiaAsas}.
ASAS\_SN provides $V$ -band amplitudes, while the amplitudes for \Gaia  are measured in the \gmag\  band.
Brightness changes are seen to be larger in the $V$ band than in the \Gaia \gmag\  band, as we described in Sect.~\ref{Sect:goodParallaxes}.
An example is shown in Fig.~\ref{Fig:lcASAS} with ASAS\_SN source \texttt{J102653.45-640058.1} crossmatched with source \texttt{\Gaia} \texttt{DR2 5936133437280978048}. 
Figure~\ref{Fig:amplitudesGaiaAsas} also reveals a sample of stars whose \Gaia amplitudes are much larger than ASAS\_SN amplitudes.
An extreme example is shown in Fig.~\ref{Fig:lcASAS_blended}. 
To understand this difference, we have to consider that the ASAS\_SN resolution element on the sky is 8". 
Therefore, multiple sources may contribute to an ASAS\_SN pixel, which reduces the amplitude observed in the light curve of the blended source.
This problem primarily occurs in crowded areas. 
Taking the condition $\log(QR_5) > \log(\mathrm{ASAS\;Amplitude})+0.15$ to select blended ASAS\_SN sources in the sample of ASAS\_SN--\Gaia crossmatches (sources above the dashed line in Fig.~\ref{Fig:amplitudesGaiaAsas}), we can then study their sky distribution in Fig.~\ref{Fig:sky_ASAS} (red symbols).
The concentration of these blended source candidates in the crowded regions of the Galactic bulge and thin disc is obvious. 
In total, we identify 1268 ASAS\_SN stars that are likely affected by blending with nearby objects, only 9 of which have been classified as Miras in ASAS\_SN. 
Therefore, the total number of Miras in the ASAS\_SN variable star catalogue should be larger than announced by the ASAS\_SN team \citep{Jayasinghe_etal18}, likely reaching $\sim$3000.

As a side analysis, the \Gaia and ASAS\_SN folded light-curves of the two example cases taken above are shown in the lower panels of Figs.~\ref{Fig:lcASAS} and \ref{Fig:lcASAS_blended}, using the periods provided in the ASAS\_SN catalogue.
They reveal a minor phase shift between the \Gaia and the ASAS\_SN light curves. 
The phase shift may be the result of small cycle-to-cycle period changes, but could also reflect a systematic phase shift between the $V$ and \gmag bands.
An additional analysis is required, but this is beyond  the scope of this catalogue presentation paper.

\subsection{Periods}
\label{Sect:ASAS_periods}

Of the 6915 ASAS\_SN--\Gaia crossmatches, a sample of 5491 have \Gaia periods.
A comparison of the ASAS\_SN and \Gaia periods of the LPVs in this sample is shown in Fig.~\ref{Fig:periodsGaiaAsas}.
A very good match is observed for the majority of them: 77\% of the sample have \Gaia periods within 15\% of the ASAS\_SN periods.
A group of crossmatched stars is visible in Fig.~\ref{Fig:periodsGaiaAsas}, however, which clump around a 2:1 relation for the periods of \Gaia versus ASAS\_SN.
They represent 9\% of the sample, and they have \Gaia periods within 15\% of twice the ASAS\_SN periods.
A preliminary analysis of these objects reveals that the corresponding ASAS\_SN stars have expected mean $V$ magnitudes that would be close to or below the magnitude limit of $V=17$~mag.
As a consequence, ASAS\_SN misses measurements over more than half the pulsation period, or only has upper limits on $V$ in these time intervals. This produces light curves that are characterised by islands of measurements at times when the brightness of the star increases above $V\simeq 17$~mag.
An example of such a case is given in Fig.~\ref{Fig:lcASAS_faint}.
An automated period-search algorithm can easily be blurred by such a time-series and lead to a period that is half, possibly even less, of the true period.
This would explain the existence of the clump around the 2:1 relation in Fig.~\ref{Fig:periodsGaiaAsas} where the ASAS\_SN periods are about half the \Gaia periods.

\section{Validation 4: Galactic bulge and extragalactic comparison with OGLE-III}
\label{Sect:OGLE}

The OGLE-III survey has published the largest LPV catalogues so far. It also has a large number of epoch photometry data for each source.
Reliable periods are thus expected to result from the analysis of their time-series, thereby characterizing multiperiodic LPVs.
The OGLE measurements are mainly performed in the $I$ band, which captures more of the red giant flux emission than the $V$ band does. The survey also extends deeper in magnitude than the ASAS\_SN survey, and easily reaches the Magellanic Clouds.
The $I$ band is close to the \Gaia $G$ band as far as LPVs are concerned.
The OGLE-III survey thus offers a nice set of data to validate \Gaia data.
It is limited to specific regions of the sky, however, including the Galactic bulge, the LMC, and the SMC.

The OGLE-III survey has recorded 6528 Miras, 33\,235 SRVs, and 192\,643 OGLE small-amplitude red giants \citep[OSARGs;][]{2004MNRAS.349.1059W} in the direction of the Galactic bulge \citep{2013AcA....63...21S};
1663 Miras, 11\,132 SRVs, and 79\,200 OSARGs in the LMC \citep{2009AcA....59..239S};
and 352 Miras, 2222 SRVs, and 16\,810 OSARGs in the SMC \citep{2011AcA....61..217S}.

In this section, we use the OGLE-III LPV catalogues to address the completeness question of our \Gaia catalogue in the regions of the Galactic bulge and the Magellanic Clouds (Sect.~\ref{Sect:OGLE_GaiaCompleteness}).
In our analysis, we do not distinguish between SRVs and OSARGs.
We instead refer to them as small-amplitude LPVs.
With our selection criteria, 1.7\% of the OGLE-III OSARGs are present in the \Gaia DR2 sample.
We also check the new sources in our \Gaia catalogue relative to the OGLE-III catalogues (Sect.~\ref{Sect:OGLE_GaiaNewSources}).
We then compare in Sect.~\ref{Sect:OGLE_periods} the periods published in our catalogue with those from OGLE-III.

\subsection{Completeness of the \Gaia catalogue relative to OGLE-III}
\label{Sect:OGLE_GaiaCompleteness}

\begin{figure}
\centering
\includegraphics[width=\hsize]{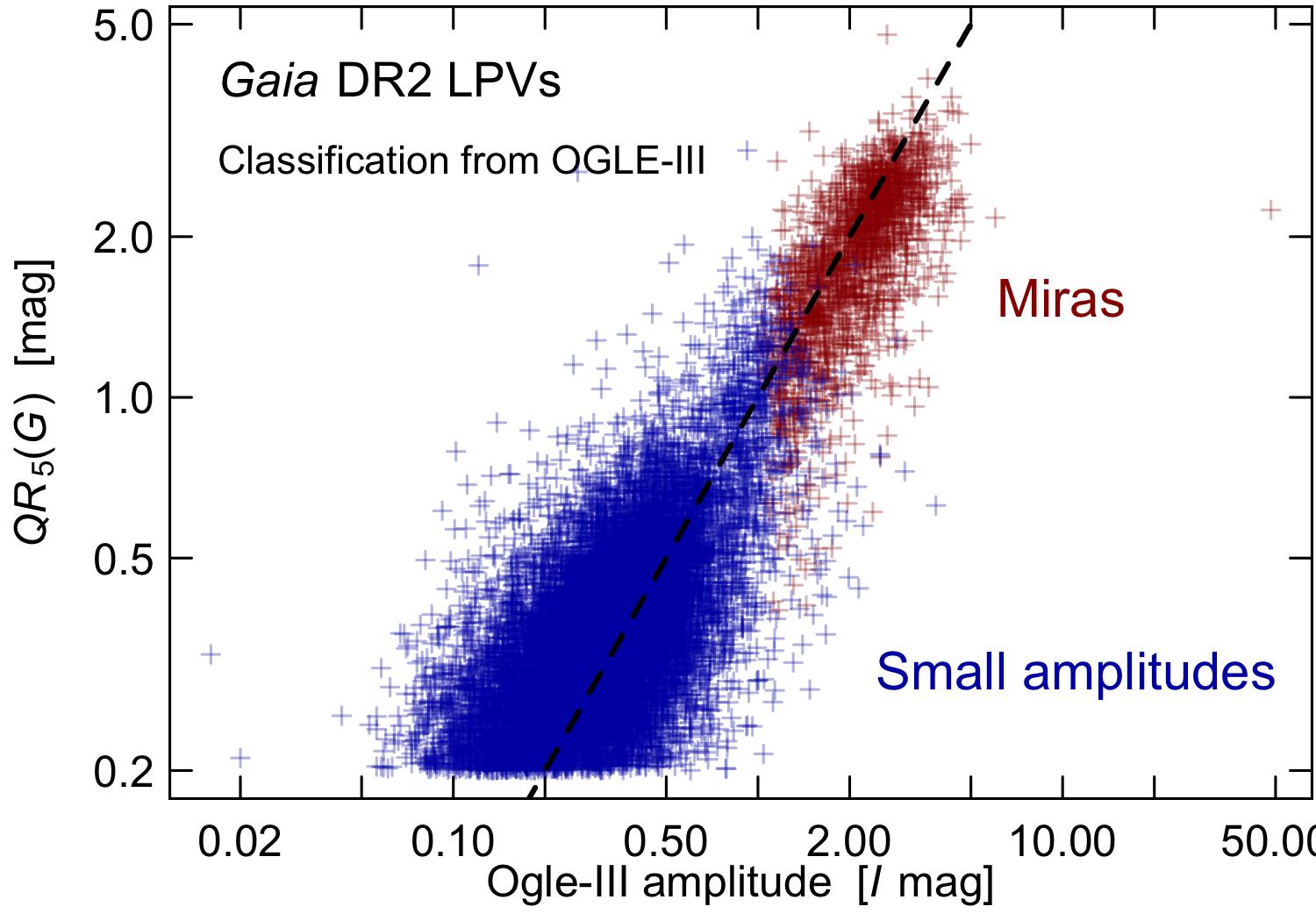}
\caption{\Gaia variability amplitude measured by the 5--95\% quantile range $QR_5$ of the cleaned \gmag time-series, vs. OGLE-III \textit{I}-band variability amplitude for all {\Gaia} DR2 LPV candidates that match a Mira (red plus) or a small amplitude LPV (blue plus) in the OGLE-III catalogue.
        }
\label{Fig:varAmpl_GaiaOgle}
\end{figure}

\begin{figure}
\centering
\includegraphics[width=\hsize]{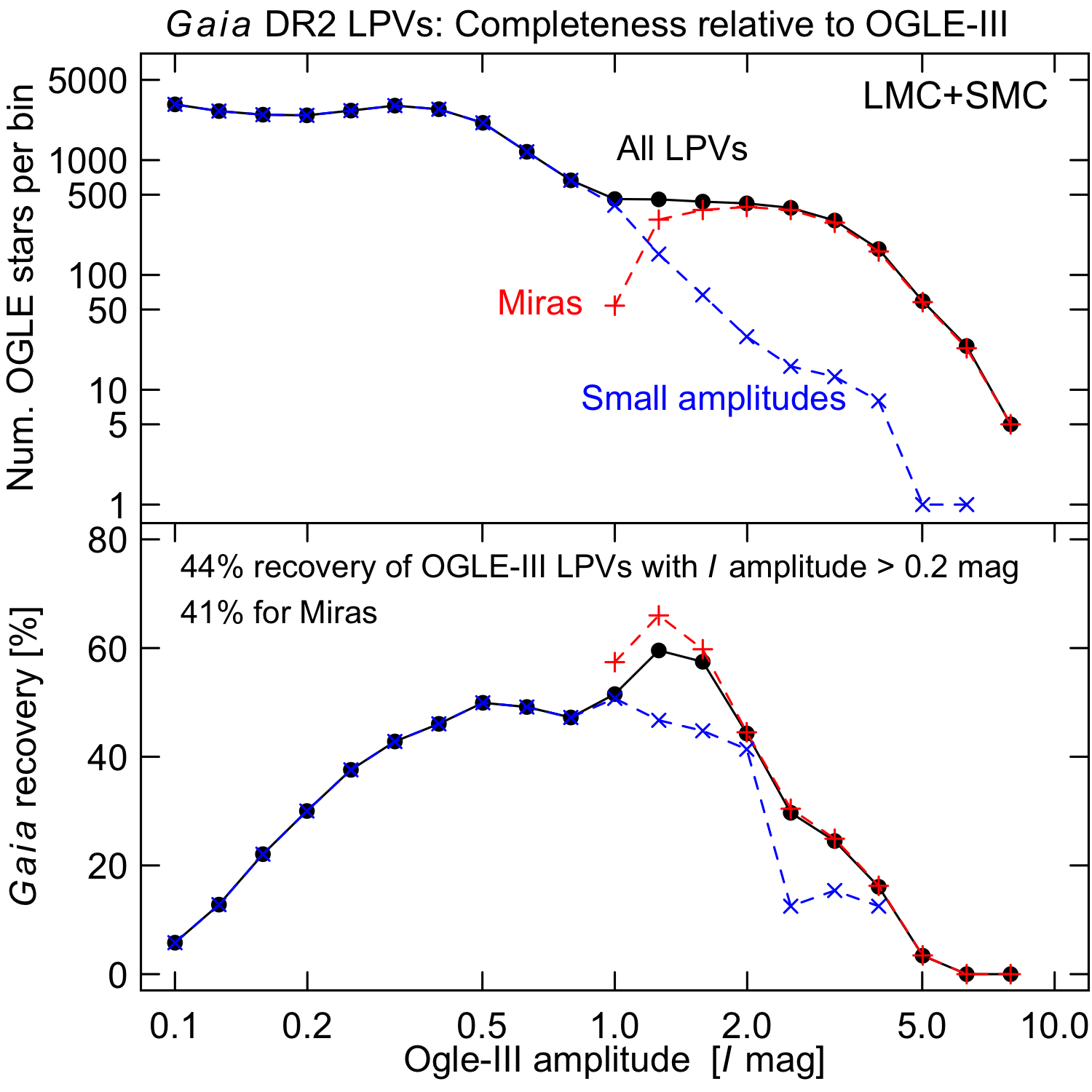}
\caption{\textbf{Top panel:} Number of LPVs reported in the OGLE-III catalogue towards the LMC and SMC as a function of OGLE $I$-band variability amplitude.
         The number of sources are given per 0.1-width bins of log amplitude.
         The group classified as Miras in the OGLE catalogue is shown as red pluses connected with a red dashed line, and the group combining all other LPVs is shown as blue crosses connected with a blue dashed line.
         The combined group of all LPVs is shown as black filled circles connected with a solid line.
         \textbf{Bottom panel:} {\Gaia} recovery fraction of OGLE-III LPVs as a function of OGLE-III $I$-band variability amplitude for the group of LPVs shown in the top panel.
         The recovery fraction is computed per log amplitude bin as the ratio of the number of OGLE-III crossmatches in {\Gaia} DR2 to the number of OGLE-III LPVs in the considered amplitude bin.
         Marker and line properties for each of the three groups are identical to those used in the top panel.
        }
\label{Fig:completeness_LMC+SMC}
\end{figure}

\begin{figure}
\centering
\includegraphics[width=\hsize]{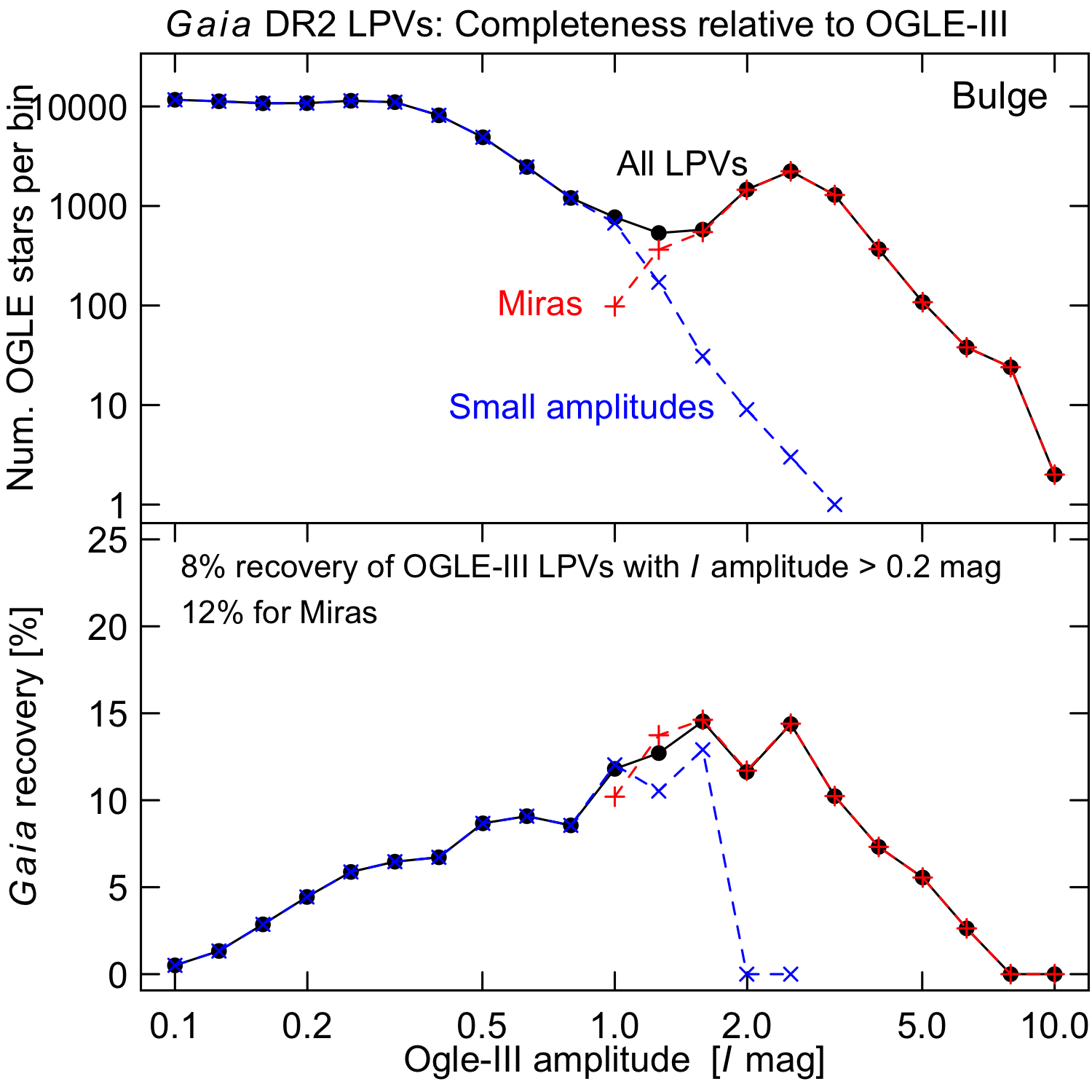}
\caption{Same as Fig.~\ref{Fig:completeness_LMC+SMC}, but in the direction of the Galactic bulge.
        }
\label{Fig:completeness_Bulge}
\end{figure}

\begin{figure}
\centering
\includegraphics[width=\hsize]{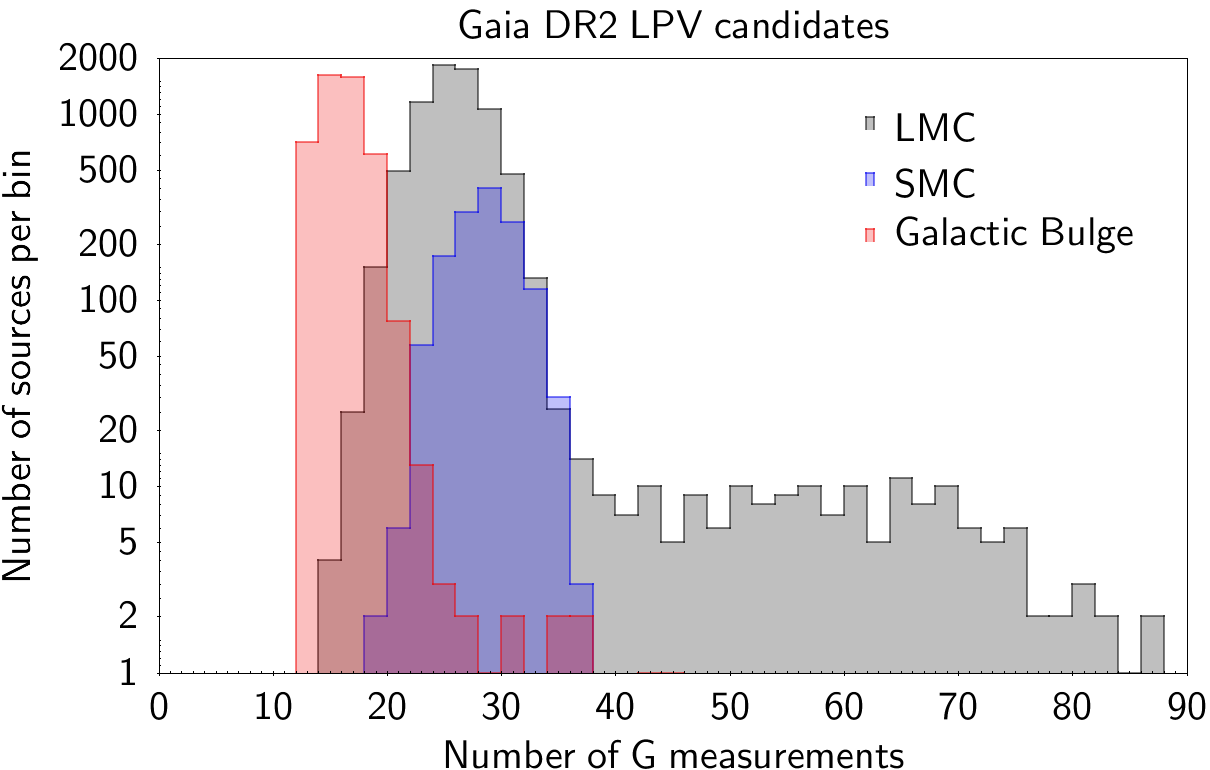}
\caption{Number of measurements in the cleaned \gmag time-series for {\Gaia} DR2 LPV candidates found by crossmatching with OGLE-III LPVs in the direction of the LMC (grey histogram), the SMC (blue histogram), and the Galactic bulge (red histogram).
        }
\label{Fig:histo_numPtsFinalG_XMs}

\end{figure}

Because of the magnitude depth of the OGLE-III survey and its extended time coverage, we may expect it to be complete in the search of LPVs in the sky regions that it covered.
We should thus be able to assess the completeness of the \Gaia LPV catalogue by checking the percentage of OGLE-III sources found in regions that are well covered by OGLE-III.
New sources in the \Gaia catalogue relative to OGLE-III catalogues, which would indicate incompleteness of the later catalogues, are addressed in Sect.~\ref{Sect:OGLE_GaiaNewSources}.

In DR2, we targeted Mira-like and semi-regular variables with variability amplitudes larger than 0.2~mag.
We therefore have to compare the DR2 set of LPV candidates to the subset of OGLE-III LPVs with the same selection criterion for the variability amplitudes.
Caution has to be taken because the light curves are in different photometric bands:~\gmag band for \Gaia, and $I$ band for OGLE-III. Moreover, the variability amplitudes are computed differently:~ using $QR_5(\gmag)$ of the \gmag times series for \Gaia, while taking for OGLE-III the sum $A_\mathrm{Ogle} = A_1 + A_2 + A_3$ of the amplitudes of the three dominant pulsation periods published in the OGLE-III catalogue.
The comparison between $QR_5(\gmag)$ and $A_\mathrm{Ogle}$ is shown in Fig.~\ref{Fig:varAmpl_GaiaOgle} for all \Gaia--OGLE-III crossmatches.
For SRVs and Miras (upper panel of Fig.~\ref{Fig:varAmpl_GaiaOgle}), the distribution of points follows a $QR_5(\gmag) \simeq A_\mathrm{Ogle}$ relation, showing the equivalence of the amplitudes in \gmag and $I$ for these stars.
A scatter is observed around the line $QR_5(\gmag) = A_\mathrm{Ogle}$, however, that is larger at small variability amplitudes.
At least three effects are at the origin of this scatter.
The first factor is the sparse sampling and short observation durations (relative to the periods of LPVs) of {\Gaia} DR2 light curves.
This leads to $QR_5(\gmag) < A_\mathrm{Ogle}$.
Its effect is most evident at large variability amplitudes (where a single long period becomes dominant), as shown in Fig.~\ref{Fig:varAmpl_GaiaOgle}.
The second factor is the limitation to three pulsation frequencies reported in the OGLE-III catalogue.
This leads to $QR_5(\gmag) > A_\mathrm{Ogle}$ if more than three significant periods are present.
Its effect is most visible in small-amplitude LPVs, as shown in the bottom panel of Fig.~\ref{Fig:varAmpl_GaiaOgle}.
A third effect is also at play for the bluest LPVs (mainly small-amplitude LPVs).
For these stars, the {\Gaia} \gmag\  band is sensitive to the small emission in the blue side of their spectra, while the OGLE-III $I$ band is not.
As a consequence, very red LPVs (mainly large-amplitude SRVs and Miras) will have relatively similar variability amplitudes in \gmag and $I$, while bluer LPVs are expected to have variability amplitudes larger in \gmag than in $I$.

After showing the near equivalence of \gmag and $I$ variability amplitudes for LPVs and identifying the origin of the scatter around $QR_5(\gmag) = A_\mathrm{Ogle}$, we now estimate the completeness of the \Gaia DR2 LPV candidates relative to the OGLE-III catalogues as a function of variability amplitude.
The number of OGLE-III LPVs as a function of $A_\mathrm{Ogle}$ is shown in the top panel of Fig.~\ref{Fig:completeness_LMC+SMC} for the LMC and SMC, and in the top panel of Fig.~\ref{Fig:completeness_Bulge} for the direction to the Galactic bulge.
There are 15\,865 OGLE-III LPVs with $A_\mathrm{Ogle}>0.2$~mag in the LMC and SMC, and 50\,579 OGLE-III LPVs towards the bulge.
The completeness of the \Gaia catalogue relative to OGLE-III is given by the percentage
of those $A_\mathrm{Ogle}>0.2$~mag OGLE-III LPVs that are present in \Gaia DR2.
This percentage is shown as a function of $A_\mathrm{Ogle}$ in the bottom panels of Figs.~\ref{Fig:completeness_LMC+SMC} and \ref{Fig:completeness_Bulge}.
Different conclusions are drawn for the bulge and clouds.

For the Magellanic Clouds, Fig.~\ref{Fig:completeness_LMC+SMC} shows a 45\% completeness level of \Gaia LPVs (the completeness levels are similar when we consider the Large and Small Cloud individually).
Part of the lack of completeness is due to the incomplete \Gaia sky coverage of sources with at least 12 observations in \gmag, which was one of the selection criteria of LPV candidates.
Figure~1 of \citet{Holl_etal18} indeed shows that towards the LMC and SMC, only $\sim$77\% of \Gaia DR2 sources have at least 12 observations.

In the direction of the Galactic bulge, the completeness of the \Gaia DR2 LPV catalogue relative to OGLE-III is only $\sim$8\% (Fig.~\ref{Fig:completeness_Bulge}).
This low level of \Gaia LPV completeness also results to a good part from the incomplete \Gaia sky coverage of sources with at least 12 observations.
Figure~1 of \citet{Holl_etal18} shows that only 67\% of the sources have at least 12 \gmag-band observations in the direction of the bulge, and as small a percentage as 14\% of sources have at least 20 \gmag-band observations.
As a result, most LPVs towards the Galactic centre have fewer than 20 observations in \Gaia DR2, while the majority of those towards the Clouds have more than 20 observations.
This is confirmed in Fig.~\ref{Fig:histo_numPtsFinalG_XMs}, which plots the histograms of the number of \gmag measurements towards the LMC (grey histogram), SMC (blue histogram) and the bulge (red histogram).
The small number of observations towards the bulge leads to a decreased identification efficiency.
See also the note at the end of Sect.~\ref{Sect:OGLE_GaiaNewSources} for the crowding issue in the bulge.

\subsection{New \Gaia DR2 LPV candidates relative to OGLE-III}
\label{Sect:OGLE_GaiaNewSources}

\begin{table}
\caption{Definition of the sample regions on the sky, in equatorial coordinates, towards the Magellanic Clouds and the Galactic bulge where the \Gaia DR2 LPV catalogue is checked for the presence of new LPV candidates relative to the OGLE-III catalogues of LPVs.
        }
\centering
\begin{tabular}{c c}
\hline\hline
Region & Sky region used for the checks \\
\hline
   LMC & $ 73.6^{\circ} < RA < 86.6^{\circ}$, $-71.3^{\circ} < Dec < -67.8^{\circ}$ \\
\hline
       & $ 11.6^{\circ} < RA < 20.5^{\circ}$, $-74.2^{\circ} < Dec < -71.7^{\circ}$ \\
   SMC & and \\
       & $ ~5.8^{\circ} < RA < 13.5^{\circ}$, $-74.6^{\circ} < Dec < -72.5^{\circ}$ \\
\hline
       & $ 267.60^{\circ} < RA < 271.30^{\circ}$, $-30.20^{\circ} < Dec < -29.22^{\circ}$ \\
   Bulge & and \\
       & $ 263.56^{\circ} < RA < 264.21^{\circ}$, $-27.46^{\circ} < Dec < -26.87^{\circ}$ \\
\hline
\end{tabular}
\label{Tab:skyRegions}
\end{table}

\begin{figure}
\centering
\includegraphics[width=\hsize]{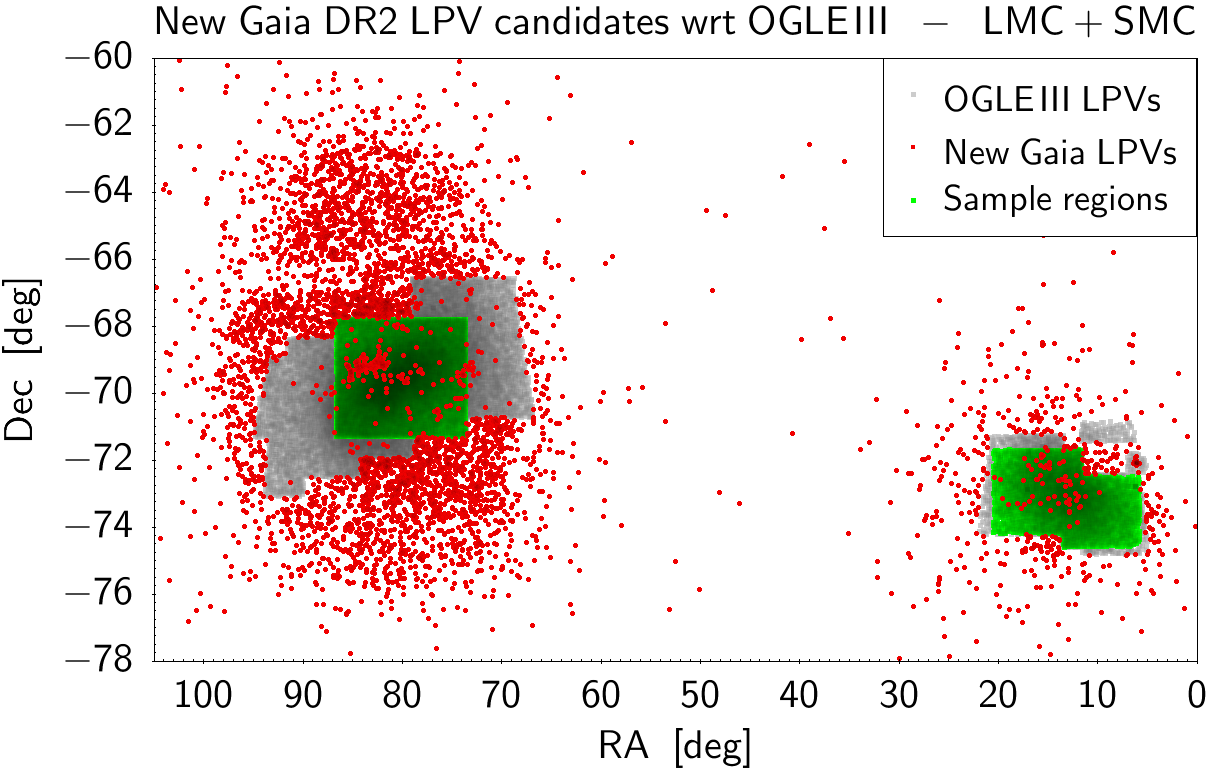}
\caption{Distribution on the sky around the Magellanic Clouds of the OGLE-III LPVs and of the new \Gaia DR2 LPVs with respect to the OGLE-III survey.
         OGLE-III sources are represented in grey and green, the green sources delimiting the regions on the sky used in the text to check the new \Gaia LPV candidates.
         The new \Gaia DR2 LPV candidates are shown in red.
         }
\label{Fig:skyRegions_Clouds}
\end{figure}

\begin{figure}
\centering
\includegraphics[width=\hsize]{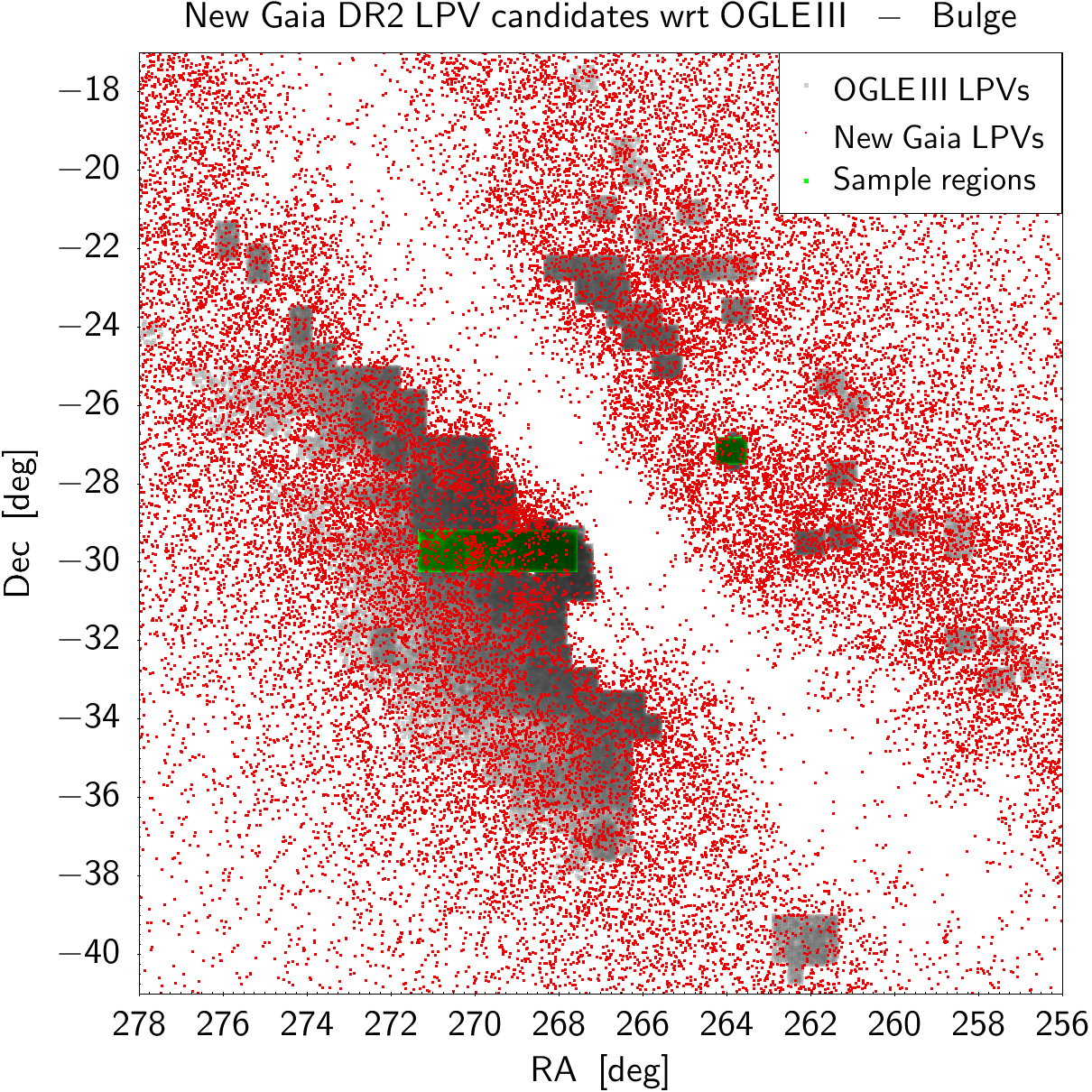}
\caption{Same as Fig.~\ref{Fig:skyRegions_Clouds} for the direction on the sky towards the Galactic bulge.
         The new \Gaia DR2 LPV candidates with respect to OGLE-III are plotted with smaller red markers than in Fig.~\ref{Fig:skyRegions_Clouds} for better visibility of the regions covered by the OGLE-III survey.
        }
\label{Fig:skyRegions_Bulge}
\end{figure}

\begin{figure}
\centering
\includegraphics[width=\hsize]{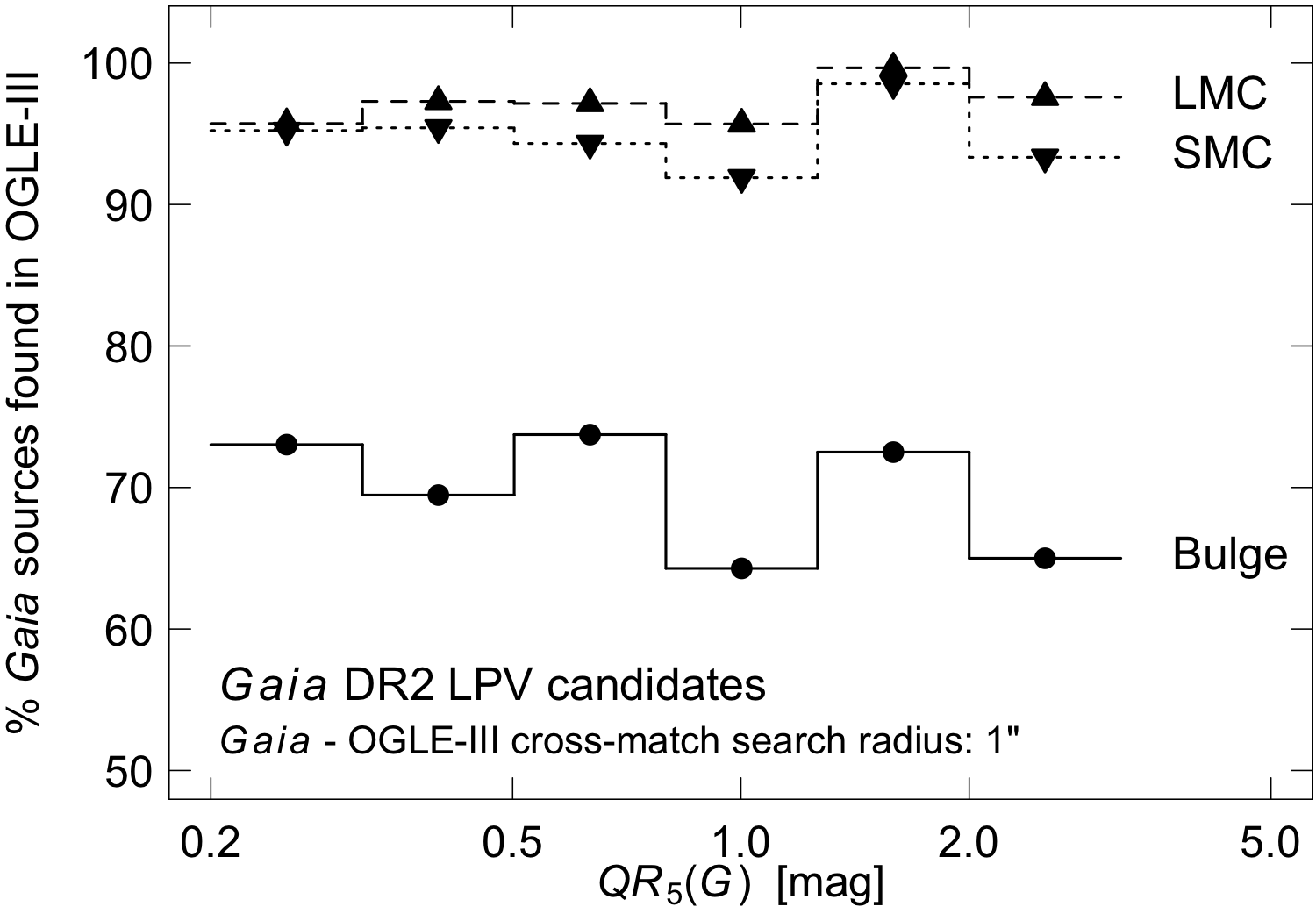}
\caption{Percentage of {\Gaia} DR2 LPV candidates for which a crossmatch is found in the OGLE-III catalogue as a function of the 5--95\% quantile range $QR_5$ of \gmag light curves towards the SMC (dotted histogram with downward-pointing triangles), the LMC (dashed histogram with upward-pointing triangles), and the Galactic bulge (solid histogram with filled circles).
         The percentages are computed in the sample sky areas shown in Figs.~\ref{Fig:skyRegions_Clouds} and \ref{Fig:skyRegions_Bulge} using a crossmatch search radius of 1~arcsec.
        }
\label{Fig:crossmatches_AmplGaia}
\end{figure}

\begin{figure}
\centering
\includegraphics[width=\hsize]{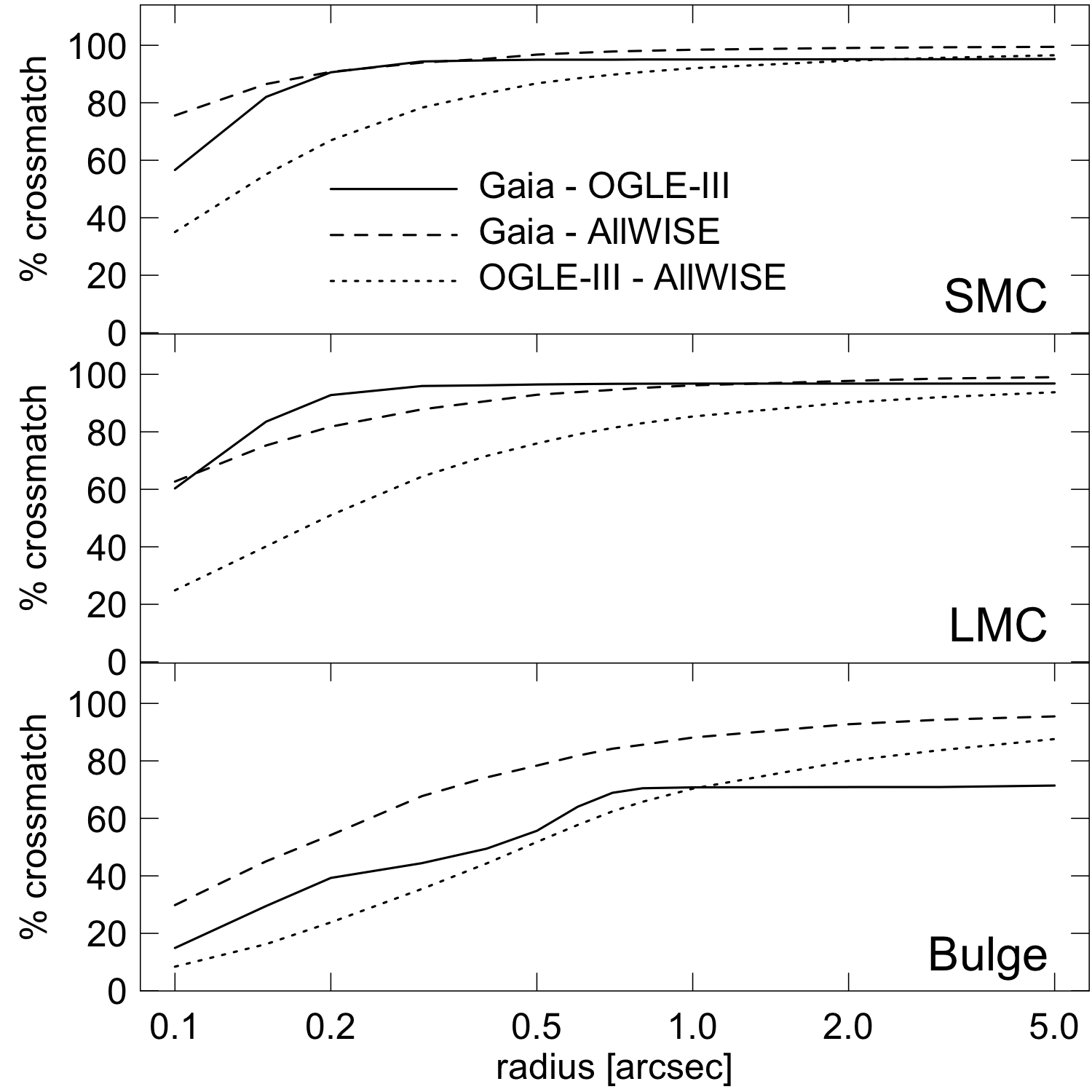}
\caption{Percentage of \textit{Gaia} DR2 LPV candidates for which a crossmatch is found with OGLE-III LPVs (solid line) in the reference sky regions towards the SMC (top panel), LMC (middle panel), and Galactic bulge (bottom panel) as a function of crossmatch radius.
The percentages of \Gaia (OGLE-III) LPV candidates for which a crossmatch is found with AllWISE sources are shown as dashed (dotted) lines in each panel.
The reference sky regions are shown in Figs.~\ref{Fig:skyRegions_Clouds} and \ref{Fig:skyRegions_Bulge}.
        }
\label{Fig:crossmatches_GaiaOgleAllwise}
\end{figure}

Despite the incompleteness of the \Gaia LPV catalogue with respect to the OGLE-III catalogues, sources can nevertheless be present in the \Gaia catalogue that are not present in the OGLE-III catalogues.
This section determines whether this is the case.

The search for the new \Gaia DR2 LPV candidates must be made in regions of the sky with full OGLE-III coverage.
We therefore defined sample regions on the sky towards the SMC, LMC, and Galactic bulge where this condition is satisfied.
The chosen sample regions are defined in Table~\ref{Tab:skyRegions} and shown in Figs.~\ref{Fig:skyRegions_Clouds} and \ref{Fig:skyRegions_Bulge} (green areas).
With a search radius of 1", we find OGLE-III crossmatches to $\sim$95\% of the \Gaia LPVs present in the sample regions towards the SMC and LMC, and to $\sim$70\% of the \Gaia LPVs found towards the Galactic bulge.
Taking these small sample regions of the sky as representative, the \Gaia catalogue of LPV candidates thus contains about 5\% new sources in the Magellanic Clouds, and about 30\% new sources towards the Galactic bulge.
These conclusions are independent of variability amplitude, as shown in Fig.~\ref{Fig:crossmatches_AmplGaia}.

The question whether these new \Gaia LPV candidates are contaminants in the DR2 catalogue or whether they are real must be assessed.
We first checked the crossmatch efficiency.
The percentage of \Gaia LPVs crossmatched to OGLE-III LPVs is shown in Fig.~\ref{Fig:crossmatches_GaiaOgleAllwise} (solid lines) as a function of crossmatch radius.
In the SMC (top panel of the figure) and LMC (middle panel), the crossmatch percentage already reaches 94\% and 96\%, respectively, at a crossmatch radius of 0.3~arcsec.
Towards the Galactic bulge (bottom panel of Fig.~\ref{Fig:crossmatches_GaiaOgleAllwise}), the percentage levels off at 71\% at a crosmatch radius of 1~arcsec.
We therefore consider that all potential OGLE-III crossmatches to the \Gaia LPVs are found using 1 arcsec.

We then determined the impact of the angular resolution of both OGLE-III and \Gaia surveys on source detection within each survey.
To determine this from an independent catalogue, we compared the list of \Gaia and OGLE-III sources in the sample regions defined above to the list of infrared sources detected by AllWISE \citep{allWISE} in these regions.
The results of the crossmatches are shown in Fig.~\ref{Fig:crossmatches_GaiaOgleAllwise}, using dashed lines for \Gaia--AllWISE crossmatches and dotted lines for OGLE-III--AllWISE crossmatches.
In all three directions, towards the SMC, LMC, and Galactic bulge, the percentage of \Gaia sources crossmatched with AllWISE is higher than the percentage of OGLE-III sources crossmatched with AllWISE.
At a search radius of 0.4~arcsec, which is the announced \Gaia angular resolution for DR2, 95\%, 91\%, and 74\% of \Gaia LPVs have an AllWISE crossmatch towards the SMC, LMC, and bulge, respectively, while these numbers are 83\%, 72\%, and 44\%, respectively, for OGLE-III.
The crossmatch success rates of \Gaia sources with AllWISE sources are thus higher than the crossmatch success rates of OGLE-III with AllWISE sources.
This consolidates the reality of \Gaia LPV candidates, even in the absence of an OGLE-III LPV crossmatch.
The low \Gaia--OGLE-III crossmatch percentage towards the Galactic bulge could then result from the lower angular resolution of OGLE-III in dense regions compared to the \Gaia angular resolution.

In conclusion, the new \Gaia LPV candidates ($\sim$5\% in the clouds and $\sim$30\% towards the bulge) are expected to be real.
They were not found by OGLE-III, possibly because the sky resolution of OGLE-III is lower than that of \Gaia.
A visual check of the light curves of the new \Gaia LPV candidates showed that they are consistent with LPV-like behaviour.
An example of a new \Gaia LPV candidate in the direction of the bulge is shown in Fig.~\ref{Fig:LcsExamples2} (bottom).

\begin{figure}
\centering
\includegraphics[width=0.94\hsize]{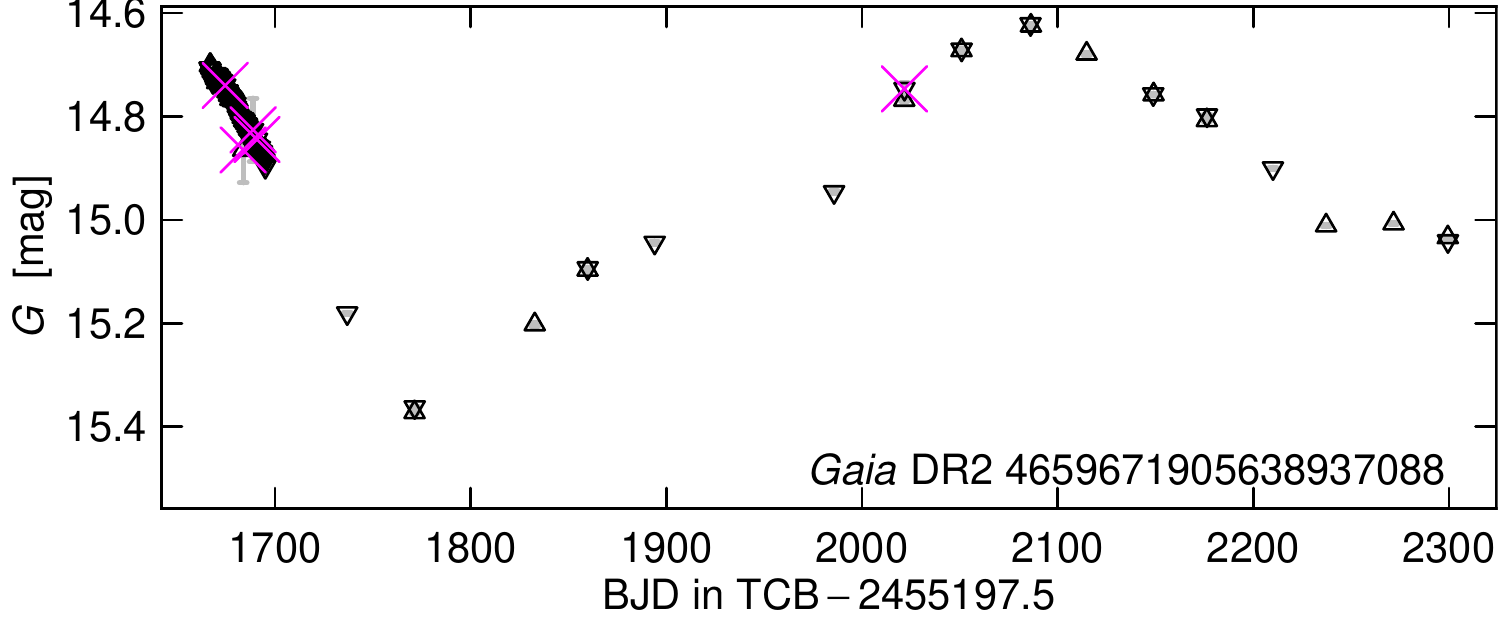}
\caption{$G$-band light curve example of an LPV in the LMC near the south ecliptic pole.
         It shows the large number of observations gathered during the first month of commissioning phase, due to the use of a specific EPSL, which is at the origin of the apparent deficiency of LPV candidates seen in Fig.~\ref{Fig:skyRegions_Clouds} in the region of the LMC near the ecliptic pole (see text).
         The black upward- and downward-pointing triangles and magenta crosses have the same meanings as in Fig.~\ref{Fig:LcsExamples1}.
        }
\label{Fig:LcExampleLMC_EPSL}
\end{figure}

\paragraph{A note on the sky distribution of \Gaia LPVs in the LMC.}
A region of the LMC at declinations around -66.5$^o$ shows an apparent deficiency of \Gaia LPV candidates in Fig.~\ref{Fig:skyRegions_Clouds} compared to neighbouring regions at higher and lower declinations.
This region corresponds to the south ecliptic pole, which has been scanned with a much higher sampling rate during the one-month commissioning phase at the beginning of the \Gaia mission (see footnote~\ref{Footnote:EPSL}).
For LPVs, this results in a clumped distribution of the magnitudes and a concomitant $QR_5(G)$ value lower than the value it would have had the measurements been homogeneously spread throughout the full observation time interval.
An example is shown in Fig.~\ref{Fig:LcExampleLMC_EPSL} with source \texttt{4659671905638937088} located at RA = $87.75948^\mathrm{o}$ and DEC = $-66.60817^\mathrm{o}$.
Its amplitude measured by $QR_5(G)$ is equal to 0.51~mag if we exclude the EPSL measurements, but drops to 0.21~mag when considering the whole light curve.
Many LPVs at the ecliptic poles are thus artificially filtered out by the criterion $QR_5(G)>0.2$~mag used for DR2, explaining the apparent deficiency of \Gaia DR2 LPVs in the LMC at declinations around $-66.5^\mathrm{o}$. 

\paragraph{A note on \Gaia sources in crowded regions.}
In crowded regions, the \Gaia on-board software has an automatic procedure for source selection based on their brightness, using a brightness threshold determined according to the number of stars detected in each field of view.
Only sources brighter than this threshold are retained for flux recording in dense regions of the sky.
Since LPVs have brightness variations that can reach several magnitudes, a given LPV may be selected at a given epoch when it is bright and not selected at another epoch when it becomes fainter than the threshold.
Therefore, only few \Gaia LPVs may have truncated light curves, whereby only the bright segments of their light curves are recorded, in a way similar to what was noted for faint ASAS\_SN LPVs (see Sect.~\ref{Sect:ASAS_periods}, and Fig.~\ref{Fig:lcASAS_faint} in particular).

\subsection{Period comparison between \Gaia DR2 and OGLE-III}
\label{Sect:OGLE_periods}

\begin{figure}
\centering
\includegraphics[width=\hsize]{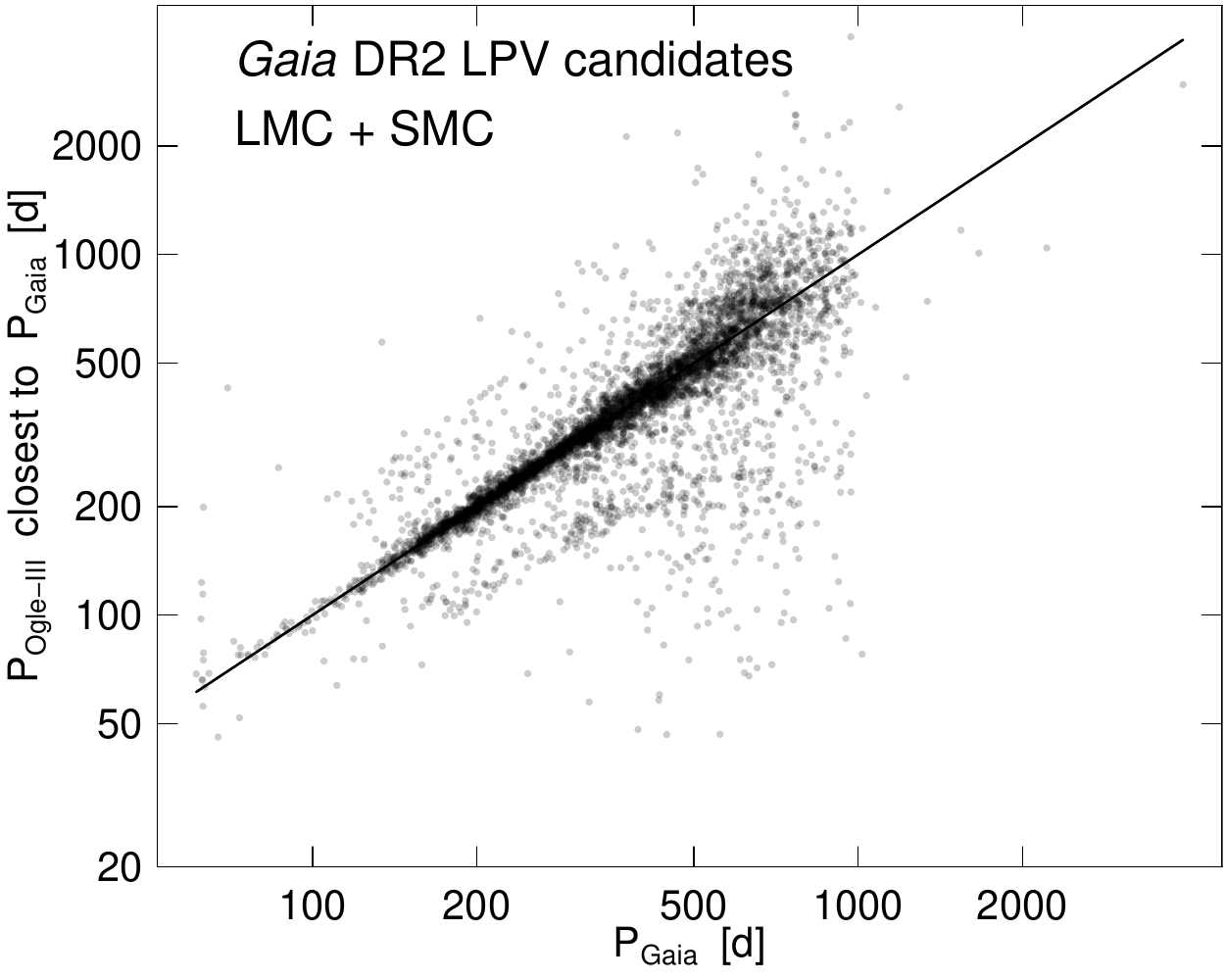}
\caption{Comparison of \Gaia\  DR2 periods (X-axis) with OGLE-III periods of LPV candidates in the Magellanic Clouds with \Gaia--OGLE-III crossmatches.
         The closest to the \Gaia period of the three periods reported in the OGLE-III catalogue for each crossmatch is considered.
        }
\label{Fig:periodsGaiaVsOgle_Clouds}
\end{figure}

\begin{figure}
\centering
\includegraphics[width=\hsize]{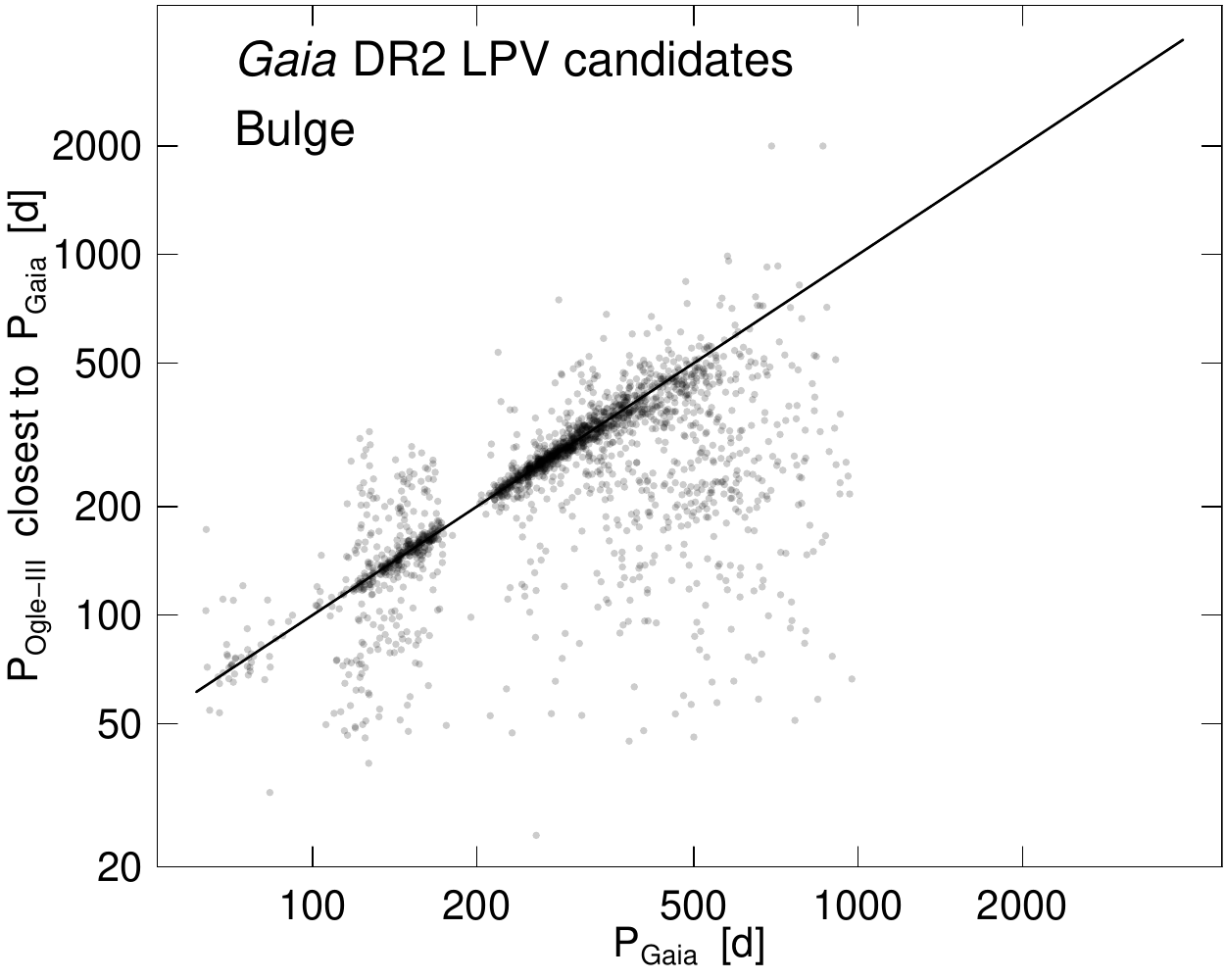}
\caption{Same as Fig.~\ref{Fig:periodsGaiaVsOgle_Clouds}, but for LPVs in the direction of the Galactic bulge.
        }
\label{Fig:periodsGaiaVsOgle_Bulge}
\end{figure}

Because most LPVs are multi-periodic, OGLE-III publishes up to three periods in their catalogues.
We therefore compare in this section the \Gaia periods to the OGLE-III periods that are closest to the \Gaia period for each source.
The comparison is shown in Fig.~\ref{Fig:periodsGaiaVsOgle_Clouds} for LPVs in the LMC and SMC, and in Fig.~\ref{Fig:periodsGaiaVsOgle_Bulge} for LPVs towards the Galactic bulge.

For sources in the Magellanic Clouds (Fig.~\ref{Fig:periodsGaiaVsOgle_Clouds}), where \Gaia sampling is very good (the mean light-curve time coverage of \gmag measurements is 82\% in the direction of the clouds), the comparison shows a good agreement between the periods in the two surveys, overall validating the \Gaia periods.
There are 71\% of \Gaia sources with a period within 15\% of the closest OGLE-III period.
Although the number of variables decreases significantly towards the short-period end, there is no indication for a decreasing period recovery efficiency at these periods.
As expected from the limited time span covered by the \Gaia observations, the agreement decreases for periods exceeding 700\,d.
A parallel sequence with $P_{Gaia} = 2 \times P_\mathrm{OGLE-III}$ is weakly visible, similar to what was found in the comparison with the ASAS\_SN dataset (see Fig.\,\ref{Fig:periodsGaiaAsas}).

The comparison of \Gaia periods with OGLE-III periods towards the Galactic bulge (Fig.~\ref{Fig:periodsGaiaVsOgle_Bulge}), on the other hand, reveals interesting features.
In this region of the sky, \Gaia sampling is rather sparse, with a mean light-curve time coverage of \gmag measurements of only 42\% (the mean is taken on the ensemble of \Gaia sources that have crossmatches with the OGLE-III catalogue of LPVs towards the bulge).
The fraction of \Gaia sources that have \Gaia periods within 15\% of the closest OGLE-III periods is 65\%, which is less good than towards the Magellanic Clouds.
The most striking absences in Fig.~\ref{Fig:periodsGaiaVsOgle_Bulge}, however, are those of \textit{Gaia} sources within well-defined period ranges, namely at $P_{Gaia} \lesssim 120$~d and at $175~\mathrm{d} \lesssim P_{Gaia} \lesssim 210~\mathrm{d}$.
At these periods, the sparse \Gaia light-curve sampling towards the Galactic bulge does not allow a proper recovery of the true period, as noted in Sect.~\ref{Sect:overview_periods}.
Period recovery also fails at periods longer than $\sim$700~d, similarly to sources in the Magellanic Clouds, due to the limited DR2 observation time range. 
Both features, lower fraction of period recovery and absence of \Gaia periods in defined period ranges below $\sim$120~d and around 190~d, result from the specific \Gaia scanning law towards the Galactic bulge.

\section{Summary}
\label{Sect:summary}

We present here a short summary of the results, which in Sects.~\ref{Sect:summary_overview} to \ref{Sect:summary_OGLE} follow the order of the analyses preformed in Sects.~\ref{Sect:overview} to \ref{Sect:OGLE}, respectively.

\subsection{Overview of the survey}
\label{Sect:summary_overview}

The sky distribution of the DR2 LPV candidates shows an expected concentration of sources in the Galactic plane, as well as in the Magellanic Clouds and in the Sagittarius dwarf galaxy.
At least 23\,405 Mira-like variables can be clearly identified from their colours, \gmag-band amplitudes, and periods, which represent about 20\% of the LPV candidates of the whole sample, or 30\% of the LPV candidates with published periods.

In addition to variables with amplitudes larger
than 1~mag, the $P>200$~d regime includes many variables with smaller amplitudes, or bluer colours, or both.
Below 200~d, on the other hand, the number of detected periods decreases significantly.
The multi-periodic nature of most of LPVs, combined with the incomplete DR2 phase coverage, decreases the probability of a reliable period detection for periods shorter than $\sim$200~d.
For sources with insufficient light-curve time coverage, period detection is particularly difficult around 190~d and below $\sim$120~d.
This is the case, in DR2, towards the Galactic centre.

\subsection{Selected individual LPVs}
\label{Sect:summary_individualLPVs}

The analysis of samples of individual LPVs has shown that a non-negligible fraction of bright LPVs are missing in the DR2 catalogue of LPV candidates (see Sect.~\ref{Sect:individualLPVs_wellStudiedCases}).
For those available in the catalogue, the \Gaia light curves match the AAVSO light curves well, with \Gaia \gmag magnitudes of LPVs being brighter than $V$ magnitudes (by about 3~mag).
The periods published in DR2 also agree very well with the ground-based derived periods.

Regarding BCs, the $m_{\rm bol}$ values derived from the BCs published in DR2 are within a few tenths of magnitudes of the SED $m_{\rm bol}$ values available in the literature for M-type stars.
This is consistent with the scatter expected from the BC relation.
For C-stars and stars showing an infrared excess from circumstellar dust, an improved BC is required.

\subsection{Sample with good parallaxes}
\label{Sect:summary_goodParallaxes}

The main conclusion of the analysis of this sample is the identification of YSOs as  main contaminants to the catalogue of LPV candidates in the solar vicinity within our Galaxy.
They constitute a very small fraction of the whole catalogue at the percent level,
but they form the majority of sources within about 500~pc around the Sun, their number decreasing to less than 5\% in a sphere of radius of $\sim$2.7~kpc.
In the sample of DR2 LPV candidates with good parallaxes ($\varpi/\varepsilon(\varpi)>10$), they amount to 671 out of 3093 objects, or 20\%.
They can easily be distinguished from LPVs in the observational HR diagram.

It has also been stressed in Sect.~\ref{Sect:goodParallaxes_selection} that a bias is introduced in samples of LPVs defined by relative parallax uncertainties.
In such samples, the reddest LPVs are under-represented.
This is due to the colour dependence of parallax precisions for these objects, whereby redder LPVs are fainter in $\gmag$ at a given distance and therefore have less precise parallaxes.

\subsection{All-sky comparison with ASAS\_SN}
\label{Sect:summary_ASAS}

About 43\% of LPV stars in the ASAS\_SN catalogue are present in the \Gaia DR2 catalogue of LPVs.
The \Gaia periods of these crossmatched stars agree very well with the published ASAS\_SN periods, 77\% of them have \Gaia periods within 15\% of the ASAS\_SN periods.
For another 9\% of the crossmatched stars with \Gaia periods, the \Gaia periods are twice as long as the ASAS\_SN periods, within 15\%.
The ASAS\_SN periods that are too short by a factor of about two are related to ASAS\_SN sources with a significant $V$ brightness below the ASAS\_SN magnitude limit of $\sim$17~mag.
Finally, the comparison of the variability amplitudes between ASAS\_SN light curves in the $V$ band and \Gaia light curves in the \gmag band enabled us to highlight the presence of blended sources in ASAS\_SN, a result inherent to the large 8" pixel scale of the survey.

\subsection{Comparison with OGLE-III}
\label{Sect:summary_OGLE}

The completeness of the \Gaia DR2 catalogue relative to OGLE-III catalogues is found to be about 45\% in the direction of the Magellanic Clouds, and of only about 8\% in the direction of the Galactic bulge.
The low recovery rate towards the Bulge is mainly due to the few observations in \Gaia DR2.
We therefore expect much higher levels of completeness in future \Gaia data releases.

Nevertheless, despite the relatively low level of completeness in DR2, about 5\% of the \Gaia DR2 LPVs in the clouds and 30\% towards the bulge are new candidates relative to the OGLE-III survey.

Finally, a comparison between the \Gaia and OGLE-III periods shows very good agreement for sufficient light-curve time coverage, even for stars at the short-period end.
This is the case for sources in the Magellanic Clouds ($\sim$70\% period recovery within 15\% of the OGLE-III periods).
For sources towards the bulge, whose light-curve time coverage is less than 50\%, the agreement is still satisfactory ($\sim$63\% period recovery within 15\% of the OGLE-III periods).
Periods below $\sim$120~d and between $\sim$175~d and $\sim$210~d cannot be recovered, however, as described in Sect.~\ref{Sect:summary_overview}.
Period recovery efficiencies will naturally improve in future \Gaia data releases with fuller time coverage.

\section{Conclusions}
\label{Sect:conclusions}

\paragraph{Main DR2 achievements for LPVs.}

The catalogue of LPV candidates published in DR2 is the first \Gaia catalogue of these objects.
Its content has been analysed in Sects.~\ref{Sect:individualLPVs} to \ref{Sect:OGLE} using successive LPV samples, from bright to faint,  a summary of which has been given in Sect.~\ref{Sect:summary}.
The main achievements in DR2 are listed below.
\begin{itemize}
    \item More than 150\,000 LPV candidates with variability amplitudes larger than 0.2~mag have been published in DR2.
    This number has to be compared with the total number of $\sim$20\,000 LPVs in the GCVS, the $\sim$16\,000 LPVs listed in ASAS\_SN, and the $\sim$68\,000 LPVs with $I$-band amplitudes larger than 0.2~mag identified in the OGLE-III surveys.
    The \Gaia DR2 catalogue of LPV candidates thus almost doubles the number of known LPVs.
    \vskip 1mm
    \item The contamination level is estimated to be at the few-percent level.
    
    \vskip 1mm
    \item LPV-specific properties, including periods and bolometric corrections, are published in the catalogue for more than half of the objects.
    
    \vskip 1mm
    \item The agreement between the periods derived from \Gaia data and the periods derived from ground-based observations is good overall.
    
    \vskip 1mm
    \item The bolometric corrections lead to apparent bolometric magnitudes that agree well overall with the values derived from the SED fit for M-type stars
    except for stars with large \gmag amplitudes.
    
    \vskip 1mm
    \item A serendipitous \Gaia sample of between 600 and 2000 YSOs in the solar vicinity is available.
    Although these objects are contaminants with respect to LPVs, they offer an interesting set of YSOs to study, with \gmag, \gbp , and \grp light curves published in DR2.
\end{itemize}

\paragraph{Limitations of the DR2 catalogue of LPVs.}

This is the first LPV catalogue produced by the \Gaia DPAC, which is delivered at an early stage of the data release scenario (many LPVs have periods longer than the time duration covered by DR2 and most of them are multi-periodic), and it was produced when the full potential of the various types of data made available by \Gaia was not yet available during the processing of variable objects. It therefore necessarily contains several limitations in this DR2.
The main limitations are listed below.
\begin{itemize}
    \item It is restricted to LPVs with amplitudes larger than 0.2~mag.

    \vskip 1mm
    \item It is incomplete because the priority for DR2 was set to limit contamination.

    \vskip 1mm
    \item The sky distribution depends on the \Gaia scanning law, which can affect sky distribution studies.  

    \vskip 1mm
    \item It is contaminated by YSOs in the solar vicinity.
    
    \vskip 1mm
    \item A large fraction of very bright LPVs ($\gmag \lesssim 7$~mag) are absent from the catalogue.

    \vskip 1mm
    \item It misses very long periods ($P>1000$~d).
    
    \vskip 1mm
    \item It is restricted to one period.

    \vskip 1mm
    \item The validity of the bolometric corrections is restricted to M-type stars with $QR_5(G)$ amplitudes smaller than 3~mag.
    
    \vskip 1mm
    \item The bolometric corrections do not take extinction into account.

    \vskip 1mm
    \item It uses a simplified prescription to identify red supergiants.
    
    \vskip 1mm
    \item It uses preliminary parallaxes available within the \Gaia consortium at the time of our processing, which affects the bolometric magnitudes published in this catalogue, as well as the identification of red supergiants and associated bolometric corrections. This limitation can be easily overcome by recomputing these quantities using the parallaxes published in \Gaia DR2, following the prescription given in Appendix~\ref{Sect:SOS}.
\end{itemize}

\paragraph{Expected improvements.}

The DR2 catalogue of LPVs must thus be considered as a first such release by the \Gaia team to give a taste of the type of results \Gaia can provide for the study of LPVs.
A non-exhaustive list of improvements for future \Gaia data releases is given below.
\begin{itemize}
    \item Increase in completeness level while keeping a low level of contamination.
    This will require improved classification and revised selection criteria, for example.
    
    \vskip 1mm
    \item The handling of low-amplitude ($<0.2$~mag) LPVs.
    
    \vskip 1mm
    \item The detection of very long periods, which will be made possible with the longer observation time intervals of future releases.
    
    \vskip 1mm
    \item The multi-periodic characterisation of LPVs, which will be made possible with the increased number of observations per time-series in future releases.
    
    \vskip 1mm
    \item The identification of the type of red giant (C-rich / O-rich), which is necessary to compute adequate bolometric corrections.
    The usage of the spectra from the \Gaia BP and RP spectro-photometers and \Gaia RVS spectrometer will be key in this characterisation.
    
    \vskip 1mm
    \item The use of radial velocity time-series to further characterise their pulsation properties.
    \vskip 1mm
    \item The availability of overall improved calibrations and \Gaia data products.
\end{itemize}

This first \Gaia catalogue of LPV candidates that has been released in DR2 represents a unique data base for the study of individual LPVs, and for the study of populations of low- to intermediate-mass stars in general, both in the Milky Way and in close extragalactic systems.
It provides an outstanding sample on which to study the full range of variability patterns, to distinguish various groups according to their distributions and dynamical behaviours in the Galaxy, and to investigate the Galactic structure using LPVs as an intermediate-age population with a significant impact on the Galaxy due to their high mass-loss rates.

\begin{acknowledgements}
We thank the referee, T. Bedding, for constructive comments.
This work has made use of data from the European Space Agency (ESA)
mission {\it Gaia} (\url{https://www.cosmos.esa.int/gaia}), processed by
the {\it Gaia} Data Processing and Analysis Consortium (DPAC,
\url{https://www.cosmos.esa.int/web/gaia/dpac/consortium}). Funding
for the DPAC has been provided by national institutions, in particular
the institutions participating in the {\it Gaia} Multilateral Agreement.\\
We acknowledge with thanks the variable star observations from the AAVSO International Database contributed by observers worldwide and used in this research.\\
This publication makes use of data products from the Wide-field Infrared Survey Explorer, which is a joint project of the University of California, Los Angeles, and the Jet Propulsion Laboratory/California Institute of Technology, funded by the National Aeronautics and Space Administration.
\end{acknowledgements}

\bibliographystyle{aa}
\bibliography{gaiaLPV_DR2}

\begin{thebibliography}{50}
\expandafter\ifx\csname natexlab\endcsname\relax\def\natexlab#1{#1}\fi

\bibitem[{{Alcock} {et~al.}(1993){Alcock}, {Allsman}, {Axelrod}, {Bennett},
  {Cook}, {Park}, {Marshall}, {Stubbs}, {Griest}, {Perlmutter}, {Sutherland},
  {Freeman}, {Peterson}, {Quinn}, \& {Rodgers}}]{1993ASPC...43..291A}
{Alcock}, C., {Allsman}, R.~A., {Axelrod}, T.~S., {et~al.} 1993, in
  Astronomical Society of the Pacific Conference Series, Vol.~43, Sky Surveys.
  Protostars to Protogalaxies, ed. B.~T. {Soifer}, 291

\bibitem[{{Aringer} {et~al.}(2016){Aringer}, {Girardi}, {Nowotny}, {Marigo}, \&
  {Bressan}}]{Aringer2016}
{Aringer}, B., {Girardi}, L., {Nowotny}, W., {Marigo}, P., \& {Bressan}, A.
  2016, \mnras, 457, 3611

\bibitem[{{Cutri} \& {et al.}(2013)}]{allWISE}
{Cutri}, R.~M. \& {et al.} 2013, VizieR Online Data Catalog, 2328

\bibitem[{{de Zeeuw} {et~al.}(1999){de Zeeuw}, {Hoogerwerf}, {de Bruijne},
  {Brown}, \& {Blaauw}}]{deZeeuwHoogerwerfdeBruijne_etal99}
{de Zeeuw}, P.~T., {Hoogerwerf}, R., {de Bruijne}, J.~H.~J., {Brown}, A.~G.~A.,
  \& {Blaauw}, A. 1999, \aj, 117, 354

\bibitem[{{Drake} {et~al.}(2017){Drake}, {Djorgovski}, {Catelan}, {Graham},
  {Mahabal}, {Larson}, {Christensen}, {Torrealba}, {Beshore}, {McNaught},
  {Garradd}, {Belokurov}, \& {Koposov}}]{DrakeDjorgovskiCatelan_etal17}
{Drake}, A.~J., {Djorgovski}, S.~G., {Catelan}, M., {et~al.} 2017, \mnras, 469,
  3688

\bibitem[{{Drake} {et~al.}(2014){Drake}, {Graham}, {Djorgovski}, {Catelan},
  {Mahabal}, {Torrealba}, {Garc{\'{\i}}a-{\'A}lvarez}, {Donalek}, {Prieto},
  {Williams}, {Larson}, {Christen sen}, {Belokurov}, {Koposov}, {Beshore},
  {Boattini}, {Gibbs}, {Hill}, {Kowalski}, {Johnson}, \&
  {Shelly}}]{DrakeGrahamDjorgovski14}
{Drake}, A.~J., {Graham}, M.~J., {Djorgovski}, S.~G., {et~al.} 2014, \apjs,
  213, 9

\bibitem[{{Edmonds} \& {Gilliland}(1996)}]{EdmondsGilliland96}
{Edmonds}, P.~D. \& {Gilliland}, R.~L. 1996, \apjl, 464, L157

\bibitem[{{Evans} {et~al.}(2018){Evans}, {Riello}, {De Angeli}, {Carrasco},
  {Montegriffo}, {Fabricius}, {Jordi}, {Palaversa}, {Diener}, {Busso},
  {Cacciari}, \& {van Leeuwen}}]{DR2-DPACP-40}
{Evans}, D.~W., {Riello}, M., {De Angeli}, F., {et~al.} 2018, ArXiv e-prints

\bibitem[{{Eyer} {et~al.}(1994){Eyer}, {Grenon}, {Falin}, {Froeschle}, \&
  {Mignard}}]{EyerGrenonFalin_etal94}
{Eyer}, L., {Grenon}, M., {Falin}, J.-L., {Froeschle}, M., \& {Mignard}, F.
  1994, \solphys, 152, 91

\bibitem[{{Eyer} {et~al.}(2017){Eyer}, {Mowlavi}, {Evans}, {Nienartowicz},
  {Ordonez}, {Holl}, {Lecoeur-Taibi}, {Riello}, {Clementini}, {Cuypers}, {De
  Ridder}, {Lanzafame}, {Sarro}, {Charnas}, {Guy}, {Jevardat de Fombelle},
  {Rimoldini}, {S{\"u}veges}, {Mignard}, {Busso}, {De Angeli}, {van Leeuwen},
  {Dubath}, {Beck}, {Aguado}, {Debosscher}, {Distefano}, {Fuchs}, {Koubsky},
  {Lebzelter}, {Leccia}, {Lopez}, {Moitinho}, {Regibo}, {Ripepi}, {Roelens},
  {Szabados}, {Tingley}, {Votruba}, {Zucker}, {Aerts}, {Barblan},
  {Blanco-Cuaresma}, {Grenon}, {Jan}, {Lorenz}, {Miranda}, {Morgenthaler},
  {Ordenovic}, {Palaversa}, {Prsa}, {Ruiz-Fuertes}, {Anderson}, {Delgado},
  {Dzigan}, {Hudec}, {Jonckheere}, {Klagyivik}, {Kutka}, {Moniez}, {Nicoletti},
  {Park}, {Van Hemelryck}, {Varadi}, {Kochoska}, {Lanza}, {Marconi},
  {Marschalko}, {Messina}, {Musella}, {Pagano}, {Sadowski}, \&
  {Schultheis}}]{DR1-DPACP-15}
{Eyer}, L., {Mowlavi}, N., {Evans}, D.~W., {et~al.} 2017, ArXiv e-prints

\bibitem[{{Gaia Collaboration} {et~al.}(2018{\natexlab{a}}){Gaia
  Collaboration}, {Brown}, {Vallenari}, {Prusti}, {de Bruijne}, {Babusiaux}, \&
  {Bailer-Jones}}]{DR2-DPACP-36}
{Gaia Collaboration}, {Brown}, A.~G.~A., {Vallenari}, A., {et~al.}
  2018{\natexlab{a}}, ArXiv e-prints

\bibitem[{{Gaia Collaboration} {et~al.}(2018{\natexlab{b}}){Gaia
  Collaboration}, {Eyer}, {Rimoldini}, {Audard}, {Anderson}, {Nienartowicz},
  {Glass}, {Marchal}, {Grenon}, {Mowlavi}, {Holl}, {Clementini}, {Aerts},
  {Mazeh}, {Evans}, {Szabados}, \& {co-authors}}]{DR2-DPACP-Eyer}
{Gaia Collaboration}, {Eyer}, L., {Rimoldini}, L., {et~al.} 2018{\natexlab{b}},
  ArXiv e-prints

\bibitem[{{Grenon}(1993)}]{Grenon93}
{Grenon}, M. 1993, in Astronomical Society of the Pacific Conference Series,
  Vol.~40, IAU Colloq. 137: Inside the Stars, ed. W.~W. {Weiss} \& A.~{Baglin},
  693--707

\bibitem[{{Heinze} {et~al.}(2018){Heinze}, {Tonry}, {Denneau}, {Flewelling},
  {Stalder}, {Rest}, {Smith}, {Smartt}, \&
  {Weiland}}]{HeinzeTonryDenneau_etal18}
{Heinze}, A.~N., {Tonry}, J.~L., {Denneau}, L., {et~al.} 2018, ArXiv e-prints

\bibitem[{{H{\"o}fner} \& {Olofsson}(2018)}]{2018A&ARv..26....1H}
{H{\"o}fner}, S. \& {Olofsson}, H. 2018, \aapr, 26, 1

\bibitem[{{Holl} {et~al.}(2018){Holl}, {Audard}, {Nienartowicz}, {Jevardat de
  Fombelle}, {Marchal}, {Mowlavi}, {Clementini}, {De Ridder}, {Evans}, {Guy},
  {Lanzafame}, {Lebzelter}, {Rimoldini}, {Roelens}, {Zucker}, {Distefano},
  {Garofalo}, {Lecoeur-Ta{\"i}bi}, {Lopez}, {Molinaro}, {Muraveva}, {Panahi},
  {Regibo}, {Ripepi}, {Sarro}, {Aerts}, {Anderson}, {Charnas}, {Barblan},
  {Blanco-Cuaresma}, {Busso}, {Cuypers}, {De Angeli}, {Glass}, {Grenon},
  {Juh{\'a}sz}, {Kochoska}, {Koubsky}, {Lanza}, {Leccia}, {Lorenz}, {Marconi},
  {Marschalk}, {Mazeh}, {Messina}, {Mignard}, {Moitinho}, {Moln{\'a}r},
  {Morgenthaler}, {Musella}, {Ordenovic}, {Ord{\'o}{\~n}ez}, {Pagano},
  {Palaversa}, {Pawlak}, {Plachy}, {Pr{\v s}a}, {Riello}, {S{\"u}veges},
  {Szabados}, {Szegedi-Elek}, {Votruba}, \& {Eyer}}]{Holl_etal18}
{Holl}, B., {Audard}, M., {Nienartowicz}, K., {et~al.} 2018, ArXiv e-prints

\bibitem[{{Hron}(1991)}]{1991A&A...252..583H}
{Hron}, J. 1991, \aap, 252, 583

\bibitem[{{Ita} {et~al.}(2004){Ita}, {Tanab{\'e}}, {Matsunaga}, {Nakajima},
  {Nagashima}, {Nagayama}, {Kato}, {Kurita}, {Nagata}, {Sato}, {Tamura},
  {Nakaya}, \& {Nakada}}]{2004MNRAS.353..705I}
{Ita}, Y., {Tanab{\'e}}, T., {Matsunaga}, N., {et~al.} 2004, \mnras, 353, 705

\bibitem[{{Jayasinghe} {et~al.}(2018){Jayasinghe}, {Kochanek}, {Stanek},
  {Shappee}, {Holoien}, {Thompson}, {Prieto}, {Dong}, {Pawlak}, {Shields},
  {Pojmanski}, {Otero}, {Britt}, \& {Will}}]{Jayasinghe_etal18}
{Jayasinghe}, T., {Kochanek}, C.~S., {Stanek}, K.~Z., {et~al.} 2018, ArXiv
  e-prints

\bibitem[{{Jordi} {et~al.}(2010){Jordi}, {Gebran}, {Carrasco}, {de Bruijne},
  {Voss}, {Fabricius}, {Knude}, {Vallenari}, {Kohley}, \&
  {Mora}}]{2010A&A...523A..48J}
{Jordi}, C., {Gebran}, M., {Carrasco}, J.~M., {et~al.} 2010, \aap, 523, A48

\bibitem[{{Jorissen} {et~al.}(1997){Jorissen}, {Mowlavi}, {Sterken}, \&
  {Manfroid}}]{JorissenMowlaviSterken_etal97}
{Jorissen}, A., {Mowlavi}, N., {Sterken}, C., \& {Manfroid}, J. 1997, \aap,
  324, 578

\bibitem[{{Kerschbaum} \& {Hron}(1996)}]{1996A&A...308..489K}
{Kerschbaum}, F. \& {Hron}, J. 1996, \aap, 308, 489

\bibitem[{{Kerschbaum} {et~al.}(2010){Kerschbaum}, {Lebzelter}, \&
  {Mekul}}]{2010A&A...524A..87K}
{Kerschbaum}, F., {Lebzelter}, T., \& {Mekul}, L. 2010, \aap, 524, A87

\bibitem[{{Kiss} \& {Bedding}(2003)}]{2003MNRAS.343L..79K}
{Kiss}, L.~L. \& {Bedding}, T.~R. 2003, \mnras, 343, L79

\bibitem[{{Lebzelter} {et~al.}(2005){Lebzelter}, {Hinkle}, {Wood}, {Joyce}, \&
  {Fekel}}]{2005A&A...431..623L}
{Lebzelter}, T., {Hinkle}, K.~H., {Wood}, P.~R., {Joyce}, R.~R., \& {Fekel},
  F.~C. 2005, \aap, 431, 623

\bibitem[{{Levesque} {et~al.}(2005){Levesque}, {Massey}, {Olsen}, {Plez},
  {Josselin}, {Maeder}, \& {Meynet}}]{Levesque2005}
{Levesque}, E.~M., {Massey}, P., {Olsen}, K.~A.~G., {et~al.} 2005, \apj, 628,
  973

\bibitem[{{Lindegren} {et~al.}(2018){Lindegren}, {Hernandez}, {Bombrun},
  {Klioner}, {Bastian}, {Ramos-Lerate}, {de Torres}, {Steidelmuller},
  {Stephenson}, {Hobbs}, {Lammers}, {Biermann}, {Geyer}, {Hilger}, {Michalik},
  {Stampa}, {McMillan}, {Castaneda}, {Clotet}, {Comoretto}, {Davidson},
  {Fabricius}, {Gracia}, {Hambly}, {Hutton}, {Mora}, {Portell}, {van Leeuwen},
  {Abbas}, {Abreu}, {Altmann}, {Andrei}, {Anglada}, {Balaguer-Nunez},
  {Barache}, {Becciani}, {Bertone}, {Bianchi}, {Bouquillon}, {Bourda},
  {Brusemeister}, {Bucciarelli}, {Busonero}, {Buzzi}, {Cancelliere},
  {Carlucci}, {Charlot}, {Cheek}, {Crosta}, {Crowley}, {de Bruijne}, {de
  Felice}, {Drimmel}, {Esquej}, {Fienga}, {Fraile}, {Gai}, {Garralda},
  {Gonzalez-Vidal}, {Guerra}, {Hauser}, {Hofmann}, {Holl}, {Jordan},
  {Lattanzi}, {Lenhardt}, {Liao}, {Licata}, {Lister}, {Loffler}, {Marchant},
  {Martin-Fleitas}, {Messineo}, {Mignard}, {Morbidelli}, {Poggio}, {Riva},
  {Rowell}, {Salguero}, {Sarasso}, {Sciacca}, {Siddiqui}, {Smart}, {Spagna},
  {Steele}, {Taris}, {Torra}, {van Elteren}, {van Reeven}, \&
  {Vecchiato}}]{DR2-DPACP-14}
{Lindegren}, L., {Hernandez}, J., {Bombrun}, A., {et~al.} 2018, ArXiv e-prints

\bibitem[{{Madore}(1982)}]{Madore82}
{Madore}, B.~F. 1982, \apj, 253, 575

\bibitem[{{Mowlavi}(2014)}]{Mowlavi14}
{Mowlavi}, N. 2014, \aap, 568, A78

\bibitem[{{Perrot} \& {Grenier}(2003)}]{PerrotGarnier2003}
{Perrot}, C.~A. \& {Grenier}, I.~A. 2003, \aap, 404, 519

\bibitem[{{Pourbaix} {et~al.}(2003){Pourbaix}, {Platais}, {Detournay},
  {Jorissen}, {Knapp}, \& {Makarov}}]{PourbaixPlataisDetournay_etal03}
{Pourbaix}, D., {Platais}, I., {Detournay}, S., {et~al.} 2003, \aap, 399, 1167

\bibitem[{{Riebel} {et~al.}(2010){Riebel}, {Meixner}, {Fraser}, {Srinivasan},
  {Cook}, \& {Vijh}}]{2010ApJ...723.1195R}
{Riebel}, D., {Meixner}, M., {Fraser}, O., {et~al.} 2010, \apj, 723, 1195

\bibitem[{Rimoldini(2018{\natexlab{a}})}]{Rimoldini_etal18}
Rimoldini, R., e.~a. 2018{\natexlab{a}}, \aap

\bibitem[{Rimoldini(2018{\natexlab{b}})}]{Rimoldini18}
Rimoldini, R. 2018{\natexlab{b}}, Gaia Documentation

\bibitem[{{Samus'} {et~al.}(2017){Samus'}, {Kazarovets}, {Durlevich},
  {Kireeva}, \& {Pastukhova}}]{2017ARep...61...80S}
{Samus'}, N.~N., {Kazarovets}, E.~V., {Durlevich}, O.~V., {Kireeva}, N.~N., \&
  {Pastukhova}, E.~N. 2017, Astronomy Reports, 61, 80

\bibitem[{{Shappee} {et~al.}(2014){Shappee}, {Prieto}, {Grupe}, {Kochanek},
  {Stanek}, {De Rosa}, {Mathur}, {Zu}, {Peterson}, {Pogge}, {Komossa}, {Im},
  {Jencson}, {Holoien}, {Basu}, {Beacom}, {Szczygie{\l}}, {Brimacombe},
  {Adams}, {Campillay}, {Choi}, {Contreras}, {Dietrich}, {Dubberley},
  {Elphick}, {Foale}, {Giustini}, {Gonzalez}, {Hawkins}, {Howell}, {Hsiao},
  {Koss}, {Leighly}, {Morrell}, {Mudd}, {Mullins}, {Nugent}, {Parrent},
  {Phillips}, {Pojmanski}, {Rosing}, {Ross}, {Sand}, {Terndrup}, {Valenti},
  {Walker}, \& {Yoon}}]{2014ApJ...788...48S}
{Shappee}, B.~J., {Prieto}, J.~L., {Grupe}, D., {et~al.} 2014, \apj, 788, 48

\bibitem[{{Soszy{\'n}ski} {et~al.}(2009){Soszy{\'n}ski}, {Udalski},
  {Szyma{\'n}ski}, {Kubiak}, {Pietrzy{\'n}ski}, {Wyrzykowski}, {Szewczyk},
  {Ulaczyk}, \& {Poleski}}]{2009AcA....59..239S}
{Soszy{\'n}ski}, I., {Udalski}, A., {Szyma{\'n}ski}, M.~K., {et~al.} 2009,
  \actaa, 59, 239

\bibitem[{{Soszy{\'n}ski} {et~al.}(2011){Soszy{\'n}ski}, {Udalski},
  {Szyma{\'n}ski}, {Kubiak}, {Pietrzy{\'n}ski}, {Wyrzykowski}, {Ulaczyk},
  {Poleski}, {Koz{\l}owski}, \& {Pietrukowicz}}]{2011AcA....61..217S}
{Soszy{\'n}ski}, I., {Udalski}, A., {Szyma{\'n}ski}, M.~K., {et~al.} 2011,
  ACTAA, 61, 217

\bibitem[{{Soszy{\'n}ski} {et~al.}(2013){Soszy{\'n}ski}, {Udalski},
  {Szyma{\'n}ski}, {Kubiak}, {Pietrzy{\'n}ski}, {Wyrzykowski}, {Ulaczyk},
  {Poleski}, {Koz{\l}owski}, {Pietrukowicz}, \&
  {Skowron}}]{2013AcA....63...21S}
{Soszy{\'n}ski}, I., {Udalski}, A., {Szyma{\'n}ski}, M.~K., {et~al.} 2013,
  \actaa, 63, 21

\bibitem[{{Spano} {et~al.}(2009){Spano}, {Mowlavi}, {Eyer}, \&
  {Burki}}]{SpanoMowlaviEyer_etal09}
{Spano}, M., {Mowlavi}, N., {Eyer}, L., \& {Burki}, G. 2009, in American
  Institute of Physics Conference Series, Vol. 1170, American Institute of
  Physics Conference Series, ed. J.~A. {Guzik} \& P.~A. {Bradley}, 324--326

\bibitem[{{Tabur} {et~al.}(2010){Tabur}, {Bedding}, {Kiss}, {Giles}, {Derekas},
  \& {Moon}}]{TaburBeddingKiss_etal10}
{Tabur}, V., {Bedding}, T.~R., {Kiss}, L.~L., {et~al.} 2010, \mnras, 409, 777

\bibitem[{{Taylor} {et~al.}(1987){Taylor}, {Dickman}, \&
  {Scoville}}]{TaylorDickman_scoville87}
{Taylor}, D.~K., {Dickman}, R.~L., \& {Scoville}, N.~Z. 1987, \apj, 315, 104

\bibitem[{{Trabucchi} {et~al.}(2017){Trabucchi}, {Wood}, {Montalb{\'a}n},
  {Marigo}, {Pastorelli}, \& {Girardi}}]{2017ApJ...847..139T}
{Trabucchi}, M., {Wood}, P.~R., {Montalb{\'a}n}, J., {et~al.} 2017, \apj, 847,
  139

\bibitem[{{von Neumann}(1941)}]{vonNeumann1941}
{von Neumann}, J. 1941, Ann. Math. Stat., 12, 367

\bibitem[{{von Neumann}(1942)}]{vonNeumann1942}
{von Neumann}, J. 1942, Ann. Math. Stat., 13, 86

\bibitem[{{Wood}(2015)}]{Wood15}
{Wood}, P.~R. 2015, \mnras, 448, 3829

\bibitem[{{Wood} {et~al.}(1999){Wood}, {Alcock}, {Allsman}, {Alves}, {Axelrod},
  {Becker}, {Bennett}, {Cook}, {Drake}, {Freeman}, {Griest}, {King}, {Lehner},
  {Marshall}, {Minniti}, {Peterson}, {Pratt}, {Quinn}, {Stubbs}, {Sutherland},
  {Tomaney}, {Vandehei}, \& {Welch}}]{1999IAUS..191..151W}
{Wood}, P.~R., {Alcock}, C., {Allsman}, R.~A., {et~al.} 1999, in IAU Symposium,
  Vol. 191, Asymptotic Giant Branch Stars, ed. T.~{Le Bertre}, A.~{Lebre}, \&
  C.~{Waelkens}, 151

\bibitem[{{Wood} {et~al.}(1983){Wood}, {Bessell}, \& {Fox}}]{Wood1983}
{Wood}, P.~R., {Bessell}, M.~S., \& {Fox}, M.~W. 1983, \apj, 272, 99

\bibitem[{{Wray} {et~al.}(2004){Wray}, {Eyer}, \&
  {Paczy{\'n}ski}}]{2004MNRAS.349.1059W}
{Wray}, J.~J., {Eyer}, L., \& {Paczy{\'n}ski}, B. 2004, \mnras, 349, 1059

\bibitem[{{Zechmeister} \& {K{\"u}rster}(2009)}]{ZechmeisterKurster2009}
{Zechmeister}, M. \& {K{\"u}rster}, M. 2009, \aap, 496, 577

\end{thebibliography}

\begin{appendix}

\section{LPV candidate selection and characterisation}
\label{Sect:SOS}

A specific module within the \Gaia variability pipeline is dedicated to the selection and characterisation of LPV candidates.
It first selects LPV candidates according to a series of source-filtering criteria, then characterises each LPV candidate with the following LPV-specific attributes:
\begin{itemize}
\item the most significant period. The majority of the LPVs are multi-periodic, but we limit this in DR2 to only one period;
\vskip 1mm
\item the absolute bolometric magnitude;
\vskip 1mm
\item a flag indicating if the source is a red supergiant;
\vskip 1mm
\item the bolometric correction;
\vskip 1mm
\item a flag indicating if the source is a C-rich red giant (not applicable for DR2):
\vskip 1mm
\item a period-luminosity classification (not applicable for DR2).
\end{itemize}

All sources satisfying the source selection criteria are published as LPV candidates in \Gaia DR2.
The LPV-specific attributes of sources with periods longer than 60 days detected in their \gmag light curves are additionally published in the \texttt{vari\_long\_period\_variable} table of DR2.
The numbers are summarised in Fig.~\ref{Fig:numLPVs}.

The source-filtering criteria are listed in Sect.~\ref{Sect:SOS_filters}, and the computation of the LPV-specific attributes is described in Sects.~\ref{Sect:SOS_period} to \ref{Sect:SOS_BCs}.

\subsection{Source filtering}
\label{Sect:SOS_filters}

Long-period variable candidates published in DR2 were selected by applying the following filtering criteria.

\begin{enumerate}
\item We first selected all sources in the 'geq2' and 'geq20' data sets \citep[see Sect.~\ref{Sect:catalog} and][]{Holl_etal18} that were classified as \texttt{MIRA\_SR} candidates according to their best classification computed by the supervised classifiers. They have at least 12 observations in their cleaned \gmag light curves and at least 9 observations in their cleaned \grp light curves.
Light-curve cleaning consists of the removal of duplicate measurements, outlier removal, and error cleaning \citep[see][]{Holl_etal18}.
\vskip 2mm
\item We then selected sources with a \gmag-band variability amplitude larger than 0.2~mag.
The variability amplitude is computed as the 5-95\% quantile range of the \gmag-band light curve, using the LEGACY strategy of commons-math to compute the percentiles\footnote{
The variability amplitude computed using the 5\% quantile range of the \gmag-band light curve is noted $QR_5(\gmag)$ in this paper.
}.
By doing this, we publish in Gaia DR2 only LPVs with large variability amplitudes.
\vskip 2mm
\item We limited our sample to candidates with a $\gbp-\grp$ colour greater than 0.5~mag (the mean values of \gbp and \grp time-series are used to compute the colour) and a Spearman correlation between \gmag and $\gbp-\grp$  larger than 0.5~mag.
The latter criterion led to a bias in the selected sample whereby LPV candidates with large uncertainties on \gbp were excluded, such as very red stars that would have \gbp values above $\sim20$~mag.

\vskip 2mm
\item Finally, we kept only sources with a maximum Abbe value \citep{vonNeumann1941,vonNeumann1942,Mowlavi14} of 0.8 on their smoothed \gmag light curve.
The smoothed \gmag light curves were computed in an iterative way by merging successive pairs of observations with time separations shorter than five days.
The criterion on Abbe is the most conservative one.
A check on the \Gaia crossmatches with ASAS\_SN LPVs showed that about 20\% of the ASAS\_SN Miras and 30\% of the ASAS\_SN SRVs were discarded from the initial \Gaia set of LPV candidates as a result of this criterion.
\end{enumerate}

\subsection{Period search}
\label{Sect:SOS_period}

A global modelling of the \gmag light curves was performed based on classical Fourier series decomposition. 
We first used the generalised Lomb Scargle algorithm \citep{ZechmeisterKurster2009} to search for the most significant period in the \gmag light curve within a frequency range of 0.001 to 0.1~d$^{-1}$, by steps of $5 \times 10^{-5}$~d$^{-1}$.
A non-linear fit to the light curve of the period and Fourier series coefficients was then performed using the Apache Commons Math implementation (\url{http://commons.apache.org/proper/commons-math/}) of  the Levenberg-Marquardt algorithm.
This led in some rare cases to periods outside the pre-defined range of [10-1000] days.
Sources with periods greater than 60 days are published in DR2 with their LPV-specific attributes as described in Sect.~\ref{Sect:catalogContent}.

\subsection{Absolute bolometric magnitude}
\label{Sect:SOS_Mbol}

The absolute bolometric magnitude $M_\mathrm{bol}$ is given by
\begin{equation}
  M_\mathrm{bol} = m_G + BC - A_G + 5 \, \log \varpi + 5 ,
\label{Eq:Mbol}
\end{equation}
where $m_G$ is the mean \gmag-band magnitude, $BC$ is the bolometric correction, $A_G$ is the interstellar extinction, and $\varpi$ is the parallax in arcseconds. When the parallax is negative, the value of $M_\mathrm{bol}$ is set to NaN.

The uncertainties on the absolute bolometric magnitudes, $\sigma_{M_\mathrm{bol}}$, are given by
\begin{equation}
   \sigma_{M_\mathrm{bol}} = \sqrt{ \sigma_{m_G}^{2} + \sigma_{BC}^{2} + \sigma_{A_G}^{2} + 4.715 \, \frac{\sigma_{\varpi}^{2}}{\varpi^{2}} } \;\; ,
\label{Eq:sigma_Mbol}
\end{equation}
where $\sigma_{m_G}$, $\sigma_{BC}$, $\sigma_{A_G}$, and $\sigma_{\varpi}$ are the uncertainties on the associated quantities.
We must note that $M_\mathrm{bol}$ values are provided in DR2 for all stars that have a positive parallax, regardless of the parallax uncertainties.

For DR2, the mean values of the \gmag-band time-series have been used in Eqs.~\ref{Eq:Mbol} and \ref{Eq:sigma_Mbol}, and the BCs are computed according to the prescription given in Sect.~\ref{Sect:SOS_BCs}.
No \Gaia interstellar extinction was available at the time of processing, and the value of $A_G$ was set to $(0.00 \pm 0.05)$~mag.
All these shortcomings will be addressed in the future releases with the improvement of the astrophysical parameters and a better definition of the light curves.

\paragraph{\textbf{Important note}:} The parallaxes available within the \Gaia consortium at the time of our processing were a preliminary version of the parallaxes published in DR2 (see Appendix~\ref{Sect:parallaxes}).
Therefore, the values of $M_\mathrm{bol}$ and $\sigma_{M_\mathrm{bol}}$ published in DR2 are not compatible with the published parallaxes. For example, users will find $M_\mathrm{bol}$ values in the \Gaia DR2 archive for sources that have a published negative-parallax value (a result that would be impossible according to Eq.~\ref{Eq:sigma_Mbol}), while other sources with positive parallaxes do not have a $M_\mathrm{bol}$ value.
Users must recompute them using Eqs.~\ref{Eq:Mbol} and \ref{Eq:sigma_Mbol}.

\subsection{Identifying supergiant stars}
\label{Sect:SOS_supergiants}

Long-period variables consist of red giant stars, mainly on the AGB, and of red supergiant stars.
According to \citet{Wood1983}, the latter population can be identified based on their absolute bolometric magnitude and their main period.
In the $P - M_\mathrm{bol}$ diagram, each population is located in a specific region,
with an upper brightness limit $M_\mathrm{bol,minAGB}(P)$ for AGB stars, at any given period, given by
\begin{equation}
    M_\mathrm{bol,minAGB}(P) = - 5.62787 - 0.00383 \, P + 1.875 \times 10^{-6} \, P^2 \;,
\label{Eq:minMbol_AGB}
\end{equation}
where $P$ is given in days.
Consequently, all LPV candidates brighter than $M_\mathrm{bol,minAGB}(P)$ are flagged as supergiant stars. When the parallax is negative, the star is not considered as supergiant.

\paragraph{Note related to the parallax issue:} Because the supergiant star identification depends on $M_\mathrm{bol}$ and hence on parallaxes, interested users should update the list of supergiant stars published in DR2.
First, updated values of $M_\mathrm{bol}$ must be computed using Eq.~\ref{Eq:Mbol} with the parallaxes published in DR2, then supergiant stars must be identified using the criterion $M_\mathrm{bol} < M_\mathrm{bol,minAGB}(P)$.

\subsection{Bolometric correction}
\label{Sect:SOS_BCs}

The computation of BCs distinguishes three different cases, depending on the \gmag-band variability amplitude and on the giant or supergiant nature of the star.
The three cases are 1) red giant stars with variability amplitudes smaller than 3~mag, 2) red giants with amplitudes larger than 3~mag, and 3) supergiant stars.

\paragraph{Case 1: Standard red giants with $QR_5(\gmag)<3$~mag.}
The BCs are computed with the following relation based on synthetic spectra of hydrostatic M-star models \citep{Aringer2016}:
\begin{eqnarray}
       BC(\mbox{\gmag}) & = & 0.2438 - 0.25155 \, (\bpminrp) \nonumber \\
                        &   & -\; 0.11433 \, (\bpminrp)^{2} \nonumber \\
                        &   & +\; 0.00154 \, (\bpminrp)^{3}
\label{Eq:BC_LPVs}
,\end{eqnarray}
where the \bpminrp values are computed using the mean values of \gbp and \grp time-series.
However, if the uncertainties in the mean values of \gbp or of \grp are larger than 4~mag, the BC is computed assuming a $\bpminrp$ colour of star equal to 3.25~mag and an uncertainty on the colour equal to 2~mag, that is, $BC(\mbox{\gmag}) = -1.72848 \pm 1.89180$~mag.

 \paragraph{Case 2: Large-amplitude red giants with $QR_5(\gmag)>3$~mag.}
A fixed value of $BC(\mbox{\gmag}) = -2.2 \pm 0.005$~mag is used in DR2, based on \citet{2010A&A...524A..87K}.
For these stars, no reliable model grids for this purpose could be constructed.
We therefore decided to use a fixed value for DR2, with the intention to using DR2 data and near-infrared photometry for a calibration in DR3.
The use of a constant value was motivated by the constancy of the $BC$-values for the reddest objects, mostly Miras, found by \citet{2010A&A...524A..87K}.

\paragraph{Case 3: Red supergiant stars.}
The supergiant nature of a red star is assessed based on the position of the star in the $P-M_{bol}$ diagram (see Sect.~\ref{Sect:SOS_supergiants}).
Because this requires knowledge of the BC, the procedure is done in two steps.
The BC is first computed assuming the star is a red giant (case 1 or 2).
If the star is a supergiant candidate, the BC is recomputed accordingly.
A mean value of $BC(\mbox{\gmag}) = -0.71 \pm 0.3$~mag is set for all red supergiants based on the compilation of $m_{\rm bol}$ values for red supergiants by \citet{Levesque2005}.
The bolometric error represents the standard deviation around the mean value of the BCs published by these authors.

It must be stressed that only relations valid for M-type stars were used in computing the BCs for DR2.
The BCs published in DR2 for S- and C-type stars are thus not reliable.
Extinction has not been taken into account either in this release, since no reddening correction was available during our processing.

\paragraph{Note related to the parallax issue:} No final DR2 parallaxes were available at the time of our processing. This problem may also affect the BC values published in DR2 for stars that would have incorrectly been identified as giant/supergiant as a result of incorrect parallaxes (see Sect.~\ref{Sect:SOS_supergiants}).
Users should check the supergiant classification when recomputing $M_\mathrm{bol}$ with the published parallaxes, and recompute the BCs for all stars whose giant/supergiant classification is different than the one published in DR2.

\section{Catalogue retrieval}
\label{Sect:catalogRetrieval}

We summarise below the ADQL queries to be used in web interface to the Gaia DR2 archive (\url{https://gea.esac.esa.int/archive/}) in order to retrieve the LPV candidates, their attributes, and their light curves.

\begin{itemize}

\item Number of sources classified as LPVs (table \texttt{vari\_classifier\_result}):
\begin{footnotesize}
\begin{verbatim}
  SELECT count(*) 
  FROM gaiadr2.vari_classifier_result
  WHERE best_class_name ='MIRA_SR'
\end{verbatim}
\end{footnotesize}

\item Number of sources with LPV attributes (table \texttt{vari\_long\_period\_variable}):
\begin{footnotesize}
\begin{verbatim}
  SELECT count(*) 
  FROM gaiadr2.vari_long_period_variable
\end{verbatim}
\end{footnotesize}

\item Number of sources in common between the two tables \texttt{vari\_long\_period\_variable} and \texttt{vari\_classifier\_result}: 
\begin{footnotesize}
\begin{verbatim}
  SELECT count(*) 
  FROM gaiadr2.vari_classifier_result as claslpv
  INNER join gaiadr2.vari_long_period_variable AS soslpv
  ON claslpv.source_id = soslpv.source_id
  WHERE best_class_name = 'MIRA_SR'
\end{verbatim}
\end{footnotesize}

\item Number of sources classified as LPV without attributes: 
\begin{footnotesize}
\begin{verbatim}
  SELECT count(*) 
  FROM gaiadr2.vari_classifier_result as claslpv
  LEFT OUTER join 
    gaiadr2.vari_long_period_variable AS soslpv
  ON claslpv.source_id = soslpv.source_id
  WHERE best_class_name='MIRA_SR' 
  AND soslpv.source_id is NULL
\end{verbatim}
\end{footnotesize}

\item LPV attributes:
\begin{footnotesize}
\begin{verbatim}
  SELECT lpv.*
  FROM gaiadr2.vari_long_period_variable AS lpv
\end{verbatim}

\end{footnotesize}

\item Light curves.
The following query retrieves the URL where the photometry of an individual \Gaia source, in the example the source has the ID equal to 484786706893959936, can be downloaded:
\begin{footnotesize}
\begin{verbatim}
SELECT gaia.epoch_photometry_url
FROM gaiadr2.gaia_source AS gaia
INNER JOIN gaiadr2.vari_long_period_variable AS lpv
ON gaia.source_id = lpv.source_id 
AND lpv.source_id = 484786706893959936
\end{verbatim}
\end{footnotesize}

\vskip 2mm
\item Light-curve statistics.
Time-series statistics on the light curves, such as the mean \gmag, \gbp , and \grp magnitudes (fields \texttt{mean\_mag\_g\_fov}, \texttt{mean\_mag\_bp} and \texttt{mean\_mag\_rp}, respectively), or the median values of these magnitudes (\texttt{median\_mag\_g\_fov}, \texttt{median\_mag\_bp} and \texttt{median\_mag\_rp}, respectively), can be retrieved from table \texttt{vari\_time\_series\_statistics}.
To retrieve the mean magnitude in \gmag together with LPV attributes, the following query can be launched:
\begin{footnotesize}
\begin{verbatim}
SELECT stat.mean_mag_g_fov, lpv.*
FROM gaiadr2.vari_time_series_statistics AS stat
INNER JOIN gaiadr2.vari_long_period_variable AS lpv
ON stat.source_id = lpv.source_id
\end{verbatim}
\end{footnotesize}

\item Source coordinates and parallaxes.
\begin{footnotesize}
\begin{verbatim}
SELECT gs.source_id, ra, dec,
       parallax, parallax_error
FROM gaiadr2.gaia_source as gs 
INNER JOIN gaiadr2.vari_long_period_variable AS lpv
ON gs.source_id = lpv.source_id
        
\end{verbatim}
\end{footnotesize}

\textbf{
\item All LPV candidates with attributes when they exist.}

\begin{footnotesize}
\begin{verbatim}
SELECT gs.source_id, gs.ra , gs.dec,
    gs.parallax, gs.parallax_error,
    gs.radial_velocity,
    gs.radial_velocity_error,
    gs.rv_nb_transits,
    gs.phot_g_mean_flux,
    gs.phot_g_mean_flux_error, 
    lpv.abs_mag_bol, lpv.abs_mag_bol_error,
    lpv.rsg_flag, lpv.bolometric_corr,
    lpv.bolometric_corr_error,
    lpv.frequency, lpv.frequency_error
FROM gaiadr2.gaia_source as gs
LEFT JOIN gaiadr2.vari_classifier_result AS claslpv 
ON gs.source_id= claslpv.source_id
LEFT JOIN gaiadr2.vari_long_period_variable
AS lpv on gs.source_id= lpv.source_id
WHERE best_class_name ='MIRA_SR' OR
lpv.source_id is not null
        
\end{verbatim}
\end{footnotesize}

\end{itemize}

\section{DR2 parallaxes of LPV candidates}
\label{Sect:parallaxes}

\begin{figure}
\centering
\includegraphics[width=\hsize]{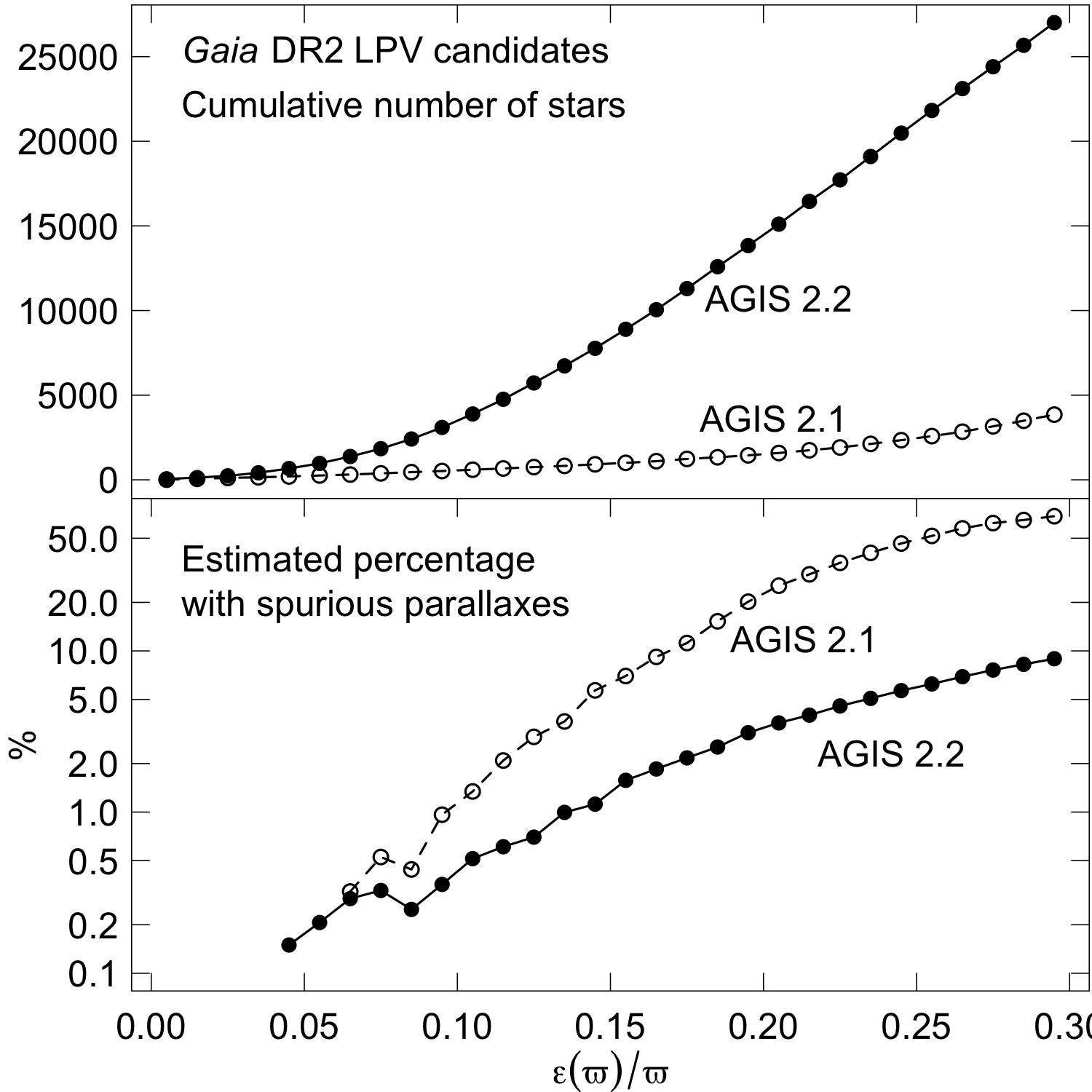}
\caption{\textbf{Top panel:} Cumulative distributions of the number of LPV candidates as a function of relative parallax uncertainties for the sample of sources with positive parallaxes.
         Open circles connected with a dashed line show the cumulative distribution for the parallaxes from AGIS~2.1 (preliminary results internal to DPAC), and filled circles connected with a solid line show the results from AGIS~2.2 (parallaxes published in DR2).
         \textbf{Bottom panel:} Same as top panel, but for the percentage of spurious positive parallaxes in the set of LPV candidates with relative parallax uncertainties below any value shown on the X-axis.
        }
\label{Fig:numStarsGoodParallaxes}
\end{figure}

The LPV attributes published in DR2 are based on a preliminary version of the parallaxes (produced by version 2.1 of AGIS, the Astrometric Global Iterative Solution) that was available within the \Gaia DPAC at the time of variable star processing.
Parallel to the variable star processing, an improved version of the parallaxes (AGIS 2.2) was being produced, which is the one published in DR2.
Because of the cyclic nature of data processing within the DPAC, these improved parallaxes could not be used to compute the LPV attributes.
The values of the LPV attributes published in DR2 are therefore based on AGIS~2.1.

The number of stars with good relative parallax uncertainties has much increased in the final version (AGIS~2.2) of the parallaxes compared to the preliminary (AGIS~2.1) version.
This is shown in the top panel of Fig.~\ref{Fig:numStarsGoodParallaxes}.
The fraction of good parallaxes as a function of relative parallax uncertainty, on the other hand, can be estimated by comparing the distribution $|\epsilon(\varpi) / \varpi|$ of the sample of stars with positive parallaxes with the similar distribution, but for the sample with negative parallaxes.
These two distributions are shown in Fig.~\ref{Fig:histoParallaxRelativeError_agis2p2} of the main body of the paper for version 2.2 of the parallaxes published in DR2.
The expected percentage of spurious parallaxes can be estimated assuming similar levels of spurious parallaxes, as a function of absolute parallax, for the distributions of positive and negative parallaxes. 
The percentage of spurious parallaxes in the published data set (version AGIS~2.2) is shown as filled circles in the bottom panel of Fig.~\ref{Fig:numStarsGoodParallaxes} as a function of relative parallax uncertainty, and as open circles for the preliminary version of the parallaxes (version AGIS~2.1) used during LPV processing. 
The percentage of spurious parallaxes is seen to have significantly decreased from AGIS~2.1 to AGIS~2.2, for example, from 20\% to 3\% in the ensemble of LPV candidates with $0 < \epsilon(\varpi) / \varpi < 0.2$.

These improved parallaxes lead to much improved bolometric magnitudes compared to those published in the DR2 catalogue of LPV candidates.
Users are therefore asked to recompute the bolometric magnitudes and derived quantities (identification of red supergiants and resulting BCs) using the prescriptions
provided in Appendix~\ref{Sect:SOS}.

\end{appendix}

\end{document}